\documentclass[12pt,letterpaper]{article}

\usepackage{amsmath}
\usepackage{amsfonts}
\usepackage{amssymb}
\usepackage{graphicx}
\usepackage{color}
\usepackage{booktabs}
\usepackage{inputenc}
\usepackage[T1]{fontenc}

\usepackage{graphicx}
\usepackage{epstopdf}
\usepackage{amsmath}
\usepackage{amsfonts}
\usepackage{amssymb}
\usepackage{color}
\usepackage{mathrsfs}
\usepackage{color}
\usepackage{enumerate}

\usepackage{xcolor,url}

\usepackage{cancel,dsfont} 
\newcommand\eea{\end{eqnarray}}
\newcommand\bea{\begin{eqnarray}}

\def\be{\begin{equation}}
\def\ee{\end{equation}}

\def\nn{\nonumber}
\def\PEFT{P^{({\rm EFT})}}
\def\Pdata{P^{({d})}}

\newcommand{\dd}{\partial}
\def\dd{\partial}
\def\knl{k_{\rm NL}}

\def\hinvMpc{h\,{\rm Mpc}^{-1}}
\def\lnAs{\ln (10^{10}A_s)}

\newcommand{\kmax}{k_{\rm max }}
\newcommand{\sigmasys}{\sigma_{\rm sys}}

\usepackage[colorlinks,bookmarks]{hyperref}
\definecolor{linkblue}{rgb}{0,0,0.8}
\definecolor{linkgreen}{rgb}{0,0.5,0}

\hypersetup{pdfpagemode=UseNone, pdfstartview=FitH, linkcolor=linkblue, %
            citecolor=linkgreen, urlcolor=linkblue}

\renewcommand{\arraystretch}{1.3}

\begin{document}

\begin{center}

{\Large \bf The Cosmological Analysis of the SDSS/BOSS data\\[0.1cm]from\\[0.25cm] the Effective Field Theory of Large-Scale Structure}
\\[0.7cm]

{\large  Guido d'Amico${}^{1,2}$,  J\'er\^ome Gleyzes${}^{3}$, Nickolas Kokron${}^{4}$,   \\[0.1cm] Dida Markovic$^{5}$, Leonardo Senatore${}^{1,4}$,  
Pierre Zhang${}^{6}$,\\[0.2cm]   Florian Beutler$^{5}$, H\'ector Gil-Mar\'in$^{7}$  }
\\[0.7cm]

\end{center}

\begin{center}

{\normalsize { \sl $^{1}$ SITP, Physics Department, Stanford University, Stanford, CA 94306}}\\
\vspace{.3cm}

{\normalsize { \sl $^{2}$ Dipartimento di SMFI dell' Universita' di Parma, Parma, Italy}}\\
\vspace{.3cm}

{\normalsize { \sl $^{3}$JPL and Astronomy Department,  California Institute of Technology, Pasadena, CA 91109}}\\
\vspace{.3cm}

{\normalsize { \sl $^{4}$ 
KIPAC, SLAC and Stanford University, Menlo Park, CA 94025}}\\
\vspace{.3cm}

{\normalsize { \sl $^{5}$ Institute of Cosmology \& Gravitation,  University of Portsmouth, Portsmouth, UK}}\\
\vspace{.3cm}

{\normalsize { \sl $^{6}$ Department of Astronomy, School of Physical Sciences, CAS Key Laboratory for Research in Galaxies and Cosmology, School of Astronomy and Space Science, \\
University of Science and Technology of China, Hefei, Anhui 230026, China}}\\
\vspace{.3cm}


{\normalsize { \sl $^{7}$ ICC, University of Barcelona, IEEC-UB, Mart\'i i Franqu\`es, 1, E08028 Barcelona, Spain}}\\
\vspace{.3cm}

\vspace{.3cm}

\end{center}

\hrule \vspace{0.3cm}
{\small  \noindent \textbf{Abstract} \\[0.3cm]
\noindent The Effective Field Theory of Large-Scale Structure is a formalism that allows us to predict the clustering of Cosmological Large-Scale Structure in the mildly non-linear regime in an accurate and reliable way. After validating our technique against several sets of numerical simulations, we perform the analysis for the cosmological parameters of the DR12 BOSS data. We assume $\Lambda$CDM, a fixed value of the baryon/dark-matter ratio, $\Omega_b/\Omega_c$, and of the tilt of the primordial power spectrum,~$n_s$, and no significant input from numerical simulations. By using the one-loop power spectrum multipoles, we measure the primordial amplitude of the power spectrum, $A_s$, the abundance of matter, $\Omega_m$, and the Hubble parameter, $H_0$, {to about $13\%$, $3.2\%$ and $3.2\%$ respectively, obtaining $\lnAs=2.72\pm 0.13$, $\Omega_m=0.309\pm 0.010$, $H_0=68.5\pm 2.2$ km/(s Mpc) at 68\% confidence level. If we then add a CMB prior on the sound horizon, the error bar on $H_0$ is reduced to $1.6\%$.} These results are a substantial qualitative and quantitative improvement with respect to former analyses, and suggest that the EFTofLSS is a powerful instrument to  extract cosmological information from Large-Scale Structure.

 \vspace{0.3cm}}
\hrule


 \vspace{0.3cm}
 \newpage
 \tableofcontents

\newpage
\section{Introduction\label{sec:intro}}
 
{\bf General Remarks:} In the past few decades, cosmology has evolved from a highly speculative science to a highly precise one, where percent-level error bars are the norm. This  has happened thanks to our capability to measure the primordial density fluctuations first in the  Cosmological Large-Scale Structure (LSS), and then, dominantly, in the Cosmic Microwave Background (CMB). Such precision in the measurements has allowed cosmology to contribute to the exploration of the fundamental laws of Nature, that is, those laws from which any other law can be derived.

In order to continue this splendid journey, it is essential that we continue to measure new primordial fluctuations. The next leading source of cosmological information will come from LSS, which will be observed either directly, or indirectly through CMB lensing. There is however a tremendous challenge from extracting information from LSS: at short distances the fluctuations are large and the dynamics is extremely complicated, with stars and galaxies forming and exploding. Because modes are coupled at non-linear level, these uncontrolled short-distance non-linearities affect the long-wavelength fluctuations. This makes the theoretical modeling very challenging, so that, in practice, either only very-long wavelength data are analyzed, as in this regime the corrections from non-linear scales become either negligible or easy to model, or also short-wavelength data are analyzed, but at the cost of adding potentially large, and often uncontrolled, systematic uncertainties. These limitations come with a huge cost in terms of information that we can extract from LSS data. In fact, in everything that we measure in cosmology, with very few exceptions, information comes from measuring a high number of modes, $N_{\rm modes}$, {\it i.e.} the error bars scale as $1/\sqrt{N_{\rm modes}}$. The number of modes scales like $N_{\rm modes}\propto \kmax^3$, where $\kmax$ indicates the highest wavenumber where the theory prediction can be trusted, and, as long as the modes are not order one nonlinear, these modes are essentially uncorrelated~\footnote{There is a well-spread argument that suggests that the information does not scale as $N_{\rm modes}\propto \kmax^3$. It is often left implicit to state that one is usually referring to the information contained {\it within} the two-point function. This is true in principle, but in practice, as long as non-linearities are perturbatively small, information is spread hierarchically to the higher $N$-point functions, which can also be analyzed, as we will show later. Something even simpler is also true: in the same mildly non-linear regime, the non-linearities represent a small correction, and therefore most of the information is still in the two-point function, and it is still scaling as $\kmax^3$. Of course this discussion does not deal with the need to include parameters in our predictions that account for uncontrolled physics and that therefore could in practice limit the amount of information available. We will discuss this explicitly later on.}. By dropping the highest wavenumbers, we drop most of the information.\\

{\bf The EFTofLSS: } There are several ways in which we might try to make progress. One is the use of numerical simulations that can simulate the formation of galaxies at all wavelengths observed by the experiments. This appears to be a prohibitive task. In practice, a more realistic approach is to try to simulate dark matter dynamics exactly, and then model in some physically-motivated and data-guided way galaxy formation physics. Doing this at all wavelengths of interest appears again to be very hard, but there is hope that one can do this at long-enough wavelengths. This is a potentially fruitful approach, it has delivered many important results, but, so far, it has allowed to analyze only the data up to a somewhat limited~$\kmax$. It should definitely be pursued more, but it is not the approach we use here.

 The approach we use here is instead much less ambitious. Instead of using all the modes that are measured by a survey,  we try to analyze the ones for which the non-linearities are non-negligible, but still smaller than order one. Though this might be a humble aim, it has the potential to bring us tremendous results. In fact, if we denote the modes for which non-linearities are negligible as the linear modes, $N_{\rm lin.}$, and the ones for which they are sizable but still less than order one, as `mildly non-linear', $N_{\rm mild.\, non-lin.}$, and if we also assume that, roughly, at redshift zero, the validity of linear theory is up to $k_{\rm max}^{\rm lin.}\sim 0.1\hinvMpc$, while the mildly non-linear regime is valid up to $k_{\rm max}^{\rm mild.\, non-lin.}\sim 0.5\hinvMpc$, then we have  that the $N_{\rm mild.\, non-lin.}\sim 100\cdot N_{\rm lin.}$: even within our extremely-rough approximations, the number of modes in the mildly non-linear regime is much larger than the linear ones.

The idea behind our approach is the following. If we restrict the analysis to wavenumbers which are in the mildly non-linear regime, then it is possible to write the equations of motion for the fluctuations, within which the effect of short-distance non-linearities is parametrized in the most general way. These equations are therefore correct, {\it i.e.} cannot be wrong, unless some basic principle of physics, such as Lorentz invariance, matter conservation, etc., are violated, because they are the most general ones compatible with the aforementioned principle of physics. In fact, given a particular galactic process, there are coefficients in the equations of motion whose numerical value can be adjusted to exactly reproduce the effect of that specific galactic process at long wavelength~\footnote{The fact that this is possible to do should not be surprising. If we think of the dielectric Maxwell equations, we know that the effect on the propagation of long-wavelength photons of any atomic structure of the dielectric material, no matter how complicated that could be, is described by a proper numerical value of the dielectric constants.}. When we apply this construction, we obtain what is called the Effective Field Theory of Large-Scale Structure~(EFTofLSS)~\cite{Baumann:2010tm, Carrasco:2012cv}.

Let us give a rough overview of the EFTofLSS, referring the reader to some lectures notes~\footnote{See \href{http://stanford.edu/~senatore/}{EFTofLSS repository}.} and some relevant literature for more detailed information. In the EFTofLSS, we derive some equations of motion for the long-wavelength fields: dark matter density~\cite{Baumann:2010tm, Carrasco:2012cv,Porto:2013qua} and velocity~\cite{Carrasco:2013sva,Mercolli:2013bsa}, baryons~\cite{Lewandowski:2014rca}, neutrinos~\cite{Senatore:2017hyk}, dark energy~\cite{Lewandowski:2016yce}, etc., which, as mentioned just above, contain some terms with coefficients that are unknown and depend on the short-distance physics. In the jargon of the EFTofLSS, these coefficients are named `counterterms'. Just to give a rough idea of these equations, if the universe is filled only with dark matter, with $\delta_\ell$ being the long-wavelength overdensity, $v_\ell$ the long-wavelength velocity and $\phi_\ell$ the long-wavelength Newtonian potential, these equations read approximately
\bea\label{eq:continuity}
&&\dot\delta_\ell+\frac{1}{a}\dd_i((1+\delta_\ell)) v_\ell^i)=0\  ,\\ \label{eq:momentum}\nn
&&\dot v_\ell^i+ H v_\ell^i+\frac{1}{a}v_\ell^j\dd_j v_\ell^i+\frac{1}{a}\dd_i\phi_\ell=\int^t dt'\; \left[c_{s,1}(t,t')\, \delta_\ell(x_{\rm fl}(\vec x, t,t'),t')\right.\\ \nn
&&\qquad\qquad\qquad\qquad\qquad\qquad\qquad\quad\quad\left.+c_{s,2}(t,t')\, \delta_\ell(x_{\rm fl}(\vec x, t,t'),t')^2+\ldots\right]\ ,
\eea
where $x_{\rm fl}(\vec x, t,t')$ represents the location at time $t'$ of a fluid element that at time $t$ is at position $\vec x$, and where $\ldots$ represents in principle an infinite number of terms. $c_{s,1},\; c_{s,2},\ldots$ represent the unknown coefficients that encode the effect of short distances at long distances.

The resulting equations are solved perturbatively in powers in the fluctuations.  Schematically, we write
\be
\delta_\ell(\vec k, t)=\sum_n \delta^{(n)}_\ell(\vec k,t)\ , 
\ee 
where $\delta^{(n)}_\ell$ is of order $(\delta^{(1)}_\ell)^n$, with $\delta^{(1)}_\ell$ being the solution to the linearized equations, and we solve eq.~(\ref{eq:continuity}) to a given order in $n$, assuming $\delta_\ell^{(1)}\ll1$. Importantly, at a given perturbative order, only a {\it finite} number of counterterms needs to be kept, as, roughly, each counterterm starts contributing at a specific order in $n$. 

The perturbative solution allows us to compute correlation functions of long-wavelength fields that depend only on a finite number of unknown coefficients. Then, long-wavelength fluctuations of galaxies are described as biased tracers of the long-wavelength fluctuations of these fields. Schematically, if $\delta_{\ell, g}$ is the long-wavelength overdensity of galaxies, we have~\cite{Senatore:2014eva}
\bea
&&\delta_{\ell,g}(\vec x,t)=\int^t dt'\;\left[\bar c_1(t,t')\, \delta_\ell(x_{\rm fl}(\vec x, t,t'),t')\right. \\ \nn
&&\quad\qquad\qquad\qquad\left.+\bar c_2(t,t')\, \dd_i v_\ell^i(x_{\rm fl}(\vec x, t,t'),t')^2+\bar c_3(t,t')\, \delta_\ell(x_{\rm fl}(\vec x, t,t'),t')^2+\ldots \right].
\eea
This means that correlation functions of galaxies can be perturbatively computed as linear combination of more complicated correlation functions of matter density and velocity fields, weighted by unknown coefficients that are named  `biases' (we will also use the name `counterterms' or `EFT parameters'). Again, at a given order $n$  in the perturbative expansion, only a finite number of biases should be included. 

Finally, the overdensity of galaxies in redshift space, $\delta_{\ell, g,r}$, is given in terms of the  galaxy fields in real space by a formula that, roughly, reads~\cite{Senatore:2014vja}
\bea\label{eq:delta_redshift_1}
&&\delta_{\ell, g,r}(\vec k,t)=\delta_{\ell,g}(\vec k,t)-i \frac{k_z}{aH} v_{\ell, g}^z(\vec k,t)+\frac{i^2}{2}\left(\frac{k_z}{aH}\right)^2 [v_{\ell,g}^z(\vec x,t)^2]_{\vec k}\\ \nonumber
&&\quad-\frac{i^3}{3!}\left(\frac{k_z}{aH}\right)^3 [v_{\ell,g}^z(\vec x,t)^3]_{\vec k}-i \frac{k_z}{aH} [v_{\ell,g}^z(\vec x,t) \delta(\vec x,t)]_{\vec k}+\frac{i^2}{2}\left(\frac{k_z}{aH}\right)^2 [v_{\ell,g}^z(\vec x,t)^2\delta_{\ell,g}(\vec x,t)]_{\vec k} +\\ \nn
&&\quad +\int dt' \left(\frac{a H}{\knl}\right)^2\left[c_{r,1}(t,t') \delta_D^{(3)}(\vec k) +\left(c_{r,2}(t,t')+c_{r,3}(t,t') \frac{k_z^2}{k^2}\right) [\delta_\ell(x_{\rm fl}(\vec x,t,t'),t')]_{\vec k}\right]+\ldots\ ,
\eea
where $\delta_D^{(3)}(\vec k)$ is the Dirac three-dimensional $\delta$-function, $[\ldots]_{\vec k}$ means that we take the Fourier transform at momentum $\vec k$ of the quantity inside the brackets, and $\hat z$ is the direction of the line of sight. $\knl$ is the wavenumber associated to the non-linear scale, ${\it i.e.}$ where $\left.k^3\delta^{(1)}(k)^2\right|_{k\to \knl}\simeq 1$. Again, there is only a finite number of  unknown numerical coefficients that needs to be used for the calculation at a given order~$n$.  

Therefore we can schematically summarize the full perturbative structure as follows. Correlation functions of the density of galaxies in redshift space are obtained as linear combination of correlation functions of galaxy density and velocities in configuration space, which, in turn, are obtained as linear combinations of correlation functions of long-wavelength matter fields satisfying eq.~(\ref{eq:continuity}). These correlation functions are computed in a perturbative expansion in powers of $\delta
_\ell^{(1)}\ll1$, or, equivalently, in powers of~$k/\knl\ll1 $, with~$k$ being the wavenumber of interest. For a given order~$n$ used in perturbation theory, only a finite, albeit by now potentially large, number of free coefficients is used. 

 Therefore, the EFTofLSS is a perturbation theory for the LSS. While there is a plethora of perturbative approaches to LSS, dating back to the time of Zeldovich~\cite{Zeldovich:1969sb}, it should be stressed that there can be only one correct perturbative approach. By this we mean that there is only one correct set of  equations that describes the long-wavelength behavior of the universe. Others can be correct only either if they are just a change of variables of the same equations, and so essentially the same equations, or if they solve the same equations expanding in different subsets of parameters.
 The novelty of the EFTofLSS with respect to the other perturbative approaches is that it has free coefficients that encode the effects of short distance physics at long distances. Their numerical value is unknown to the EFTofLSS and must be provided by either a direct fit to the data, or by some knowledge of the short distance physics, {\it i.e.} of the galaxy formation physics. But it is thanks to these coefficients that the EFTofLSS can be made arbitrarily accurate\footnote{In the sense of asymptotic series such as Quantum Electrodynamics.} at long distances. Explicitly, for a given order $n$, the error associated with the lack of inclusion of the higher order terms scales as $(k/\knl)^{\gamma \cdot n}$, with $\gamma$ being an order one number. This is known as {\it systematic theoretical error}~\cite{Carrasco:2012cv,Carrasco:2013sva,Senatore:2014vja,Foreman:2015uva,Baldauf:2016sjb}. So, for a given required precision and a given wavenumber, one has an idea, albeit rough, of what order $n$ to choose and how large the mistake of the prediction can be. It is the presence of the unknown numerical coefficients that makes these bold-sounding claims correct, though it comes at the cost of harming at some level the predictivity of the theory. But these coefficients are as non-vanishing as the viscosity for honey is a non-vanishing coefficient. One could use a more predictive theory to simulate honey's motion: a fluid without viscosity. The theory would be more predictive, as it would have one less parameter to measure from experiment, but would be wrong. This is why the EFTofLSS should be used.

Though the EFTofLSS provides the only correct set of equations for a perturbative approach to the large-scale clustering of LSS, it should be thought of as the result of an evolution of many attempts to  formulate a consistent set of equations and a consistent perturbative approach to solve them and predict the clustering of LSS. Here we name, as an example, a few of the perturbative approaches that were developed prior to the EFTofLSS and that were essential for the development of the EFTofLSS, and we list a very-small sample of the relevant references: Zeldovich approximation~\cite{Zeldovich:1969sb}, SPT~\cite{Goroff:1986ep,Bernardeau:2001qr},  LPT~\cite{Matsubara:2008wx}, RPT~\cite{Crocce:2005xy}, MPTbreeze~\cite{Bernardeau:2008fa,Crocce:2012fa}, RegPT~\cite{Bernardeau:2011dp}, CLPT~\cite{Carlson:2012bu},  combined Perturbation Theory, Halo Model and Simulations~\cite{Hand:2017ilm}, etc..\\

{\bf Application to data:} In this paper, we apply the EFTofLSS to the cosmological data. More specifically, we compare the EFTofLSS prediction for the monopole and quadrupole of the power spectrum of galaxies in redshift space at one-loop order ({\it i.e.} at perturbative order $n=3$), and for the monopole of the bispectrum of galaxies in redshift space at tree level ({\it i.e.} $n=2$), to the DR12 sample of galaxies as measured in the BOSS/SDSS data~\cite{Alam:2016hwk,Beutler:2016ixs,Beutler:2016arn,Gil-Marin:2014sta}. We use such a comparison to infer a measurement of {\it each} of the following cosmological parameters: the amplitude of the primordial power spectrum, $A_s$, the fraction of matter of the universe, $\Omega_m$, and the present value of the Hubble constant,~$h$. For convenience, we will work in $\Lambda$CDM cosmology and fix the tilt $n_s$ and the ratio of baryons versus dark matter. A reason of concern for the predictive power of the EFTofLSS is  the potentially-large number of parameters its predictions depend on. Nevertheless, we believe our results show that very substantial qualitative and quantitative gains in extracting information from the LSS data can be achieved by employing the EFTofLSS in an analysis.

 This paper is the completion of a long journey for the EFTofLSS, which started from the initial development of the theory~\cite{Baumann:2010tm, Carrasco:2012cv,Porto:2013qua} and involved many important steps in order for it to be compared to the data. Here we mention a few of them in order to give a broad orientation within the relevant literature to the interested reader.  The dark matter power spectrum has been computed at one-, two- and three-loop orders and compared with simulations in~\cite{Carrasco:2012cv, Carrasco:2013sva, Carrasco:2013mua, Carroll:2013oxa, Senatore:2014via, Baldauf:2015zga, Foreman:2015lca, Baldauf:2015aha, Cataneo:2016suz, Lewandowski:2017kes,Konstandin:2019bay}. This also involved some additional theoretical developments, such as a careful understanding of renormalization~\cite{Carrasco:2012cv,Pajer:2013jj,Abolhasani:2015mra} (including some rather-subtle aspects such as lattice-running~\cite{Carrasco:2012cv}) and of the way of extracting the numerical value of the counterterms from simulations~\cite{Carrasco:2012cv,McQuinn:2015tva}, of the non-locality in time~\cite{Carrasco:2013sva, Carroll:2013oxa,Senatore:2014eva}, and of some subtleties with the velocity field~\cite{Carrasco:2013sva,Mercolli:2013bsa}. These theoretical explorations also include a study in 1+1 dimensional spacetime~\cite{McQuinn:2015tva}. In order to correctly describe the baryon acoustic oscillation (BAO) peak, an IR-resummation of the long displacement fields had to be performed. This has led to the so-called IR-resummed EFTofLSS~\cite{Senatore:2014vja,Baldauf:2015xfa,Senatore:2017pbn,Lewandowski:2018ywf,Blas:2016sfa}.  A method to describe baryons and baryonic effects has been developed in~\cite{Lewandowski:2014rca}, and was shown to work extremely well when compared with simulations. Concerning the computation and comparison with simulations of higher $n$-point functions, the dark-matter bispectrum has been computed at one-loop in~\cite{Angulo:2014tfa, Baldauf:2014qfa} (and compared with simulations), the one-loop trispectrum in~\cite{Bertolini:2016bmt}, 
the dark matter displacement field in~\cite{Baldauf:2015tla} (and compared with simulations). Understanding the dark matter dynamics has allowed us to compute the first observable, lensing~\cite{Foreman:2015uva}. Passing to biased tracers, they have been studied in the context of the EFTofLSS in~\cite{ Senatore:2014eva, Mirbabayi:2014zca, Angulo:2015eqa, Fujita:2016dne, Perko:2016puo, Nadler:2017qto} (see also~\cite{McDonald:2009dh}), the halo power spectrum and bispectrum including all cross correlations with the dark matter field have been studied and compared with simulations in~\cite{Senatore:2014eva, Angulo:2015eqa}. Redshift space distortions have been first developed for dark matter~\cite{Senatore:2014vja, Lewandowski:2015ziq}, and then applied to biased tracers in~\cite{Perko:2016puo}. Some potentially present subleading ingredients of the universe have also been included in the EFTofLSS, such as clustering dark energy \cite{Lewandowski:2016yce,Lewandowski:2017kes,Cusin:2017wjg,Bose:2018orj}, primordial non-Gaussianities \cite{Angulo:2015eqa, Assassi:2015jqa, Assassi:2015fma, Bertolini:2015fya, Lewandowski:2015ziq, Bertolini:2016hxg}, and neutrinos~\cite{Senatore:2017hyk,deBelsunce:2018xtd}. 
 
 With the completion of~\cite{Perko:2016puo}, the one-loop power spectrum of biased tracers in redshift space in $\Lambda$CDM cosmology had been computed and compared to simulations. At that point, the EFTofLSS had become ready to be compared with observational data of galaxy clustering. This is where the journey of this paper begins.\\

 \section{Summary of the Theory model\label{sec:modeltheory}}

\font\BF=cmmib10
\def\k{{\hbox{\BF k}}}
\def\q{{\hbox{\BF q}}}
\def\z{{\hbox{\BF z}}}

\subsection{Biased tracers in redshift space}

With a handful of parameters, the EFTofLSS can describe the gravitational clustering of galaxies in redshift space at sufficiently long wavelengths with arbitrary accuracy. This is done by adding suitable counterterms, whose form is dictated by the symmetries obeyed by LSS, that are able to correct for the sensitivity to uncontrolled short-distance wavelengths that affect the perturbative expression. For completeness, we here review the basic equations of the EFTofLSS that we use  to analyze the data.

 In the standard very accurate approximation where the time dependence of the diagrams is approximated by the growth factors, at one loop, the galaxy power spectrum in redshift space depends on the cosmology and the redshift but also on $10$ free parameters \cite{Perko:2016puo}. There are $4$ galaxy biases $b_i$ entering in the one-loop expression~\footnote{While at third order in perturbations the galaxy fields depend initially on $8$ bias parameters \cite{Angulo:2015eqa}, $4$ are enough to describe the galaxy power spectrum at one loop due to  some accidental degeneracies in the $P_{13}$ diagram \cite{Perko:2016puo}.}. Those biases are introduced when the galaxy density and velocity fields are expanded in terms of the underlying dark matter density and velocity fields that satisfy the effective `smoothed' equations of motions for the long wavelength modes \cite{Senatore:2014eva}. These biases and $3$ additional counterterms absorb the  sizeable sensitivity of the one-loop expressions to the uncontrolled short-distance non-linearities that enter in the kernel of the convolution integrals. One captures the nonlinear effects of the short-scale physics on long-wavelength dark matter modes, parametrized by the so-called effective `sound speed' of dark matter $c_\text{ct}$~\cite{Carrasco:2012cv}  and that, at this order, is degenerate with  higher-derivative terms of the bias derivative expansion. Two additional counterterms, parametrized by $c_{r,i}$, with $i=\{1,2\}$, cancel the UV-dependence arising from redshift-space distortions \cite{Senatore:2014vja}. Finally, there are $3$ stochastic terms $c_{\epsilon,i}$ to account for the difference between the actual realization of the galaxy fields in our universe and their expectation values taken in the computation of the observables~\cite{Perko:2016puo} (the first of them is the well-known shot-noise term). They also correct the UV-dependence of some perturbative diagrams.
 
 The one-loop redshift-space galaxy power spectrum reads~\footnote{In the following we make the time-dependence implicit in the notation when it is clear from the context. In practice all observables need to be computed at all the effective redshifts, $z_\text{eff}$'s, of the survey sample.}~\cite{Perko:2016puo}:
\begin{align}\label{eq:powerspectrum}
P_{g}(k,\mu) & =  Z_1(\mu)^2 P_{11}(k) \nonumber \\
& + 2 \int \frac{d^3q}{(2\pi)^3}\; Z_2(\q,\k-\q,\mu)^2 P_{11}(|\k-\q|)P_{11}(q) + 6 Z_1(\mu) P_{11}(k) \int\, \frac{d^3 q}{(2\pi)^3}\; Z_3(\q,-\q,\k,\mu) P_{11}(q)\nonumber \\
& + 2 Z_1(\mu) P_{11}(k)\left( c_\text{ct}\frac{k^2}{{ k^2_\textsc{m}}} + c_{r,1}\mu^2 \frac{k^2}{k^2_\textsc{m}} + c_{r,2}\mu^4 \frac{k^2}{k^2_\textsc{m}} \right) + \frac{1}{\bar{n}_g}\left( c_{\epsilon,1}+c_{\epsilon,2}\frac{k^2}{k_\textsc{m}^2} + c_{\epsilon,3} f\mu^2 \frac{k^2}{k_\textsc{m}^2} \right).
\end{align}
Here we set  $k^{-1}_\textsc{m}$, which controls the bias derivative expansion, to be $\lesssim k^{-1}_\textsc{nl}$, which is the scale controlling the expansion of the dark matter  derivative expansion. $\bar{n}_g$ is the mean galaxy density. Providing estimates for these physical parameters, all free parameters in the EFT are expected to be order $\mathcal{O}(1)$ \footnote{ More exactly, for the type of galaxies that are of interest here, all EFT parameters are expected to be of order of the linear bias $b_1$, which is $\sim \mathcal{O}(2)$, see for example~\cite{Perko:2016puo, Nadler:2017qto}.}. This fact is of some importance when one is only interested in the cosmological parameters and wants to marginalize over the others. We will build on this consideration for the partial marginalization procedure described in section~\ref{bias:marginalziation}.

The redshift-space galaxy density kernels are given by: 
\begin{align}\label{eq:redshift_kernels}
    Z_1(\q_1) & = K_1(\q_1) +f\mu_1^2 G_1(\q_1) = b_1 + f\mu_1^2,\\ \nonumber
    Z_2(\q_1,\q_2,\mu) & = K_2(\q_1,\q_2) +f\mu_{12}^2 G_2(\q_1,\q_2) + \, \frac{1}{2}f \mu q \left( \frac{\mu_2}{q_2}G_1(\q_2) Z_1(\q_1) + \text{perm.} \right),\\ \nonumber
    Z_3(\q_1,\q_2,\q_3,\mu) & = K_3(\q_1,\q_2,\q_3) + f\mu_{123}^2 G_3(\q_1,\q_2,\q_3) \nonumber\\ \nonumber
    & + \frac{1}{3}f\mu q \left(\frac{\mu_3}{q_3} G_1(\q_3) Z_2(\q_1,\q_2,\mu_{123}) + \frac{\mu_{23}}{q_{23}}G_2(\q_2,\q_3)Z_1(\q_1)+ \text{cyc.}\right),
\end{align}
where here $\mu= \q \cdot \hat{\z}/q$, $\q = \q_1 + \dots +\q_n$, and $\mu_{i_1\ldots  i_n} = \q_{i_1\ldots  i_n} \cdot \hat{\z}/q_{i_1\ldots  i_n}$, $\q_{i_1 \dots i_m}=\q_{i_1} + \dots +\q_{i_m}$, with $\hat{\z}$ being the unit vector in the direction of the line of sight, and $n$ is the order of the kernel $Z_n$. $K_i$ and $G_i$ are the galaxy density and velocity kernels, respectively, given in Appendix~\ref{sec:galaxykernels}, and $f$ is the logarithmic growth rate, which can be calculated by first solving for the linear growth factor, $D$, yielding:
\begin{equation}
D(a)=\frac{5}{2} \Omega_m H_0^2 H(a) \int_0^a \,  \frac{da'}{a'^3H(a')^3} \ ,
\end{equation} 
from which $f$ follows:
\begin{equation}\label{eq:fdef}
f(a) \equiv \frac{d \log D(a)}{ d \log a} = \frac{(5a-3D(a))\Omega_m}{2D(a) (\Omega_m+a^3(1-\Omega_m))}\ .
\end{equation} 
The form used here in~(\ref{eq:redshift_kernels}) is a re-writing with different notations of the eexpressions in~\cite{Perko:2016puo} (see eq.~(3.23) there).

The bispectrum in redshift space is a function of five kinematic variables. Three describe the shape of the triangle, such as two sides $k_1$, $k_2$ and the angle between them $x = \cos \theta = \hat{\k}_1 \cdot \hat{\k}_2 = -\tfrac{k_1^2+k_2^2-k_3^2}{2 k_1 k_2}$. The orientation of the triangle with respect to the line of sight $\hat{\z}$ is characterized by two variables, such as the polar angle $\omega$ and the azimuthal angle $\phi$ of $\hat{\k}_1$, or equivalently by $\mu_1 = \cos \omega = \hat{\k}_1 \cdot \hat{\z} $ and $\mu_2 = {\hat{\k}_2 \cdot \hat{\z} }=\mu_1 \cos \theta - \sqrt{1-\mu_1^2}\sin \theta \cos \phi$. 

The tree-level galaxy bispectrum in redshift space reads
\begin{equation}\label{eq:treelevelbispectrum}
    B_g(k_1,k_2,k_3,\mu_1,\mu_2) = 2 Z_2(\k_1,\k_2,\mu_{12}) Z_1(\k_1) Z_1(\k_2) P_{11}(k_1) P_{11}(k_2) + \frac{c_{\epsilon,4}}{\bar{n}_g}P_{11}(k_1)  + \text{cyc.} + \frac{c_{\epsilon,1}^2}{\bar{n}_g^2},
\end{equation}
where one needs to introduce only one extra (stochastic) parameter, $c_{\epsilon,4}$, with respect to the power spectrum. Indeed, using a basis of renormalized bias coefficients allows one to identify the constant piece of the bispectrum as the square of the constant piece of the power spectrum~\cite{Angulo:2015eqa}. 

For an efficient exploration of the cosmology dependence of the power spectrum, one can expand the EFTofLSS prediction as a sum of terms that are bias- and redshift-independent. Symbolically, the EFTofLSS power spectrum can be written as:
\begin{equation}\label{eq:factorizedpowerspectrum}
    P_g(k,z) = \sum_n  \mu^{2\alpha_n} f(z)^{\beta_n} b_{i_n}(z)^{\gamma_n} b_{j_n}(z)^{\delta_n} D(z)^{2\rho_n} P_{n}(k)\ ,
\end{equation} 
where $\{\alpha_n, \beta_n\} \in\{ 0, 1,2,3,4\}$ (redshift-space distortions), $\{\gamma_n,\delta_n \}\in\{ 0,1\}$ (biasing), $\rho_n \in\{ 1,2\}$ (linear or loop term), with the pieces $P_n(k)$ depending only on the cosmological parameters. 
While the linear power spectrum and the counterterms are straightforward to obtain, efficient evaluations of the loop integrals are needed to explore the cosmological parameter space in a timely fashion.
To this extent, we adopt two computational strategies, a numerical one \cite{Cataneo:2016suz} and an analytical one relying on the FFTLog decomposition of the linear power spectrum \cite{Simonovic:2017mhp}, described in appendix \ref{sec:Loopeval}. Proceeding in this way allows us to keep robust control over numerical instabilities while adequate performances can be reached for practical purposes.
The description of biased tracers in redshift space would not be complete without resumming the long wavelength displacements whose effect can be accurately described in the Eulerian perturbative treatment outlined above only at very high perturbative order. We adopt the original IR-resummation scheme developed in \cite{Senatore:2014via}, further extended to redshift space with controlled approximations in \cite{Senatore:2014vja,Lewandowski:2015ziq}. Details on the IR-resummation can be found in Appendix~\ref{sec:IR-resummation}~\footnote{With this publication we release a C++ code that computes the IR-resummed one-loop power spectrum multipoles and tree-level bispectrum monopole of biased tracers in the EFTofLSS, called CBiRd: Code for Biased tracers in Redshift space at 
\href{https://github.com/pierrexyz/cbird}{CBiRd GitHub} 
(see also for a repository of all EFTofLSS codes \href{http://stanford.edu/~senatore/}{EFTofLSS repository}).}.

Finally, we make a comment on additional higher order terms. In the EFTofLSS, each term contributes with a typical size controlled, roughly, by an expansion in~$k/\knl$ (see for example~\cite{Senatore:2014vja} for details). Given that the power spectrum, around the $k$'s of interest for us, scales approximately as $P_{11}(k)\sim\frac{1}{\knl^3}\left(k/\knl\right)^{-2}$, a term of order $\sim k^4P_{11}(k)$ is roughly comparable to the $k^2$ stochastic term, and could therefore be included in our model prediction. We will do this in this paper to test the dependence on this additional counterterm. We find that the numerical and observational data prefer the stochastic $k^2$ term, while the $k^4P_{11}$ is hardly measurable. We therefore do not include it in the final analysis.

\subsection{Alcock-Paczynski effect}

To estimate the galaxy spectra from data, a reference cosmology (usually denoted as `fiducial' cosmology) is assumed to transform the measured redshifts and celestial coordinates into three-dimensional cartesian coordinates. The difference between the reference cosmology and the true cosmology (usually denoted as `observed' cosmology) produces a geometrical distortion known as the Alcock-Paczynski (AP) effect \cite{Alcock:1979mp}. The wavenumbers transverse $k_\perp$ and parallel $k_\parallel$ to the direction of the line-of-sight get distorted, such that the measured wavenumbers assuming a reference cosmology, denoted with a subscript `ref', relate to the true ones as~(see for example \cite{Hand:2017ilm}):
\begin{equation}
    k^{\rm ref.}_\perp = q_\perp k_\perp, \qquad k^{\rm ref.}_\parallel = q_\parallel k_\parallel,
\end{equation}
where the distortion parameters are given by:
\begin{equation}
    q_\perp = \frac{D_A(z_\text{eff}) H(z=0)}{D^{\rm ref.}_A(z_\text{eff}){H^{\rm ref}}(z=0)}, \qquad q_\parallel = \frac{{H^{\rm ref}}(z_\text{eff}) /{H^{\rm ref}}(z=0)}{H(z_\text{eff})/H(z=0)}.
\end{equation}
and wavenumbers are measured in units of $h/{\rm Mpc}$. Here $H(z)$ is the Hubble parameter, and the angular-diameter distance $D_A(z)$ is defined as:
\begin{equation}
    D_A(z) = \frac{1}{1+z}\int_0^z \, \frac{dz'}{H(z')}.
\end{equation}

When exploring the cosmological parameters, the distorsion parameters are calculated at each probed cosmology using the usual formula for $H$ and $D_A$~(\footnote{The procedure we describe here is quite different from other analyses present in the literature that usually assume a fixed cosmology for their template and measure the cosmology and redshift-space distortions through the AP effect (see e.g. \cite{Hand:2017ilm}, \cite{Beutler:2016arn}, \cite{Gil-Marin:2016wya}), where $q_\perp$ and $q_\parallel$ are taken as free parameters to fit. We rather choose to evaluate our theoretical predictions in each step of the regression such that the AP effect is a completely deterministic effect. The exploration of the parameter space stems from the cosmology and redshift dependence of the observables, for which no unnecessary degeneracies are introduced by an approximate modeling.}). Note that since the wavenumbers $k$ are measured in [$h$/Mpc], the AP parameters $q_\perp$ and $q_\parallel$ are independent of $h$.

The power spectrum in eq.~\eqref{eq:factorizedpowerspectrum} is expanded in Legendre polynomials $\mathcal{P}_{\ell}(\mu)$ and its multipole moments read: 
\begin{equation}
    P^\ell_n(k^{\rm ref})=\frac{2\ell+1}{2q_\parallel q_\perp^2}\int_{-1}^1\, d\mu^{\rm ref}\, \mu( \mu^{\rm ref})^{2 \alpha_n} P_n\left(k(k^{\rm ref}, \mu^{\rm ref})\right)\mathcal{P}_{\ell}(\mu^{\rm ref})\ ,
\end{equation}
where the true $({k},{\mu})$ are related to the reference $(k^{\rm ref},\mu^{\rm ref})$ by \cite{Ballinger:1996cd}:
\begin{align}
    {k} & = \frac{ k^{\rm ref}}{q_\perp}\left[ 1 + (\mu^{\rm ref}){}^2\left( \frac{1}{F^2} - 1 \right) \right]^{1/2} , \label{eq:distordedkAP}\\
    {\mu} & = \frac{ \mu^{\rm ref}}{F}\left[ 1 + (\mu^{\rm ref})^2\left( \frac{1}{F^2} - 1 \right) \right]^{-1/2},\label{eq:distordedmuAP}
\end{align}
where $F = q_\parallel/q_\perp$.

For the bispectrum, we focus on the monopole, given by~\cite{Gil-Marin:2016wya}:
\begin{equation}\label{eq:bispectrummonopole}
    B_0( k_1^{\rm ref}, k_2^{\rm ref}, k_3^{\rm ref}) = \frac{2 \cdot 0 +1 }{2 q_\parallel^2 q_\perp^4}\int_{-1}^{1} \, d\mu^{\rm ref}_1 \, \int_{0}^{2\pi} \ d \phi^{\rm ref} \; B_g({k}_1,{k}_2,{k}_3,{\mu_1},{\mu_2}),
\end{equation}
where the true components $(k_i,\mu_i)$ are again related to the respective fiducial ones   $(k^{\rm ref}_i,\mu^{\rm ref}_i)$ by  eq.~\eqref{eq:distordedkAP} and~\eqref{eq:distordedmuAP}~\cite{Song:2015gca}.

 \section{Technical Aspects of Analysis\label{sec:technical-likl}}
 
 Given our theory model for the clustering of LSS as specified by the EFTofLSS, it is, at least in principle, a well-established procedure to perform the analysis of the data. We sample the posterior distribution using an MCMC routine, in a standard way that we review later in the section. However, the use of the EFTofLSS enables us to perform some interesting manipulations that simplify the analysis. We therefore describe them in some detail. In this section we focus on the analysis that includes only the power spectrum. The manipulations when we include the bispectrum are very similar, and we describe them in detail in App.~\ref{sec:appendix_bispectrum_lik}.
 
 We denote the data by $d$ and the model parameters by $\{\vec\Omega,\vec b\}$, $\vec \Omega$ being the cosmological parameters, $\vec b$ being the EFT parameters. Explicitly, 
 \be
 \vec b=\{b_1,b_2,b_3,b_4,c_{\rm ct},c_{r,1},c_{r,2},c_{\epsilon ,1},c_{\epsilon ,2},c_{\epsilon ,3}, c_{\epsilon ,4}\}\ .
 \ee
 Similarly, we denote  the multipole $\ell$ of the power spectrum of the data as $\Pdata_{\ell}$.
 The Likelihood of the data given the model parameters is assumed to be a Gaussian distribution in the residuals:
 \bea\label{eq:likelyhoodone}
&&\!\!\!\!{\cal L}(d|\{\vec\Omega,\vec b\})=\\ \nn
&&\!\!\!\!={\rm Exp}\left[-\frac{1}{2}(\PEFT_{\ell}{}^{(W)}(k,\{\vec\Omega,\vec b\})-\Pdata_\ell (k)) \cdot C^{-1}(k,k')_{\ell,\ell'}\cdot (\PEFT_{\ell'}{}^{(W)}(k',\{\vec\Omega,\vec b\})-\Pdata_{\ell'} (k'))\right], 
 \eea
 where repeated indexes are summed over.
 Here $C^{-1}(k,k')_{\ell,\ell'}$ is the inverse of the covariance matrix of the data, and  $\PEFT_{\ell}{}^{(W)}(k)$ is the EFT power spectrum after being convolved with the window function of the survey:
 \be\label{eq:windowzation}
 \PEFT_{\ell}{}^{(W)}(k)=W(k,k')_{\ell,\ell'} \cdot \PEFT_{\ell'}(k') \ ,
\ee
where $W(k,k')_{\ell,\ell'} $ is the window function in Fourier space. 

\subsection{Window function\label{sec:window}}

The real space correlation function multipoles $\xi_\ell(s)$ after application of the survey window function are given in terms of the configuration-space window functions, $Q_{\ell,\ell'}(s)$, by \cite{Wilson:2015lup,Beutler:2016arn,Beutler:2018vpe}:
 \be\label{eq:windowzation2}
 \xi_{\ell}{}^{(W)}(s)=Q_{\ell,\ell'}(s) \cdot \xi_{\ell'}(s) \ .
\ee
Explicitly, $Q_{\ell,\ell'}(s) \equiv C_{\ell,\ell',\ell''}\, Q_{\ell''}(s)$, where:
\begin{equation}
C_{0,\ell',\ell''} = 
\begin{bmatrix}
    1 & 0 & 0 \\
    0 & \tfrac{1}{5} & 0 \\
    0 & 0 & \tfrac{1}{9}
\end{bmatrix}_{\ell',\ell''}, \quad
C_{2,\ell',\ell''} = 
\begin{bmatrix}
    0 & 1 & 0 \\
    1 & \tfrac{2}{7} & \tfrac{2}{7} \\
    0 & \tfrac{2}{7} & \tfrac{100}{693}
\end{bmatrix}_{\ell',\ell''}, \quad
C_{4,\ell',\ell''} = 
\begin{bmatrix}
    0 & 0 & 1 \\
    0 & \tfrac{18}{35} & \tfrac{20}{77} \\
    1 & \tfrac{20}{77} & \tfrac{162}{1001}
\end{bmatrix}_{\ell',\ell''}\ ,
\end{equation}
 and the $Q_{\ell}(s)$ are given in~\cite{Wilson:2015lup,Beutler:2016arn,Beutler:2018vpe}. The window functions for the power spectrum, $W(k,k')_{\ell,\ell'}$ of~(\ref{eq:windowzation}), are then given by:
\be\label{eq:windowfourierr}
W(k,k')_{\ell,\ell'}=\frac{2}{\pi}(-i)^\ell i^{\ell'} k'^2\int ds \,s^2\; j_\ell(k s)\, Q_{\ell,\ell'}(s)\cdot  j_{\ell'}(k' s) \ . 
\ee
We defer to App.~\ref{sec:appendix_bispectrum_lik} to discuss how we apply the window function for the bispectrum. 

\begin{figure}[h]
\centering
\includegraphics[width=0.495\textwidth,draft=false]{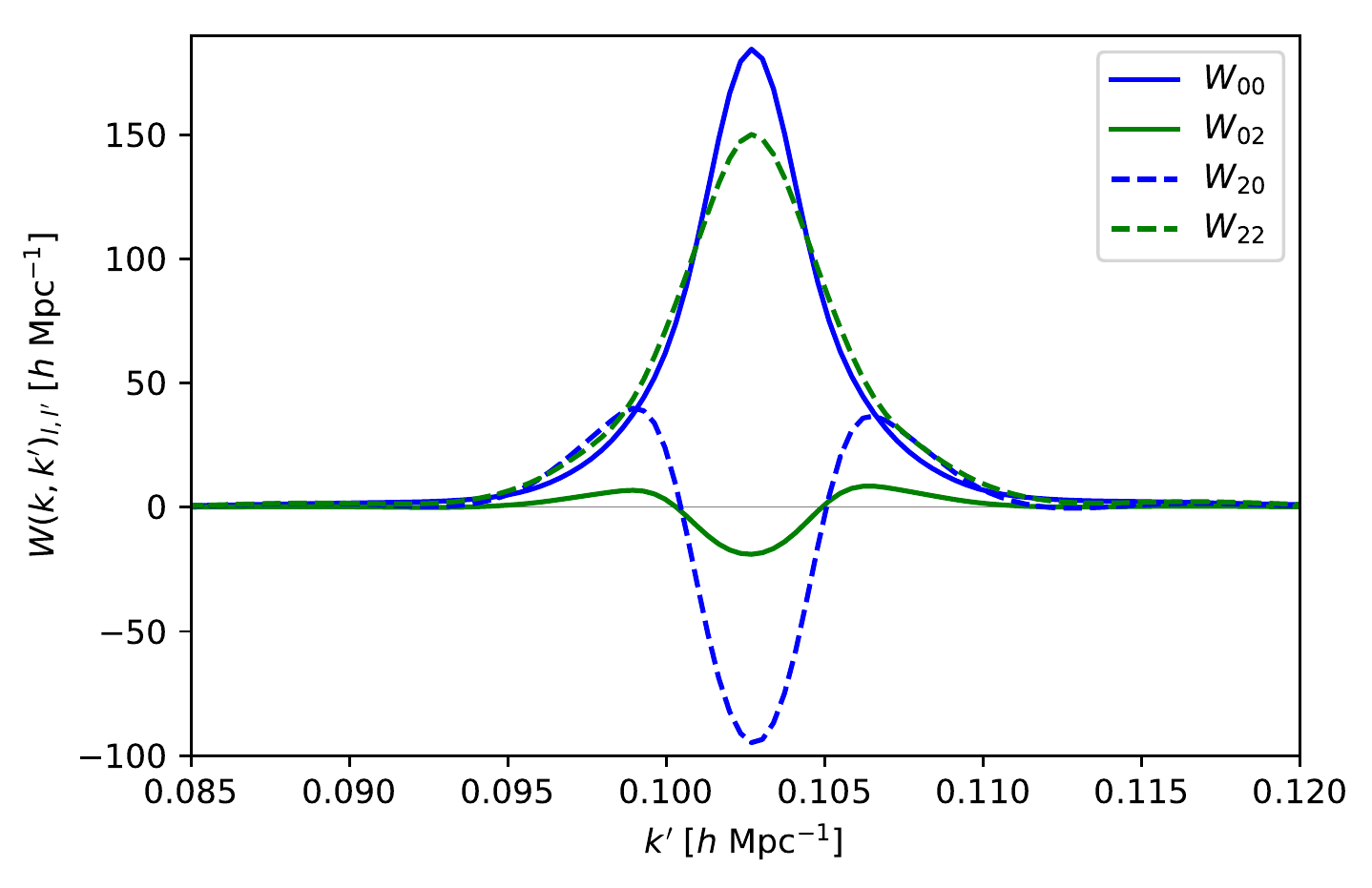}
\caption{\small Fourier-space window function $W(k,k')_{\ell,\ell'}$, for $k=0.1023\hinvMpc$. One easily sees that indeed, being the window function a multiplication in real space, it is a convolution in Fourier space. One can also see that the computation of $W(k,k')_{\ell,\ell'}$ does not seem to present ringing or other numerical issues.}
\label{fig:windowFourier}
\end{figure}

Computationally, it is quite straightforward to pre-compute $W(k,k')_{\ell,\ell'}$ for the values of interest. In fact, for each couple $\{k,\ell\}$, eq.~(\ref{eq:windowfourierr}) can be computed with a simple Bessel-transform of $j_\ell(k s)\, Q_{\ell,\ell'}(s)$. Therefore, $W(k,k')_{\ell,\ell'}$ can be computed by evaluating $N_\ell\times N_k$ Bessel transforms, with $N_\ell$ being  the number of multipoles, and $N_k$ being the one of $k$'s ($N_\ell=3,\; N_k=68$ in the case of the analysis presented in this paper). We plot $W(k,k')_{\ell,\ell'}$ for $k\simeq0.1\hinvMpc$ in Fig.~\ref{fig:windowFourier}. One easily sees that indeed, being the window function a multiplication in real space, it is a convolution in Fourier space. One can also see that the computation of $W(k,k')_{\ell,\ell'}$ does not present  numerical challenges, and, on the linear power spectrum, we have checked that  we obtain the same result on the power spectrum as the standard procedure which acts on configuration space.

\begin{figure}[h]
\centering
\includegraphics[width=0.495\textwidth,draft=false]{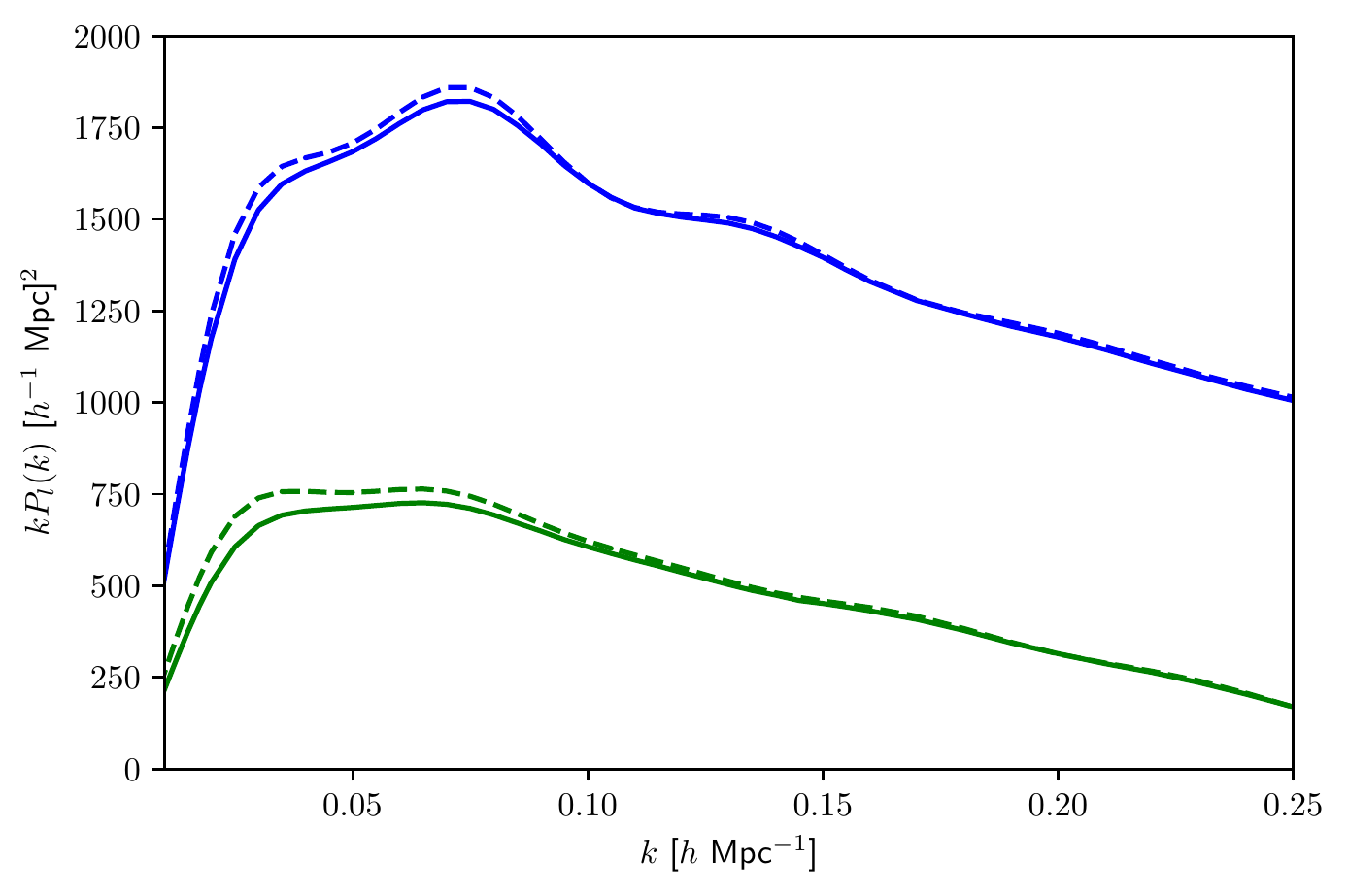}
\caption{\small Typical power spectra before ({\it dashed}) and after ({\it continous}) the application of the window function, using the Fourier-space formula of~(\ref{eq:windowfourierr}). Blue is monopole, green is quadrupole. As expected, the window function suppresses the power spectra at low wavenumbers. We also see that there seem not to appear any numerical artifacts, such as oscillations and ringing.}
\label{fig:windowPk}
\end{figure}

Now that we have $W(k,k')_{\ell,\ell'}$ at our disposal, we can follow two approaches. If one uses a grid of cosmological predictions, it is possible to apply directly (\ref{eq:windowzation}) to the points of the grid, so that the evaluation-time in each step in the MCMC is shortened as the  window function is applied only once, directly at the level of the grid, for a given choice of the model parameters. The result of the application of the window function to $P_{\ell}(k)$ is given in Fig.~\ref{fig:windowPk}. This is the approach we will use in this paper. 

Alternatively, one can notice that one can expand eq.~(\ref{eq:likelyhoodone}) and write:
\bea\label{eq:likelyhoodtwo}
&&{\cal L}(d|\{\vec\Omega,\vec b\})={\rm Exp}\left[- \frac{1}{2}\PEFT_{\ell}(k,\{\vec\Omega,\vec b\}) \cdot C^{-1}_{WW}(k,k')_{\ell,\ell'}\cdot \PEFT_{\ell'}(k',\{\vec\Omega,\vec b\})+\right.\\ \nn
&&\  \ \qquad \quad\quad\left.+\PEFT_{\ell}(k,\{\vec\Omega,\vec b\}) \cdot C^{-1}_{W}(k,k')_{\ell,\ell'}\cdot \Pdata_{\ell'}(k')-\frac{1}{2}\Pdata_{\ell}(k) \cdot C^{-1}(k,k')_{\ell,\ell'}\cdot \Pdata_{\ell'}(k')\right] \ ,
\eea
 where
 \bea\label{eq:windowedcovmatrix}
&&  C^{-1}_{WW}(k_1,k_2)_{\ell_1,\ell_2}=W(k_1',k_1)_{\ell_1',\ell_1}{}^{\rm \dag} \cdot  C^{-1}(k_1',k_2')_{\ell_1',\ell_2'} \cdot W(k_2',k_2)_{\ell_2',\ell_2}\ , \\ \nn
&&  C^{-1}_{W}(k_1,k_2)_{\ell_1,\ell_2}=W(k_1',k_1)_{\ell_1',\ell_1}{}^{\rm \dag} \cdot  C^{-1}(k_1',k_2)_{\ell_1',\ell_2}\ ,
\eea
where $\dag$ means hermitian conjugate and applies to both the $\{k,k'\}$ and $\{\ell,\ell'\}$ indexes. In practice, $C^{-1}_W$ is obtained by a left multiplication of $C^{-1}$ with $W^\dag$, while $C^{-1}_{WW}$ is obtained by a left and right multiplication of $C^{-1}$ with $W^\dag$ and $W$ respectively.  Notice that eq.~(\ref{eq:likelyhoodtwo}) has drastically simplified the implementation of the window function. Once $W(k,k')_{\ell,\ell'}$ has been computed, $C^{-1}_W$ and $C^{-1}_{WW}$ in eq.~(\ref{eq:windowedcovmatrix}) can be pre-computed. So, there is no need to apply the window function at each step of the MCMC. 

In terms of computational cost, this advantage compensates for the fact that now to evaluate eq.~(\ref{eq:likelyhoodtwo}) at each step of the MCMC, we have to evaluate two contractions of the power spectra with the covariance matrix: one for the term in $C^{-1}_W$ and one for the term in $C^{-1}_{WW}$. The term in $C^{-1}$ does not depend on the cosmological parameters and so can be precomputed. This is in contrast with the evaluation of eq.~(\ref{eq:likelyhoodone}), which, after $\PEFT_{\ell}{}^{(W)}$ has been computed, has only one contraction. However, the computational advantage of eq.~(\ref{eq:likelyhoodtwo}) is quite large.

In the case of the EFTofLSS, the advantage of using eq.~(\ref{eq:windowzation}) or eq.~(\ref{eq:likelyhoodtwo}) is particularly important  also for another reason. In the standard approach of applying the window function, one takes $\PEFT_{\ell}(k)$, Bessel transforms it to obtain $\xi^{(\rm EFT)}_\ell(s)$, applies $Q_{\ell,\ell'}(s)$, that acts diagonally in $s$ space, and Bessel transforms it back to obtain  $\PEFT_{\ell}{}^{(W)}(k)$. The EFTofLSS has the peculiarity that it is able to reach arbitrarily large accuracy at low wavenumbers, and this is achieved somewhat at the cost of producing large mistakes at large wavenumbers, where the theory is not reliable. These mistakes  take the form of steep functions that make it particularly hard to Fourier transform $\PEFT_{\ell}(k)$. It is therefore challenging to go back-and-forth from $\PEFT_{\ell}(k)$ to $\xi^{(\rm EFT)}_\ell(s)$. Being able to use the window function in Fourier space avoids this difficulty.
 
\subsection{Redshift selection effects\label{sec:redshiftselection}}
 
\begin{figure}[h]
\centering
\includegraphics[width=0.495\textwidth,draft=false]{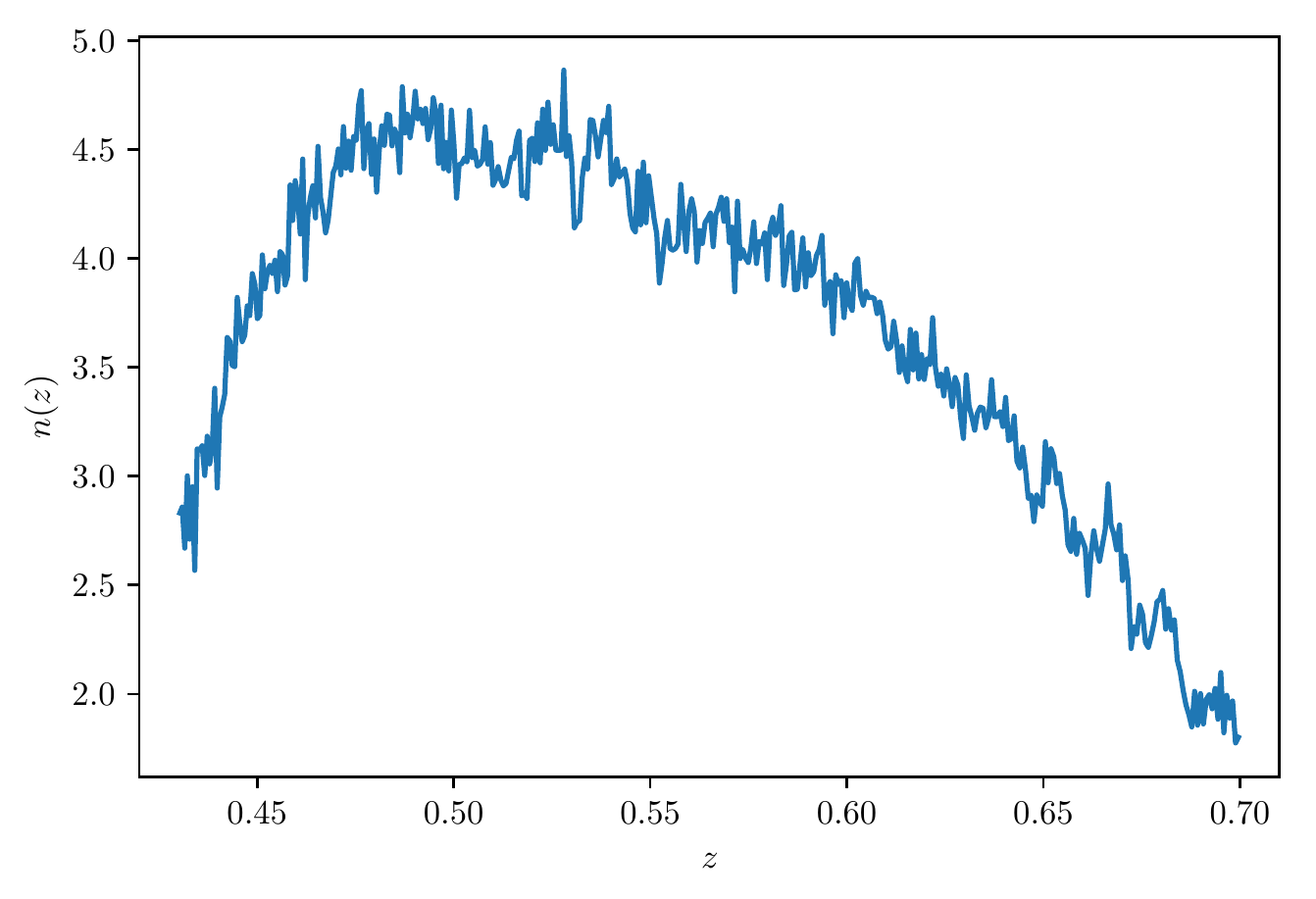}
\caption{\small Redshift distribution, up to a normalization factor, of the galaxies in the CMASS NGC sample.}
\label{fig:NGCredshift}
\end{figure}

The BOSS data we analyze in this paper have been binned in broad redshift bins. For example, the largest data set, the high-$z$ North Galactic Cap (NGC), has the redshift distribution $n(z)$ given in Fig.~\ref{fig:NGCredshift}, up to a normalization factor. In principle, it is possible to account for this distribution exactly in the EFTofLSS. In fact, it is enough to average the EFTofLSS prediction over this redshift distribution: 
 \be\label{eq:zbinning}
 \PEFT_{\ell,\rm binned} (k)=\int dz\; n(z)\, \PEFT_{\ell}(k,z)\ .
 \ee
 
\begin{figure}[!]
\centering
\includegraphics[width=0.495\textwidth,draft=false]{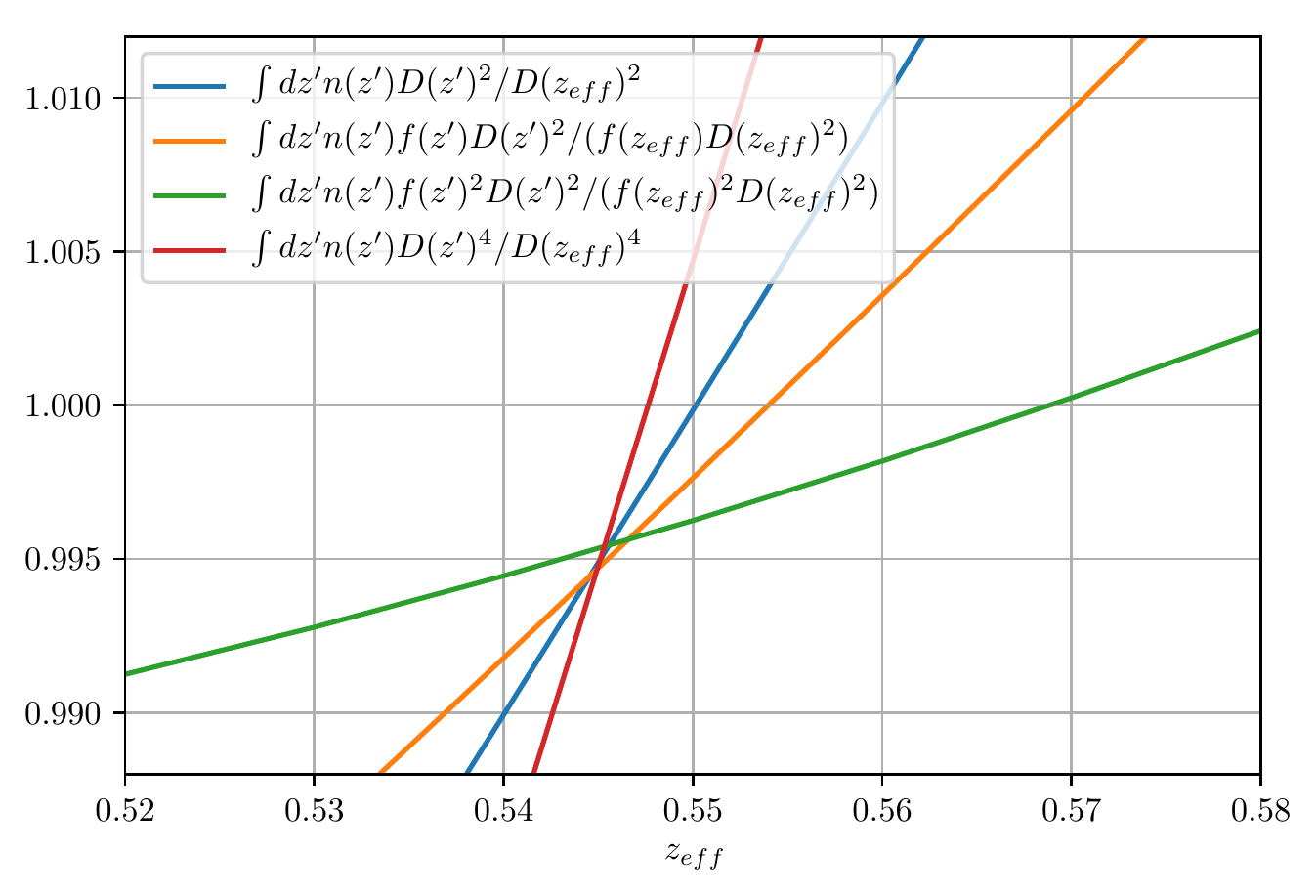}
\caption{\small We plot the ratios $\frac{1}{D(z_{\rm eff})^2} \int dz\; n(z) \, D(z)^{2}\ ,  \frac{1}{f(z_{\rm eff}) D(z_{\rm eff})^2} \int dz\; n(z) \,f(z)\, D(z)^{2}$, $\frac{1}{f(z_{\rm eff}) ^2D(z_{\rm eff})^2} \int dz\; n(z) \,f(z)^2\, D(z)^{2}$ and  $\frac{1}{D(z_{\rm eff})^4} \int dz\; n(z) \, D(z)^{4}$ as a function of $z_{\rm eff}$. We can see that the choice of $z_{\rm eff}=0.55$ for all the integral is very accurate, better than 0.5\%. In particular, notice that the loop terms will contribute by a smaller amount to the final answer (about 10\%), so one can afford a larger inaccuracy in those. Notice also that the contribution of the term in $f^2$ to the monopole, which is the better measured quantity, is suppressed by several numerical factors. We conclude that we can simply evaluate the EFTofLSS prediction directly at $z_{\rm eff}$.}
\label{fig:zeff}
\end{figure}

 The $z$-dependence of $ \PEFT_{\ell}(k,z)$ is rather straightforward, as it factorizes in the sum of terms with different $z$-dependence, and the $z$-dependence is factorized from the $k$-dependence. The dark matter power spectra depend on $z$ as $D(z)^2$, $f(z)D(z)^2$ and $f(z)^2D(z)^2$ at linear level, according if we consider matter-matter, matter-velocity or velocity-velocity power spectra. At one loop level, one has, respectively,  $D(z)^4$, $f(z)D(z)^4$ and $f(z)^2D(z)^4$. On top of this, the EFT-parameters depend on $z$ in a way that is expected to be similar to the one of powers of the growth factors. Therefore, implementing eq.~(\ref{eq:zbinning}) amounts to pre-compute several integrals of the form
 \be\label{eq:zbinning2}
 \int dz\; n(z)\, b_i(z)^{n_1} b_j(z)^{n_2} \, f(z)^{n_3} \, D(z)^{n_4}\ .
 \ee
  In the limit that the redshift bin is much smaller than the scale of variation of $D(z)$ and the EFT-parameters, we expect that for all these quantities the binning corresponds to evaluating all the quantities at the same effective redshift. Let us start by discussing  the terms with $n_1=n_2=0$. In this case, the integral in eq.~(\ref{eq:zbinning2}) can be performed, and one can set it equal to a $f(z_{\rm eff}(n_3,n_4))^{n_3}D(z_{\rm eff} (n_3,n_4))^{n_4}$. Mathematically, $z_{\rm eff}$ depends on $n_{\rm 3,4}$. However, in Fig.~\ref{fig:zeff}, we show, as an example, that for $n_4=2,4$ and $n_3=0,2$, the choice of $z_{\rm eff}=0.55$ is extremely accurate for all the terms. In particular, notice that the loop terms will contribute by a smaller amount to the final answer (about 10\%), so one can afford a larger inaccuracy in those. Notice also that the contribution of the term in $f^2$ to the monopole, which is the better measured quantity, is suppressed by several numerical factors.  Since the dependence on $z$ of the EFT-parameters is expected to be comparable, we extend this conclusion to all the terms. This allows us to conclude that we can safely approximate $ \PEFT_{\ell,\rm binned} (k)\simeq \PEFT_{\ell} (k,z_{\rm eff})$.

 \subsection{Fiber Collisions\label{sec:fibercollisions}}

 In the BOSS survey, fiber collisions affect the determination of the redshift of nearby galaxies, systematically affecting the power spectrum at $k\gtrsim 0.1\hinvMpc$. Several methods have been proposed to correct for this systematic error. In this paper, we will employ the `effective window method' of~\cite{Hahn:2016kiy}. This method has been shown to be extremely accurate in cancelling the systematic error induced by fiber collisions up to very small scales, well beyond the ones we will be interested in this paper: $k\gtrsim 0.3\hinvMpc$ and, possibly, even much higher than this~\cite{Hahn:2016kiy}. The method consists of adding to the model power spectrum a $\Delta P(k)$~(defined in eq. (24) of~\cite{Hahn:2016kiy}), which is computed in terms of the model and in terms of the window function of the fiber collision, plus two nuisance parameters that  correspond exactly to the shot noise of the monopole and the $k^2$ stochastic term in the quadrupole that already appear in the EFTofLSS prediction. The resulting power spectrum  has to be compared to the data measured with the method dubbed `Nearest Neighbor' (NN), which already takes into partial account the effect of the fiber collisions. 
 
 In order to get a sense of how much, potentially-undetected, residual inaccuracies of this method might affect the determination of cosmological parameters, we compare the results of the Effective Window Method with the ones we obtain if we instead use the less-accurate (but still, as we will see, sufficiently so) method of~\cite{Gil-Marin:2015sqa}, dubbed `Nearest Neighbor Method+adjustable shot noise' (NN\&shot-noise).
 
\begin{figure}[!]
\centering
\includegraphics[width=0.495\textwidth,draft=false]{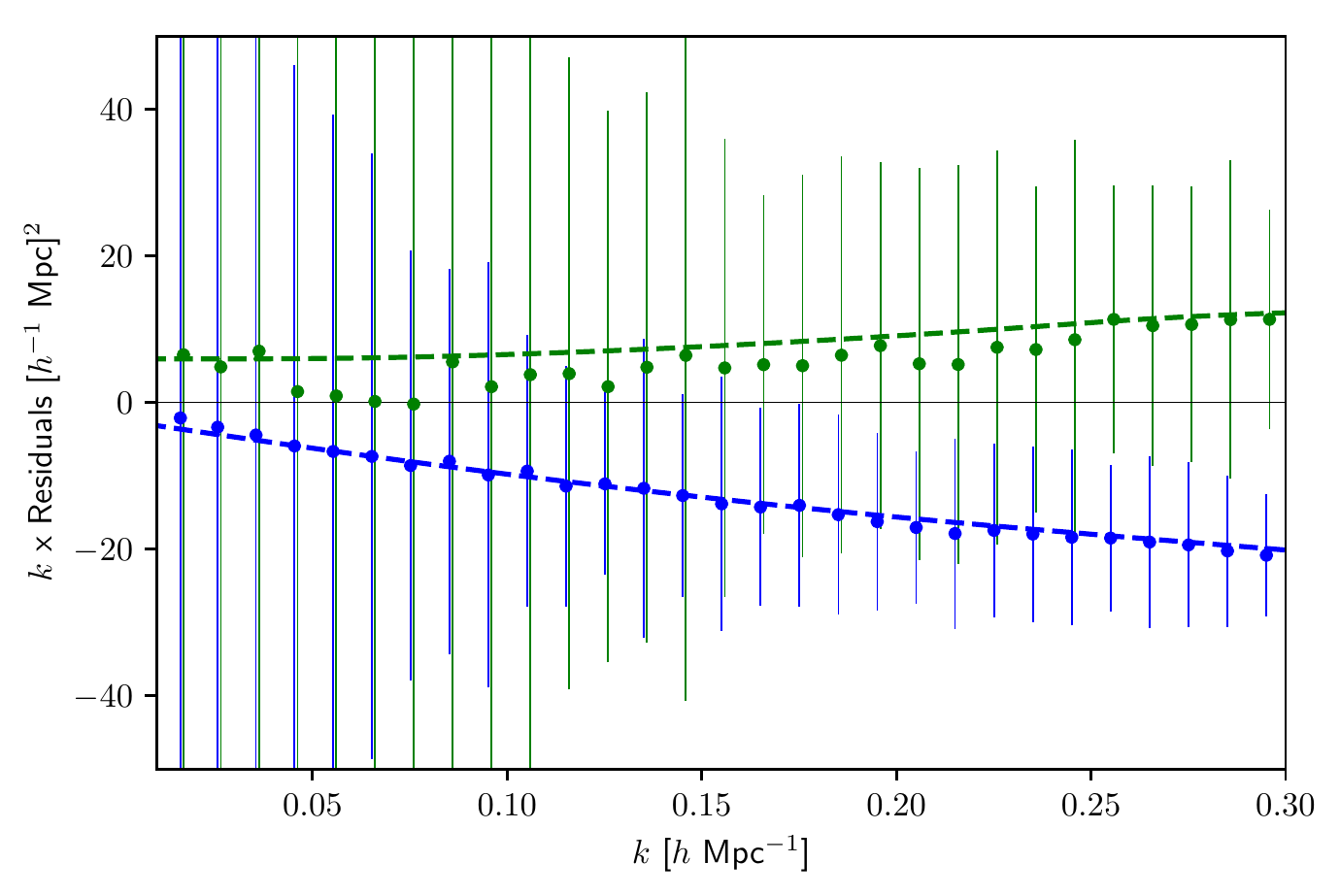}
\caption{\small Residuals between the monopole ({\it dotted, blue}) and quadrupole ({\it dotted, green}) power spectra and the ones reconstructed using the  NN\&shot-noise method, measured by the mean of 30 BOSS-like Patchy mocks, together with the error bars of one box. The dashed lines represents the correction given by the `Effective Window Method' of~\cite{Hahn:2016kiy} with $k_{\rm trust}=0.25\hinvMpc$. In the figure we use the nuisance parameters of~\cite{Hahn:2016kiy} equal to $-43$ for the monopole shot-noise and $-9$ for the quadrupole $k^2$ stochastic term (Notice that these values are much smaller than the natural ones we have for the stochastic EFT-parameters). We see that the signal is reconstructed with extreme accuracy, allowing us to neglect any residual systematic error from fiber collision effects. }
\label{fig:fiber-residual}
\end{figure}

 The result of the implementation of the corrections for fiber collisions can be seen in Fig.~\ref{fig:fiber-residual}, where we plot the residuals between the true power spectra and the ones reconstructed using the  NN\&shot-noise method, measured in BOSS-like Patchy mock catalogues whose characteristics we describe later in Sec.~\ref{sec:simulations}, together with the error bars of the data. On the same plot, we show the reconstructed signal obtained with the Effective Window Method. We see that, after the implementation of the Effective Window Method, the residuals are negligible, safely guaranteeing that we have fully accounted for the effect of fiber collisions. As an extra precautionary note, we analyze our data also using the NN\&shot-noise method, finding that the mildly inaccurate implementation shifts our $A_s$ parameter by less than $1/6$ of a $\sigma$, and unmeasurably so the other parameters. Given that the Effective Window Method is much more accurate than the NN\&Shot-noise method, we therefore conclude that fiber collisions present no residual non-negligible source of systematic error.

 \subsection{EFT-parameter marginalization\label{bias:marginalziation}}
 
The prediction $ \PEFT_{\ell}$ depends on several counterterms and galaxy bias parameters (EFT parameters). Their measurement is interesting, as they contain information about galaxy formation: it is the way different galaxy formation mechanisms affect large scales. However if somebody is only interested in the cosmological parameters, as we are in this paper, one will eventually marginalize over them.
 
 In this paper we take two approaches. We will explore the Likelihood in eq.~(\ref{eq:likelyhoodtwo}), which depends on $10$ EFT parameters and the cosmological parameters, by sampling with MCMC's, and then marginalize over them. We will use this as a consistency check of a more efficient approach, which is the following.  If we are ultimately interested in marginalizing over the bias parameters, it means that we are interested in the Likelihood of the cosmological parameters, which is the Likelihood  after integration over the EFT parameters:
 \be\label{eq:likelyhoodthree}
{\cal L}(d|\{\vec\Omega\})=\int d b\; {\cal L}(d|\{\vec\Omega,\vec b\})\ .
 \ee
We now split the EFT parameters as:
 \be
 \vec b=\{\vec b_G,\vec b_{NG}\}\ ,
 \ee
 where the distinction between $\vec b_G$, and $\vec b_{NG}$ is that the $\vec b_G$'s appear in the exponent of eq.~(\ref{eq:likelyhoodtwo}) only up to quadratic level. This means that  the $b_G$'s appear in $ \PEFT_{\ell}(k,\{\vec\Omega,\vec b\})$ only up to linear level. In our notation, $\vec b_G=\lbrace b_3, c_{\rm ct}, c_{r,1}, c_{r,2}, c_{\epsilon ,1},c_{\epsilon ,2},c_{\epsilon ,3} \rbrace$, and $\vec b_{NG}= \lbrace b_1, b_2, b_4 \rbrace$. We therefore can write~\footnote{Explicitly, 
\be
\PEFT_{\ell,\, {\rm const}}(k,\{\vec\Omega,\vec b_{NG}\})=\left.\PEFT_{\ell}(k,\{\vec\Omega,\vec b\})\right|_{\vec b_G \to 0}\ , \qquad  \PEFT_{\ell,\, {\rm lin,\, } i}(k,\{\vec\Omega,\vec b_{NG}\})=\left.\frac{\partial \PEFT_{\ell}(k,\{\vec\Omega,\vec b\})}{\partial b_{G,i}}\right|_{\vec b_G \to 0}\ .
 \ee}:
\be
\PEFT_{\ell}(k,\{\vec\Omega,\vec b\})\equiv \sum_i b_{G,i}\; \PEFT_{\ell,\, {\rm lin,\, } i}(k,\{\vec\Omega,\vec b_{NG}\})+\PEFT_{\ell,\, {\rm const}}(k,\{\vec\Omega,\vec b_{NG}\})\ .
\ee 
Plugging back in eq.~(\ref{eq:likelyhoodthree}), we realize that the integral over the $b_G$'s can be done analytically, as it is a Gaussian integral, without the  need to explore the Likelihood numerically. In order to make the Gaussian structure more apparent, we can define:
 \bea\label{eq:likelyhoodfour}
&&{\cal L}(d|\{\vec\Omega,\vec b_G,\vec b_{NG}\})={\rm Exp}\left[-\frac{1}{2}\sum_{ij}  b_{G,i} \, F_{2,ij}\left( \PEFT_{\ell,\, {\rm lin}}(\{\vec\Omega,\vec b_{NG}\}),C^{-1}_{WW}\right) b_{G,j}+\right. \\ \nn
&&\qquad\qquad\qquad\qquad\left.+\sum_i  b_{G,i} \, F_{1,i}\left( \PEFT_{\ell,\, {\rm lin}}(\{\vec\Omega,\vec b_{NG}\}),\PEFT_{\ell,\, {\rm const}}(\{\vec\Omega,\vec b_{NG}\}),\Pdata_{\ell}, C^{-1}_{WW},C^{-1}_{W},\Pdata_{\ell}\right)\right.\\ \nn
&&\qquad\qquad\qquad\qquad\left.+  \, F_{0}\left( \PEFT_{\ell,\, {\rm const}}(\{\vec\Omega,\vec b_{NG}\}),\Pdata_{\ell},C^{-1}_{WW},C^{-1}_{W}\right)\right]\ .
\eea
where
\bea
&&F_{2,ij} =\PEFT_{\ell,\,{\rm lin,\, } i}(k,\{\vec\Omega,\vec b_{NG}\}) \cdot C^{-1}_{WW}(k,k')_{\ell,\ell'}\cdot \PEFT_{\ell',\,{\rm lin,\, }j}(k',\{\vec\Omega,\vec b_{NG}\})\ , \\ \nn
&&F_{1,i}=-\PEFT_{\ell,\,{\rm const} }(k,\{\vec\Omega,\vec b_{NG}\}) \cdot C^{-1}_{WW}(k,k')_{\ell,\ell'}\cdot \PEFT_{\ell',\,{\rm lin,\, }i}(k',\{\vec\Omega,\vec b_{NG}\})+\\ \nn
&&\qquad\qquad\qquad\qquad+\PEFT_{\ell, \,{\rm lin,\,} i}(k,\{\vec\Omega,\vec b_{NG}\}) \cdot C^{-1}_{W}(k,k')_{\ell,\ell'}\cdot \Pdata_{\ell'}(k')\ , \\ \nn
&&F_{0}=-\frac{1}{2}\PEFT_{\ell,\,{\rm const} }(k,\{\vec\Omega,\vec b_{NG}\}) \cdot C^{-1}_{WW}(k,k')_{\ell,\ell'}\cdot \PEFT_{\ell'{\rm const }}(k',\{\vec\Omega,\vec b_{NG}\})+\\ \nn
&&\qquad\qquad\qquad\qquad+\PEFT_{\ell, \,{\rm const}}(k,\{\vec\Omega,\vec b_{NG}\}) \cdot C^{-1}_{W}(k,k')_{\ell,\ell'}\cdot \Pdata_{\ell'}(k')\\ \nn
&&\qquad\qquad\qquad\qquad-\frac{1}{2}\Pdata_{\ell}(k) \cdot C^{-1}(k,k')_{\ell,\ell'}\cdot \Pdata_{\ell'}(k')\ ,
\eea
and then we can perform a Gaussian integral to define the partially marginalized Likelihood:
\bea
\label{eq:likelyhoodfour}
&&{\cal L}(d|\{\vec\Omega, b_{NG}\})=\int d b_G\; {\cal L}(d|\{\vec\Omega,\vec b_G,\vec b_{NG}\})=\\ \nn
&&\quad ={\rm Exp}\left[\frac{1}{2} F_{1,i}(\{\vec\Omega,\vec b_{NG}\})\cdot F_{2}(\{\vec\Omega,\vec b_{NG}\})^{-1}{}_{ij} \cdot F_{1,j} (\{\vec\Omega,\vec b_{NG}\})\right.\\ \nn\
&&\qquad\qquad\left.+F_0(\{\vec\Omega,\vec b_{NG}\})-\frac{1}{2}\log\left[\det\left(F_2(\{\vec\Omega,\vec b_{NG}\})\right)\right]\right] \ ,
 \eea
 where in the last passage we neglected an irrelevant constant.
 ${\cal L}(d|\{\vec\Omega, b_{NG}\})$ depends on only 3 EFT parameters, so it is quite easier to explore numerically with an MCMC. This is compensated, but only in part, by the fact that this Likelihood is slower to evaluate numerically for each choice of the cosmological parameters, as the functional form is more complex than a simple $\chi^2$-one. We will refer to this Likelihood as the partially marginalized Likelihood. The partially marginalized Likelihood in the case in which we include the bispectrum is discussed in App.~\ref{sec:appendix_bispectrum_lik}.  Notice finally that this procedure does not require to include priors on the EFT-parameters, but these can always be simply included before performing the integral in~eq.~(\ref{eq:likelyhoodfour}).

\subsection{Details on the  MCMC analysis \label{sec:MCMCdetails}}

We obtain the parameter Likelihood described above using the Python-based Monte-Carlo Markov Chain (MCMC) sampler {\it emcee}~\cite{ForemanMackey:2012ig}. We run four separate chains in parallel and use the Gelman-Rubin convergence criteria~\cite{Gelman:1992zz} with a threshold of $\epsilon \leq 0.005$ on the scale-reduction parameters, to determine convergence.

Beyond the Gelman-Rubin convergence criteria, we have performed several additional convergence tests. Here we list a few. We have checked that the non-marginalized Likelihood of eq.~(\ref{eq:likelyhoodtwo}) and the partially marginalized Likelihood of eq.~(\ref{eq:likelyhoodfour}) give consistent results. Similarly, we obtain consistent results if we impose the Gelman-Rubin test threshold to be 0.005 instead of 0.003, or if we change other settings of the MCMC (such as the number of walkers).

At each step of the likelihood evaluation, we take the model from a pre-computed grid of the cosmology-dependent pieces of eq.~\eqref{eq:factorizedpowerspectrum}. The dependence on the EFT parameters is instead analytic, and so evaluation for the model concerning the variation of these parameters can be done directly at the level of the MCMC. This setup allows for an extremely rapid evaluation of the likelihood. For this first analysis of the BOSS data using the EFTofLSS, we limit ourself to a three-dimensional grid scanning over the cosmological parameters $\lnAs$, $\Omega_m$ and $h$, where $A_s$ is the amplitude of the primordial power spectrum, $\Omega_m$ is the matter abundance, and $h$ is the present day Hubble parameter measured in units of 100\,km/(sec Mpc). We will therefore fix the ratio of the baryons versus dark-matter abundance, $\Omega_b/\Omega_c$, and the tilt of the primordial power spectrum, $n_s-1$, to the Planck2018 best-fit values. However, there should be no difficulty in extending our analysis to include these additional cosmological parameters as well as others, such as, for example, the dark-energy equation of state, $w$. In fact, one possible option is the following. Since the cosmological parameters are already quite well-determined, we are interested only in small relative variations of them. Therefore, the predictions for all models of interest can be obtained by Taylor expanding the prediction around the best cosmology (say, for example, Planck2018+{BOSS DR12}), to the desired order. In this way, the dependence on all the parameters would be analytical, allowing for a rapid evaluation of the likelihood directly at the level of the chain, without building grids, and potentially for some additional simplifications. This approach has been developed in~\cite{Cataneo:2016suz} for the two-loop dark-matter power spectrum in the EFTofLSS, and was found to give sub-percent accuracy. {We foresee no obstacles in doing this also for the power spectrum of biased tracers in redshift space. We leave this to future work~\cite{leoprogressTaylorSDSS}.} 

Finally, in the EFTofLSS, the EFT parameters are supposed to be order one, as in this case all operators become strongly coupled at around the same scale. We will therefore impose the following priors on the parameters that 
encompass the physically-motivated range: 
\bea 
&& b_1\in [0, 4]_{\rm flat}\ , \quad
c_2 \in [-4, 4]_{\rm flat}\ , \quad
c_4 \in 2_{\rm gauss}\ ,\quad b_3 \in 2_{\rm gauss}\ , \\ \nonumber
&& 
c_{\rm ct} \in 2_{\rm gauss}\ , \quad
c_{r,1} \in 4_{\rm gauss}\ , \quad c_{r,2}\in 4_{\rm gauss}\ , \\ \nonumber
&& 
c_{\epsilon,1}/\bar{n}_g \in \left(\frac{400}{(\hinvMpc)^{3}}\right)_{\rm gauss}\ , \quad c_{\epsilon,\rm mono} \in  2_{\rm gauss}\ , \quad c_{\epsilon,\rm quad} \in  2_{\rm gauss}\ , \quad
\eea
where $b_2 = \frac{1}{\sqrt{2}} (c_2 + c_4),\ \ b_4 = \frac{1}{\sqrt{2}} (c_2 - c_4)$, $c_{\epsilon,\rm mono}= c_{\epsilon,1}+\frac{1}{3} c_{\epsilon,2}$,  $c_{\epsilon,\rm quad}= \frac{2}{3} c_{\epsilon,2}$. We choose $\knl=k_M = 0.7\hinvMpc$. Notice that $\bar{n}_g = 4.5 \cdot 10^{-4} (\hinvMpc)^3$ for CMASS and $\bar{n}_g = 4.0 \cdot 10^{-4} (\hinvMpc)^3$ for LOWZ~(\footnote{\label{footnote:shotnoisechallenge}  For the Challenge boxes, we use $\bar{n}_g=3\cdot 10^{-4} (\hinvMpc)^3$.}). Here the subscript `flat' meanst that we impose a flat prior with the stated boundaries, while the subscript `gauss' means that we impose a gaussian prior centered at zero with that standard deviation.
 
We perform the change of variables between $(b_2, b_4)$ and $(c_2, c_4)$ as we find for all simulations and in the observational data that the former are almost completely anti-correlated (at $\sim 99\%$). Thus, $c_4$ that parametrizes the difference between the contributions of $b_2$ and $b_4$ in the 1-loop term, which are highly degenerate, is practically undetermined, and we put a gaussian prior with standard deviation of $2$ on it. Then, upon comparison with the numerical and observational data, we find that for the signal-to-noise ratios that are relevant in our study (for BOSS-like volume), the functions that are multiplied by $c_4$ and $ c_{\epsilon,\rm mono} $ are too small to significantly affect the results. We therefore set $c_4$ and $ c_{\epsilon,\rm mono}$ to zero. 
As we do not analyze the hexadecapole in our analyses, we set $c_{r,2} =0$ and put a prior $8_{\rm gauss}$ on $c_{r,1}$ to absorb the main contribution of $c_{r,2}$ in the quadrupole. Also, we impose a prior on the shot noise consistent with fiber collisions and selection effects, since (most of) the Poisson shot noise is subtracted from the data. Our choice of the priors plays no role as long as all parameters are well determined by the data, as we will find to be the case in simulations with volume as large as the Challenge boxes that we describe later, and whose volume is about 16 times the BOSS volume. We have checked for those boxes that significantly enlarging the priors does not affect our results, and all parameters stay of order of their expected physical size. For volumes of the size of the BOSS data, the priors make certain that the fit lies within the physical region of the EFT parameters, as potential unbroken degeneracies allowed by the large error bars of the data might lead to undesirable systematic errors. We will test our prior choice using the Patchy mocks that have the same volume as the data we want to analyze.
 When we fit the bispectrum, we do not set $c_4$ to zero, as $b_2$ and $b_4$ now enter  in the observables in a less degenerate way. Furthermore, as we fit the bispectrum only up to a low $\kmax$, the bispectrum shot noise terms are neglibly small, especially since most of them are subtracted from the data. We therefore put them to zero.

 \section{Tests on Simulations\label{sec:simulations}}

 While there is no question that the EFTofLSS is accurate at low enough wavenumbers, it is unclear what is the wavenumber at which, given the error bars on the data, the EFTofLSS becomes too inaccurate and measurements begin to be biased. We call this error the `theory-systematic', to be distinguished from the observational and instrumental systematic errors, and from the bias parameters of the EFTofLSS. As discussed above, it is possible to have a rough idea of what this wavenumber is, because, in the EFTofLSS, one can estimate the theoretical error associated with the higher-order terms that have not been computed~\cite{Carrasco:2012cv,Carrasco:2013sva,Senatore:2014vja,Foreman:2015uva,Baldauf:2016sjb}. However, this procedure is quite approximate, and, given the importance of going to higher wavenumbers to gain more information, it is preferable to have additional methods to measure the theory systematic error in the measurements. 
 
 Testing the EFTofLSS against simulations can provide a systematic way to quantify these different errors, even though it is a bit unclear if the systematic error associated with the simulations is much smaller than percent level (see for example~\cite{Schneider:2015yka})~\footnote{Not all systematic errors in the simulations are very problematic for this test. As long as the simulations implement some mechanism that preserves the laws of physics that are used to derive the EFTofLSS, such as for example coordinate reparametrization invariance, the EFTofLSS should be able to describe it, notwithstanding that the actual physics of galaxy formation in our universe might be different than the simulated one.}. We analyze two sets of simulations.

 \subsubsection*{{BOSS DR12 Challenge boxes}} 
 
  The BOSS DR12 analysis~\cite{Alam:2016hwk,Beutler:2016arn} performed a blind mock challenge on a set of high-resolution N-body simulations populated with galaxies using Halo Occupation Distribution (HOD) modeling. There are seven different catalogs with different HOD prescriptions, constructed from periodic simulation boxes with varying cosmologies, and they do not include any sky-geometry selection function effect.  The seven catalogs are labeled A through G. Several of the boxes are based on the Big MultiDark simulation~\cite{Riebe:2011gp}. The catalogs are constructed out of simulation boxes with a range of 3 underlying cosmologies, and boxes with the same cosmology have varying galaxy bias models, with resulting variance of the overall galaxy bias by about 5\%. The redshift of the boxes ranges from $z = 0.441$ to $z = 0.562$. We analyze the simulations at high redshift, which are relevant for our analysis, {\it i.e.} A, B, F, G and D. These cases were designed to quantify the sensitivity of RSD models to the specifics of the galaxy bias model over a reasonable range of cosmologies. The cosmology and relevant simulation parameters for each of the boxes that we use are given in Table~\ref{tab:simulspec}. A comparison of the results from the mock challenge in the context of the BOSS DR12 results is presented in~\cite{Alam:2016hwk}, and individual Fourier-space clustering results from the challenge are discussed in~\cite{Beutler:2016arn,Grieb:2016uuo}.

  \subsubsection*{Patchy Mocks}
 
 The BOSS collaboration provides a set of mock catalogs, which mimic the clustering properties and geometry of the observed dataset~\cite{Kitaura:2015uqa}. These mock catalogues have been produced using approximate gravity solvers and analytical-statistical biasing models. The catalogues have been calibrated to an N-body based reference sample extracted from one of the BigMultiDark simulations~\cite{Klypin:2014kpa}, which were performed using Gadget-2~\cite{Springel:2005mi} with $3840^3$ particles in a volume of $[2500 h^{-1}{\rm Mpc}]^3$ assuming a $\Lambda$CDM cosmology with $\Omega_m = 0.307115,\, \Omega_b = 0.048206,\, \sigma_8 = 0.8288,\, n_s = 0.9611$, and a present-day Hubble constant of $H_0 = 67.77 {\rm km}\, {\rm s}^{-1}{\rm Mpc}^{-1}$. Here 
 $\sigma_8$ is the root mean square of the linearly-evolved matter power spectrum within a top-hat radius of $8\hinvMpc$ at redshift zero. Halo abundance matching is used to reproduce the observed BOSS 2- and 3-point clustering measurements~\cite{Kitaura:2015uqa}. This technique is applied at different redshift bins to reproduce the BOSS DR12 redshift evolution. These mock catalogues are combined into lightcones, also accounting for the selection effects and survey mask of the BOSS survey. While they are usually called `Patchy lightcones mocks' in the literature, we will refer to them simply as `Patchy mocks'.
In total we have 2048 mock catalogues available for each patch of the BOSS dataset that we analyze: CMASS NGC and SGC and LOWZ NGC.
We analyze the power spectra and bispectra of 16 NGC boxes only, as they are sufficient for our purposes, but we use all of them to estimate the covariance matrices.\\
 
\begin{table}[]
\centering
\footnotesize
\begin{tabular}{|l|l|l|l|l|l|l|l|} 
\hline
Simulation name            & $V_{\rm eff}$ (Gpc$^3$) & $\Omega_m$ & $\Omega_b h^2$ & $h$    & $\ln (10^{10} A_s)$ & $n_s$  & $z_{\rm eff}$  \\ \hline
Challenge A, B, F, G & 50.2                  & 0.307    & 0.022       & 0.6777 & 3.0824         & 0.96   & 0.5617        \\ \hline
Challenge D & 50.2                  & 0.286   & 0.023       & 0.7 & 3.096         & 0.96   & 0.5       \\ \hline
Patchy lightcone NGC      & 3.0  ($\times 16$)                         & 0.307       & 0.022       & 0.6777 & 3.091        & 0.96 & 0.55          \\ \hline
\end{tabular}
     \caption{\small Cosmological parameters and specifications of the N-body simulations analyzed in this work.      $V_{\rm eff}$ is the effective volume of the box, while $z_{\rm eff}$ is the effective redshift, accounting for the window and redshift selection. The factor of `$\times 16$' in the $V_{\rm eff}$ of the Patchy mocks represents the fact that we will use 16 independent boxes, each with effective volume equal 3 Gpc$^3$.}
     \label{tab:simulspec}
\end{table}

 In this section, we present the results of the fits of the monopole and quadrupole of the power spectrum with the partially marginalized likelihood of eq.~(\ref{eq:likelyhoodfour}) on measurements from these simulations. We do not analyze the hexadecapole because we find it adds very marginal information and it is more subject to measurement systematics. We analyze five challenge boxes A, B, F, G and D, and 16 Patchy boxes for the North Galactic Cap (NGC) region. The characteristics of these simulations are given in Table~\ref{tab:simulspec}. Besides asserting the validity range of the EFTofLSS and its ability to determine the cosmological parameters, these tests provide interesting complementary results, following from the specificity of the simulations. The challenge boxes ABFG have all the same underlying cosmology but different HOD models, while box D represents both a different cosmology and a different HOD model. This allows us to study the response of the EFTofLSS to the UV physics ({\it i.e.} galaxy formation physics), not only to see if it introduces significant biases in the determination of the cosmological parameters, but also to explore the sensitivity of the EFT parameters to different HOD populations. In fact, as we will see shortly, our results suggest that the measurement of the EFT-parameters can potentially be used to learn which   formation mechanism was actually dominant for the galaxy sample under consideration.  The Patchy mocks reproduce important observational aspects, such as the effects of a mask and the redshift selection. Again, it is of interest to keep control of the possible induced systematics from these geometrical factors. Furthermore, since measurements of the bispectrum from the Patchy mocks are available, we will use it to test the performance of the EFTofLSS when the bispectrum is included. 
 
After discussing how one can estimate and measure the theory-systematic error from tests on simulations in Sec.~\ref{sec:loopterms} and Sec.~\ref{sec:theorysys}, we explain in Sec.~\ref{sec:physicalconsiderations} how within the EFTofLSS it is possible to determine with precision the cosmological parameters without introducing unnecessary degeneracies and uncontrolled systematics from approximate treatments. We also explore in Sec.~\ref{sec:planckprior} the impact on the fit of the inclusion of a prior on the sound horizon at decoupling from CMB measurements, and in~\ref{sec:bispectrum} of the inclusion of the tree-level bispectrum monopole. The reader uninterested in the details can directly go to Sec.~\ref{sec:simulsummary} where we summarize our findings from the tests.

 All the MCMC are run with the baryons-to-dark-matter ratio $\Omega_b/\Omega_c$ and the tilt of the primordial power spectrum $n_s$ fixed and given in Table~\ref{tab:simulspec}, while we freely scan over $A_s,\; \Omega_m$ and $h$. We have checked on Fisher forecasts of the CMASS NGC dataset that removing the priors on $\Omega_b/\Omega_c$ and $n_s$ to about ten times the errors from Planck2018 degrades our constraints just roughly by about 18\% for~$\ln A_s$, 27\% for~$\Omega_m$, and 53\% for $h$~(\footnote{\label{gootnote:firstfisher}If instead we put an extremely wide prior on $n_s$ to roughly the interval $[0.70-1.22]$, which is about 60 times the error bar from Planck2018, and put a prior on $\Omega_b/\Omega_c$ of the order of four times the percentage error on $\Omega_b h^2$ obtained by Big Bang Nucleosynthesis (BBN) (which is about 20$\sigma$'s of Planck2018)~\cite{Tanabashi:2018oca}, the error bars are degraded by about  
 $117\%,\, 158\%$ and $267\%$ for $\ln A_s, \, \Omega_m$ and $h$ respectively. Given that we will measure these quantities to 
  to $12\%,\, 3.2\%,$ and $2.9\%$ respectively, we still would have a significant and valuable measurement.  }). 

\subsection{Size of the loop terms\label{sec:loopterms}}

\begin{figure}[h]
\centering
\includegraphics[width=0.495\textwidth,draft=false]{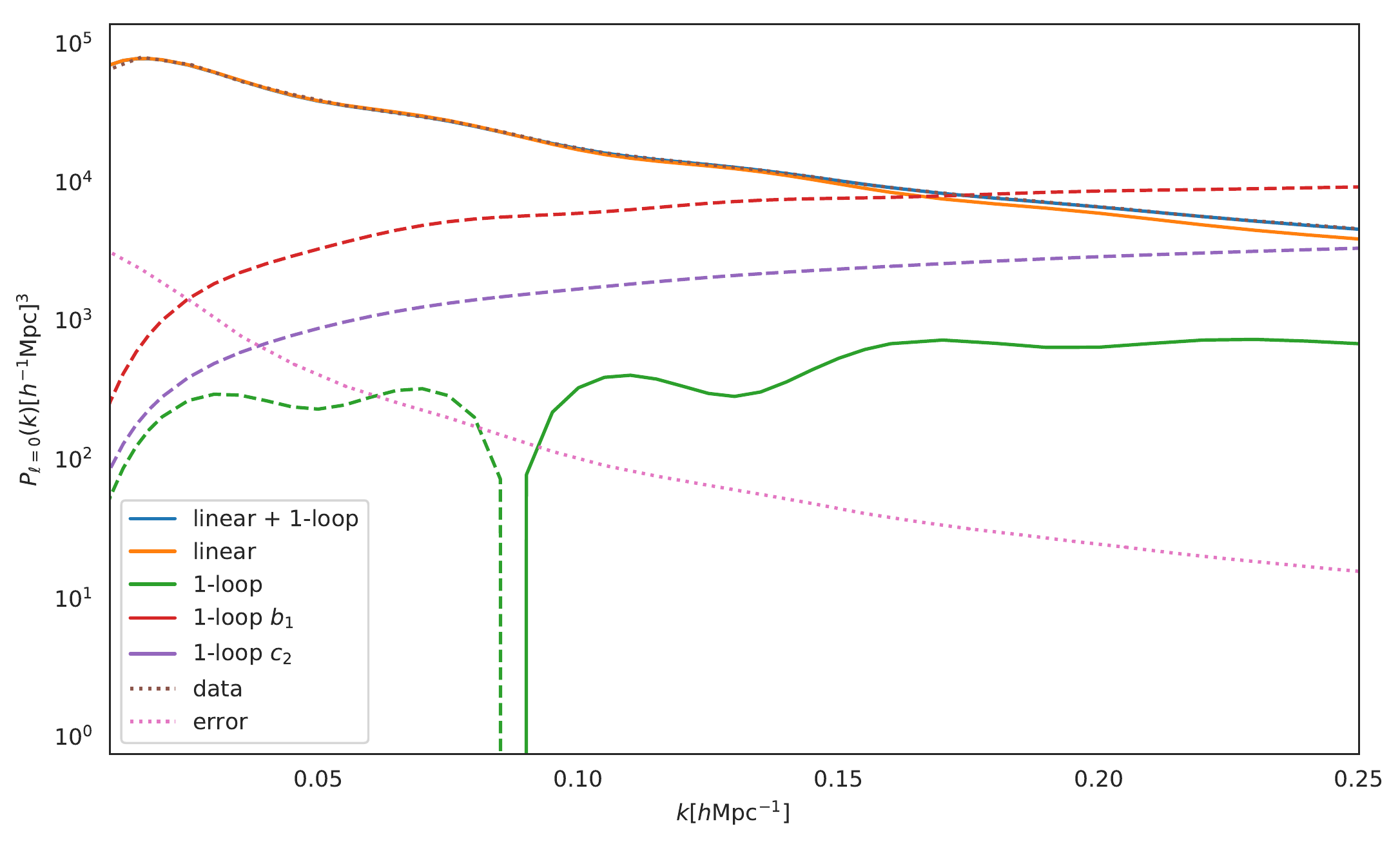}
\includegraphics[width=0.495\textwidth,draft=false]{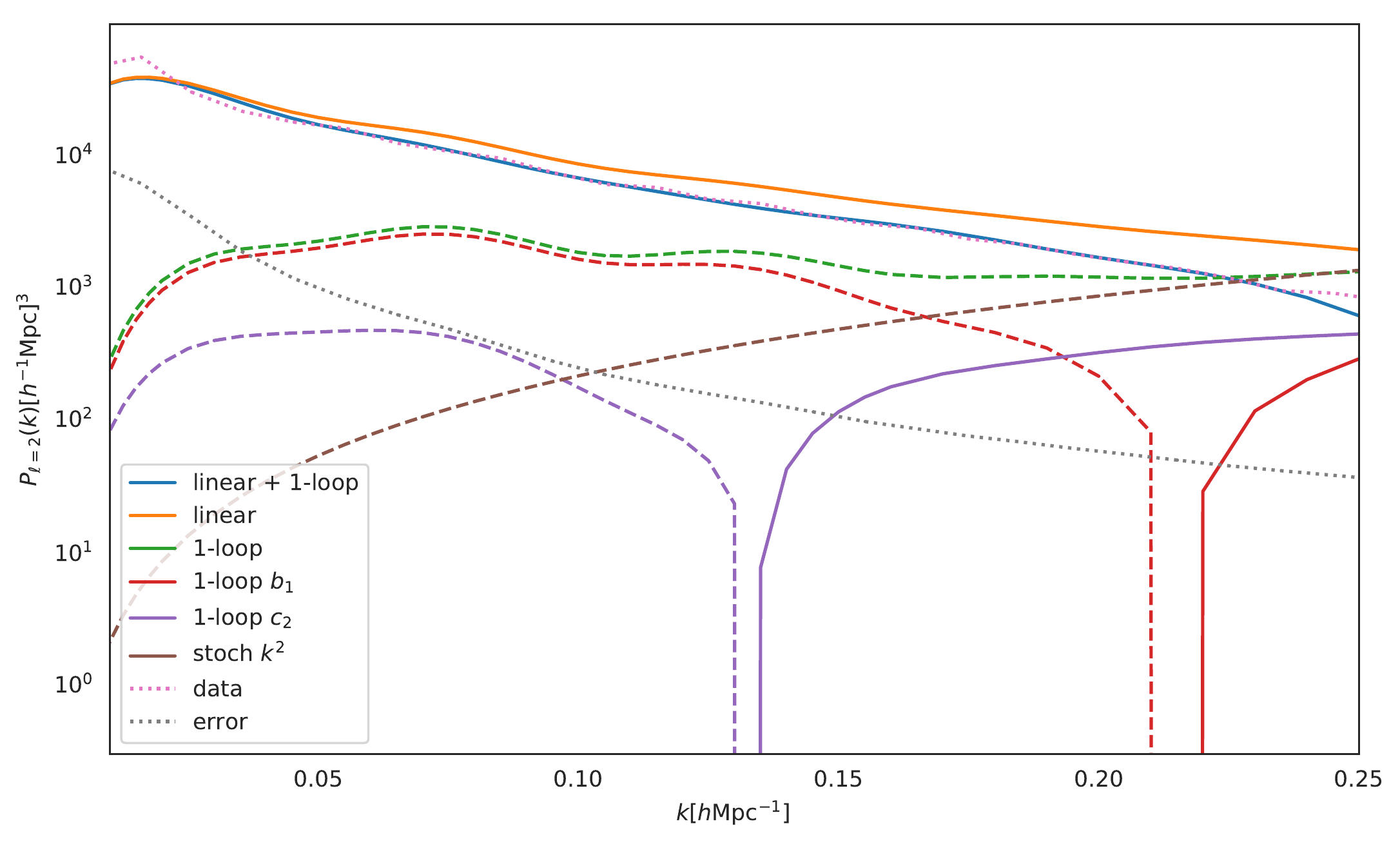}
\includegraphics[width=0.495\textwidth,draft=false]{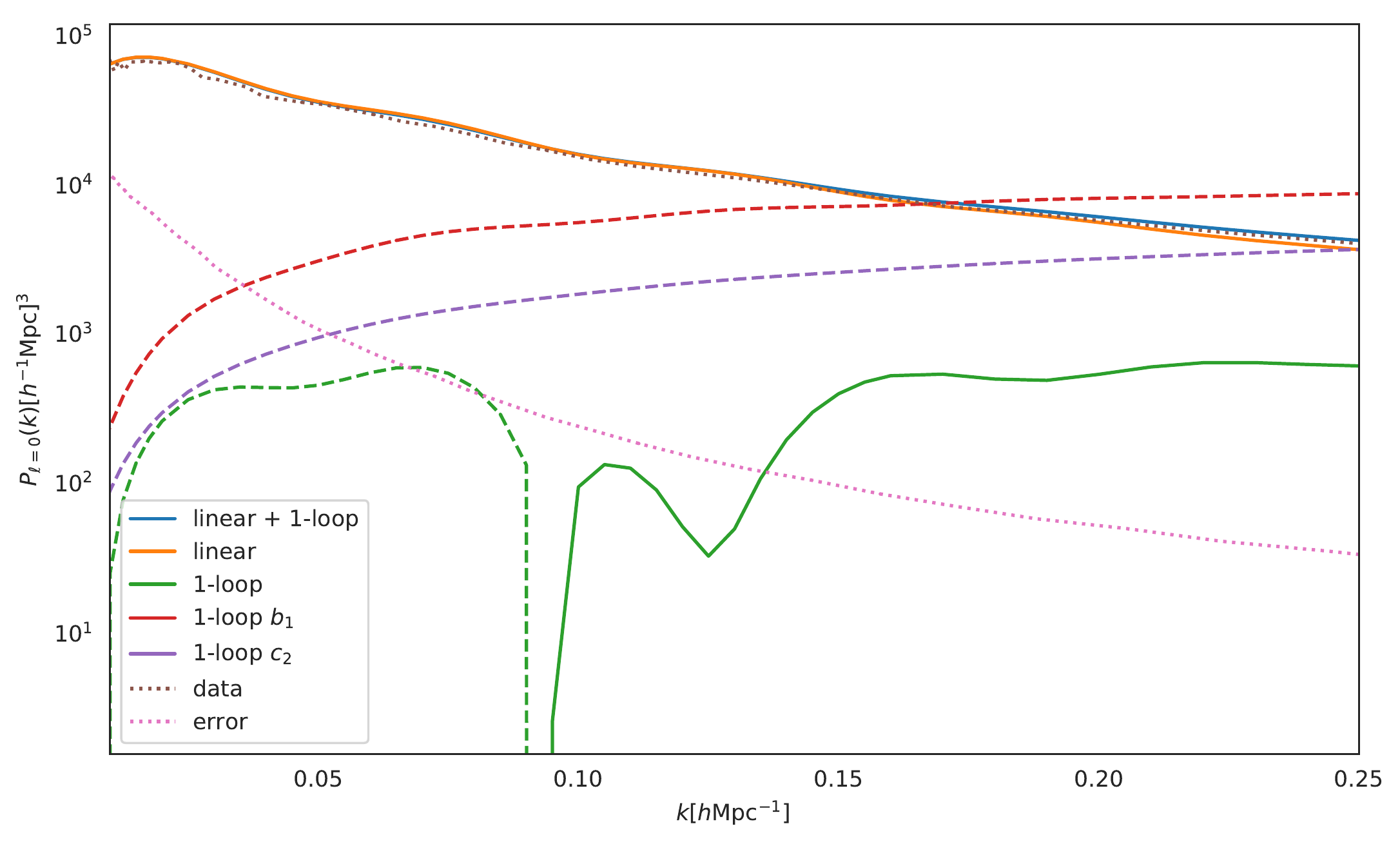}
\includegraphics[width=0.495\textwidth,draft=false]{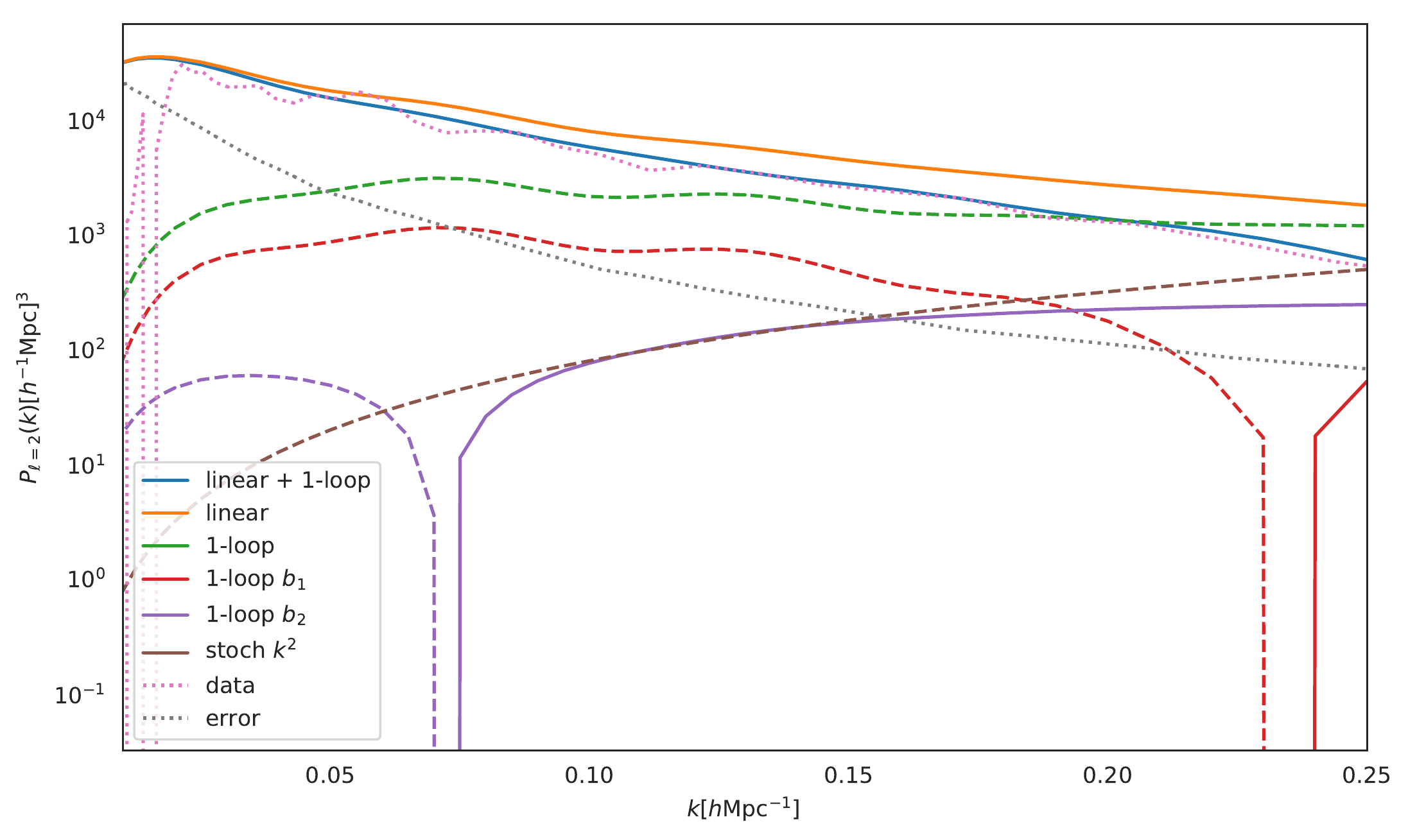}
\caption{\small Size of various contributions to the monopole and quadrupole of the one-loop power spectrum of eq.~\eqref{eq:powerspectrum} evaluated at the best fit cosmology at $\kmax=0.23\hinvMpc$ for the challenge boxes ({\it top}) and for the Patchy mocks ({\it bottom}). Dashed stays for negative value. We also plot, in dotted, the  size of the statistical error bars. 
\label{fig:loopsize} }
\end{figure}

Before analyzing the measurements in a systematic way, a simple $\chi^2$-fit at the best fit of the simulations can provide insights on how much the various terms in the power spectrum of eq.~\eqref{eq:powerspectrum} contribute overall (one could perform this study equally well by fixing the fit to the fiducial cosmology). One can get a sense of the applicability range of the EFTofLSS along with other interesting features that we detail in the following. In Fig.~\ref{fig:loopsize} are shown various contributions to the monopole and quadrupole for the challenge boxes and Patchy mocks. 

First, at high wavenumbers, $\kmax\simeq 0.23\hinvMpc$, on the monopole, we notice that, on the Challenge A box, the linear part contributes to {$\sim 85 \%$} of the total while the one-loop contributes to {$\sim 15 \%$}. On the quadrupole, the contribution of the loops is instead larger, {and reaches order half of the linear one}. On Patchy, on the monopole, the loop contribution is about {$10 \%$ , while on the quadrupole the loop contribution is about half of the linear one}.  This allows us to very roughly  estimate the theoretical error which corresponds to the size of the next-to-leading contribution, namely, the two-loop. The size of the $L$-loop $P^{(L)}(k)$ can be roughly estimated as $P^{(L)}(k) \sim \alpha^L P_{11}(k)$, where $P_{11}$ is the linear power spectrum and $\alpha$ is a small number representing the size of the loop contribution. It follows that around {$k \sim 0.25 \hinvMpc$}, the theoretical error is about $1 \%$ of the power spectrum for the monopole, which is of the same order as the measurement statistical error, and a similar conclusion holds, approximately, for the quadrupole, where the statistical error is larger. This suggests, if not completely justifies, that we can use the EFTofLSS at one-loop order up to $\kmax$ in the range $0.20\hinvMpc$ to $0.25\hinvMpc$ for surveys with volume similar to the BOSS CMASS one. 
 We will discuss further the $k$-range accessible by the theory at this order in the following, when concluding our systematic analysis. 
 
Second, representative contributions are also plotted in Fig.~\ref{fig:loopsize}. We first observe that keeping only one galaxy bias $b_i$ to its best-fit value ($\sim \mathcal{O}(2)$) while setting all other EFT parameters to zero gives the typical size of the contributions entering in the one-loop contribution. We notice that there are some cancellation at play in loops in the monopole, but they are not extremely large. Additionally, in the quadrupole, the $k^2$ stochastic term seems to be the larger contribution.

To finish on preliminary considerations, we would like to make a comment on the constant stochastic term. In principle, it is of size $\bar{n}_g^{-1}$, which is of the order of the typical significant contributions to the one-loop. In practice, the shot-noise contribution is often subtracted from the measurements.
However, this procedure relies on estimators for weighting and selecting the galaxies to account for the spatial variance of the number density. 
For these reason, we fit without shot noise the Challenge boxes as the shot-noise subtraction is reliable in simulations where the number density is known, while we fit with a shot noise the data to account for small inaccuracies in the estimation of $\bar n_g$. We indeed find that keeping the constant stochastic term with reasonable priors (a few percents of $\bar{n}_g^{-1}$) in regards of the above discussion helps the performance of the fit. Furthermore, we remind the reader that the `Effective Window Method' correcting fiber collisions contains a nuisance parameter in the form of a shot noise.

\subsection{Theory-systematic error \label{sec:theorysys}}

\begin{figure}[h!]
\centering
\includegraphics[width=0.496\textwidth,draft=false]{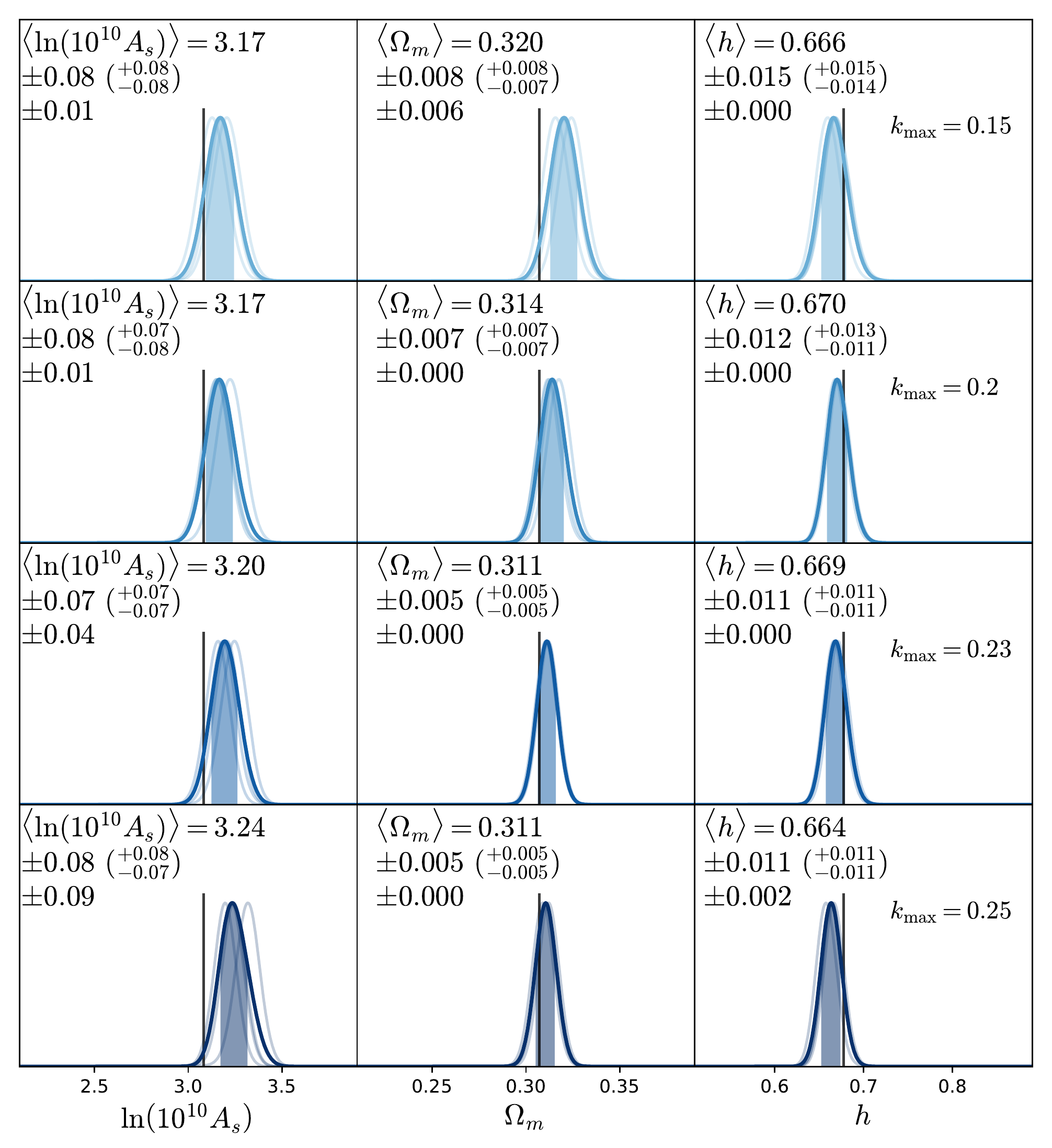}
\includegraphics[width=0.496\textwidth,draft=false]{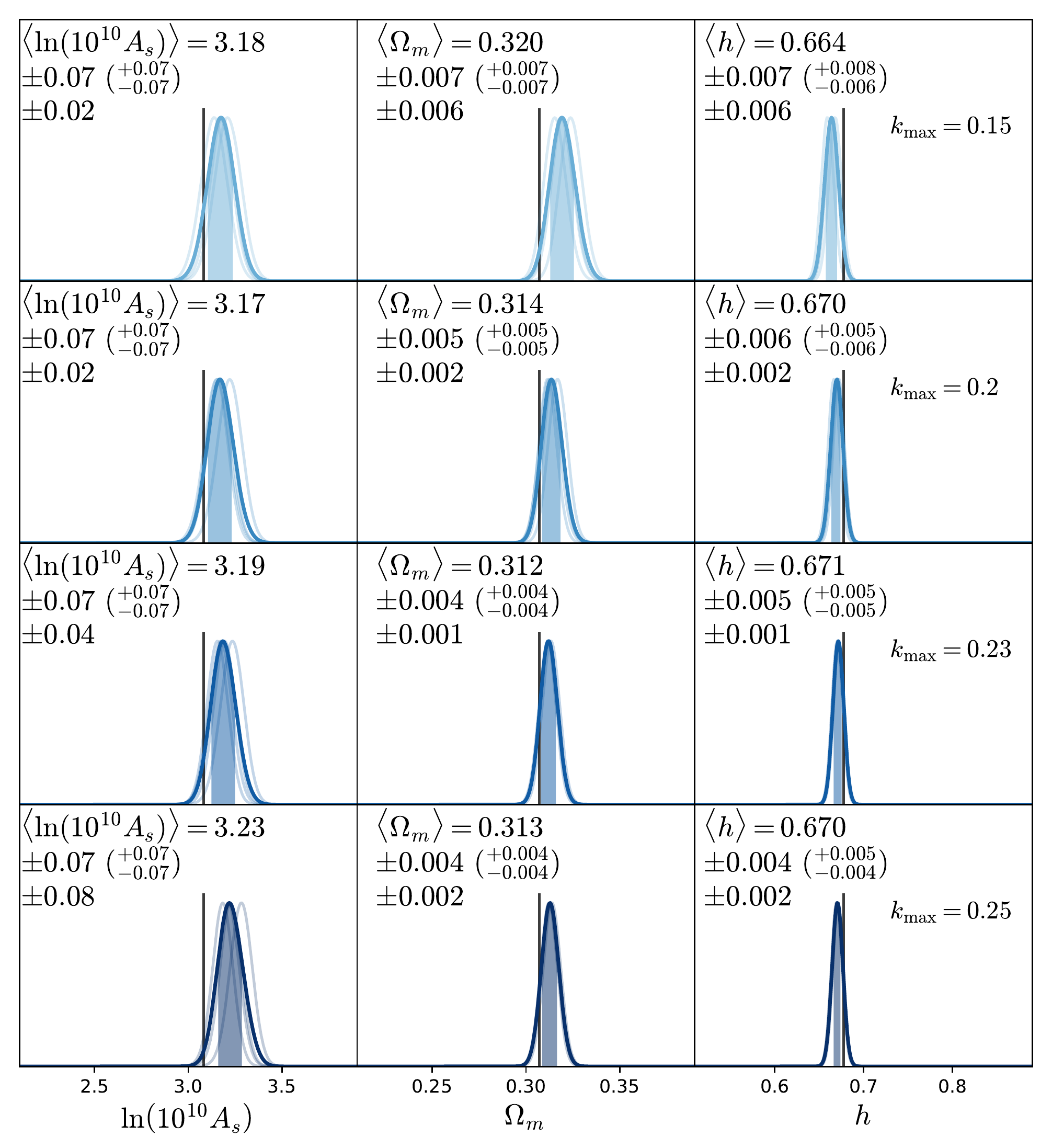}
\includegraphics[width=0.496\textwidth,draft=false]{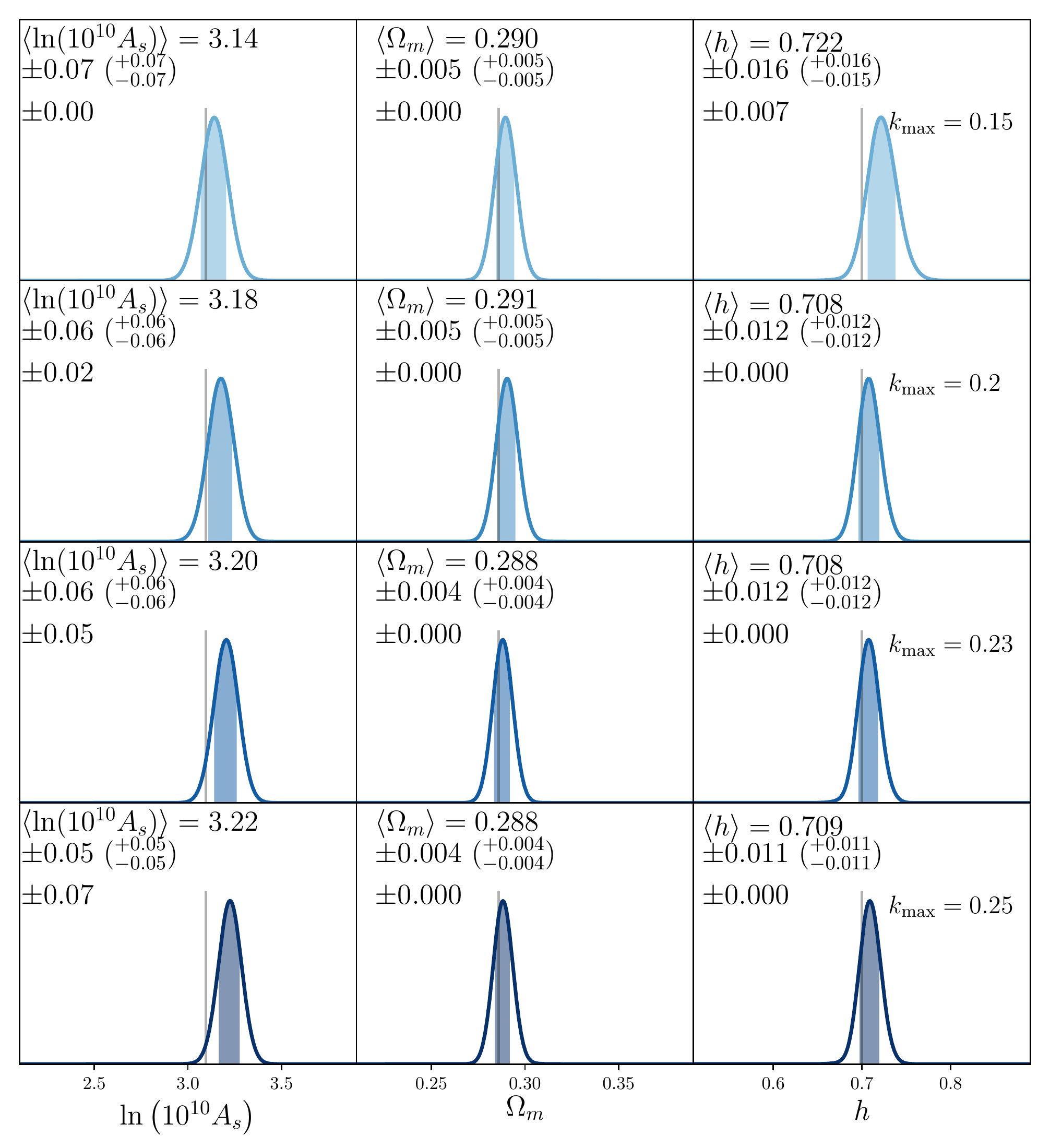}
\includegraphics[width=0.496\textwidth,draft=false]{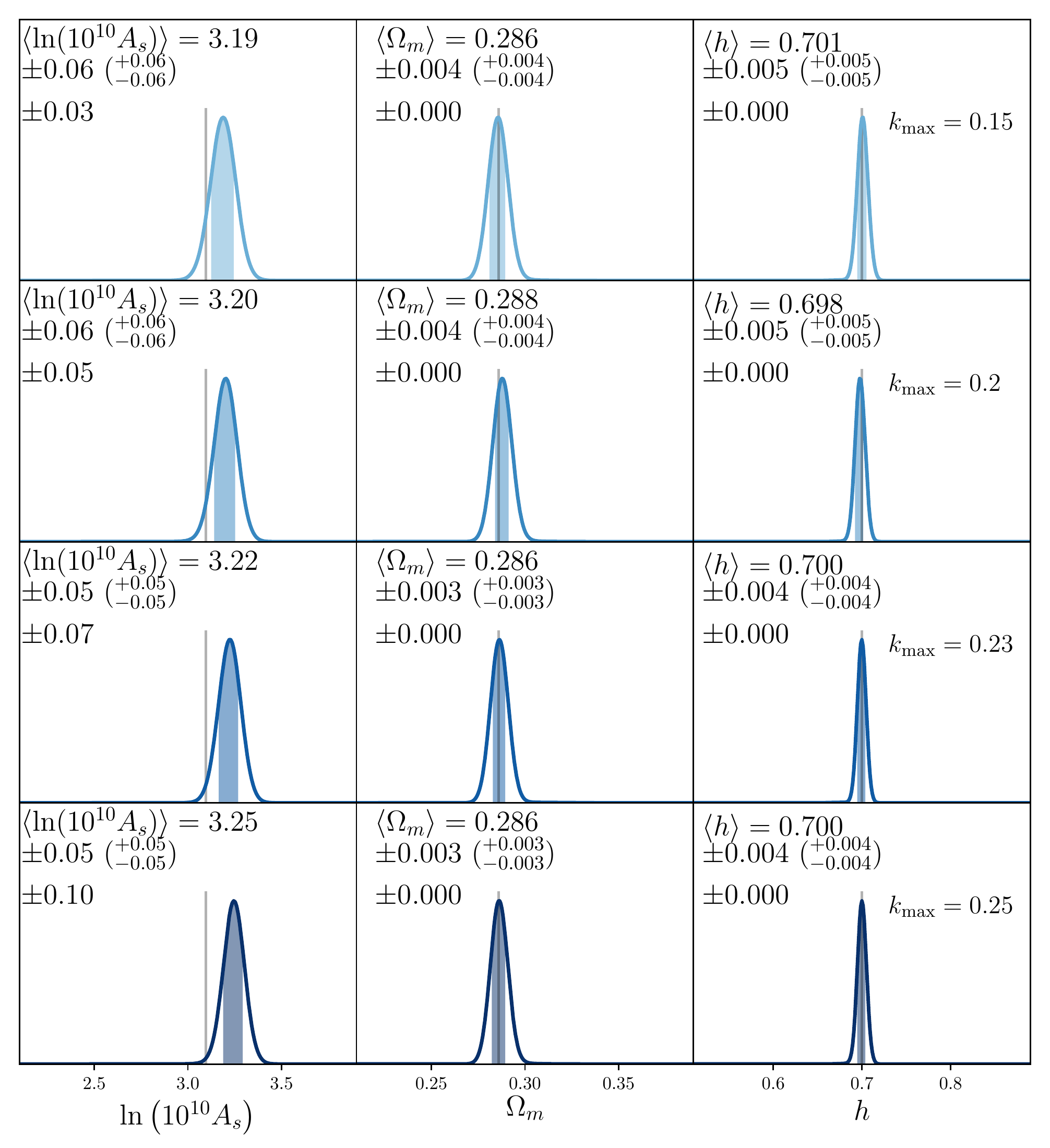}
\caption{\small Marginalized posterior distributions of the cosmological parameters obtained from fitting the power spectrum of the four Challenge boxes ABFG and their average ({\it top}) and for box D~({\it bottom}), without ({\it left}), and with  ({\it right})  the inclusion of Planck sound horizon prior. For each cosmological parameter, the expectation value is given together with the statistical error and the theory-systematic error (from top to bottom in each subplot). The vertical lines represent the fiducial cosmology of the simulation.
 \label{fig:posteriorchallenge} }
\end{figure}

\begin{figure}[h!]
\centering
\includegraphics[width=0.49\textwidth,draft=false]{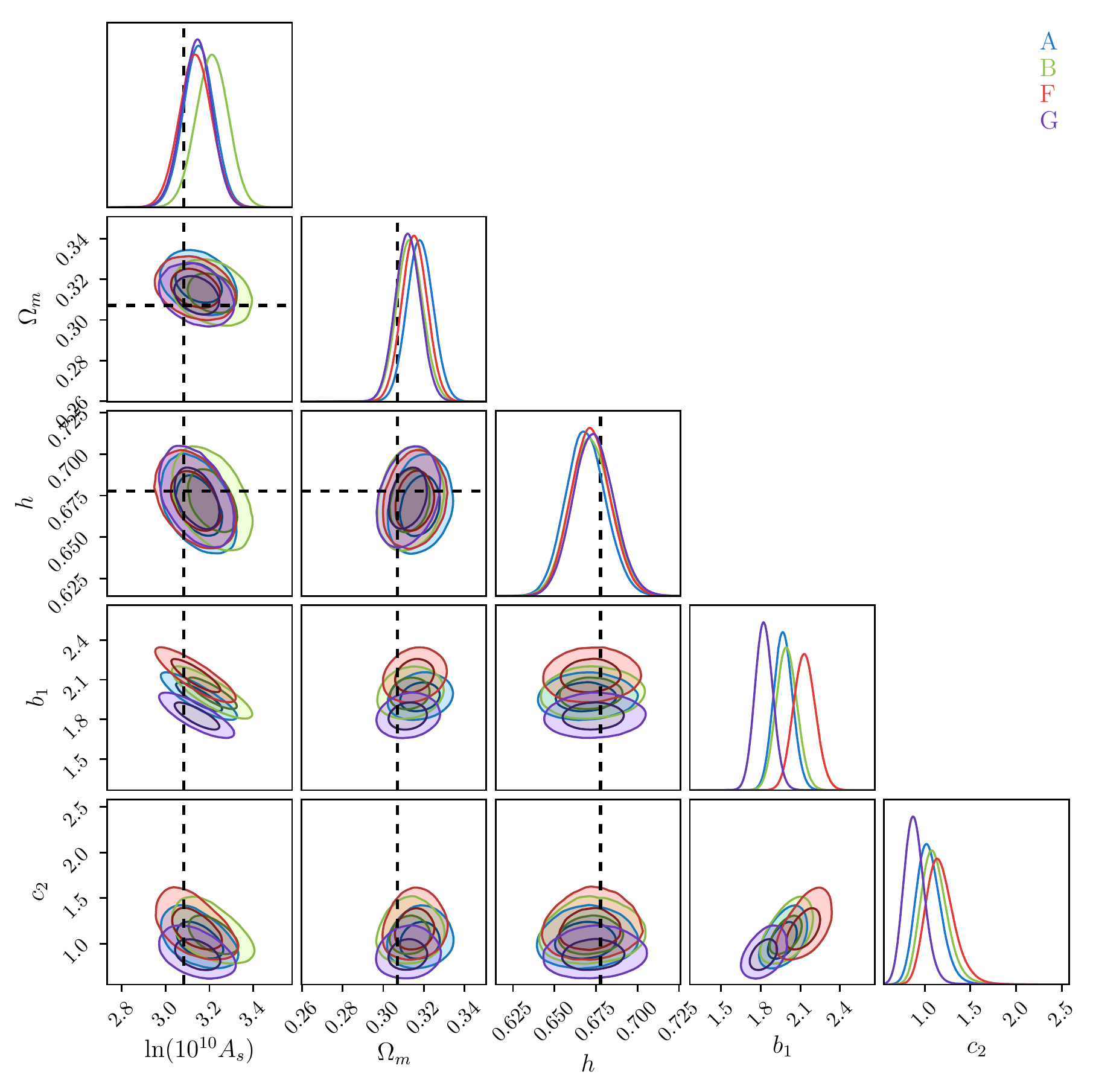}
\includegraphics[width=0.49\textwidth,draft=false]{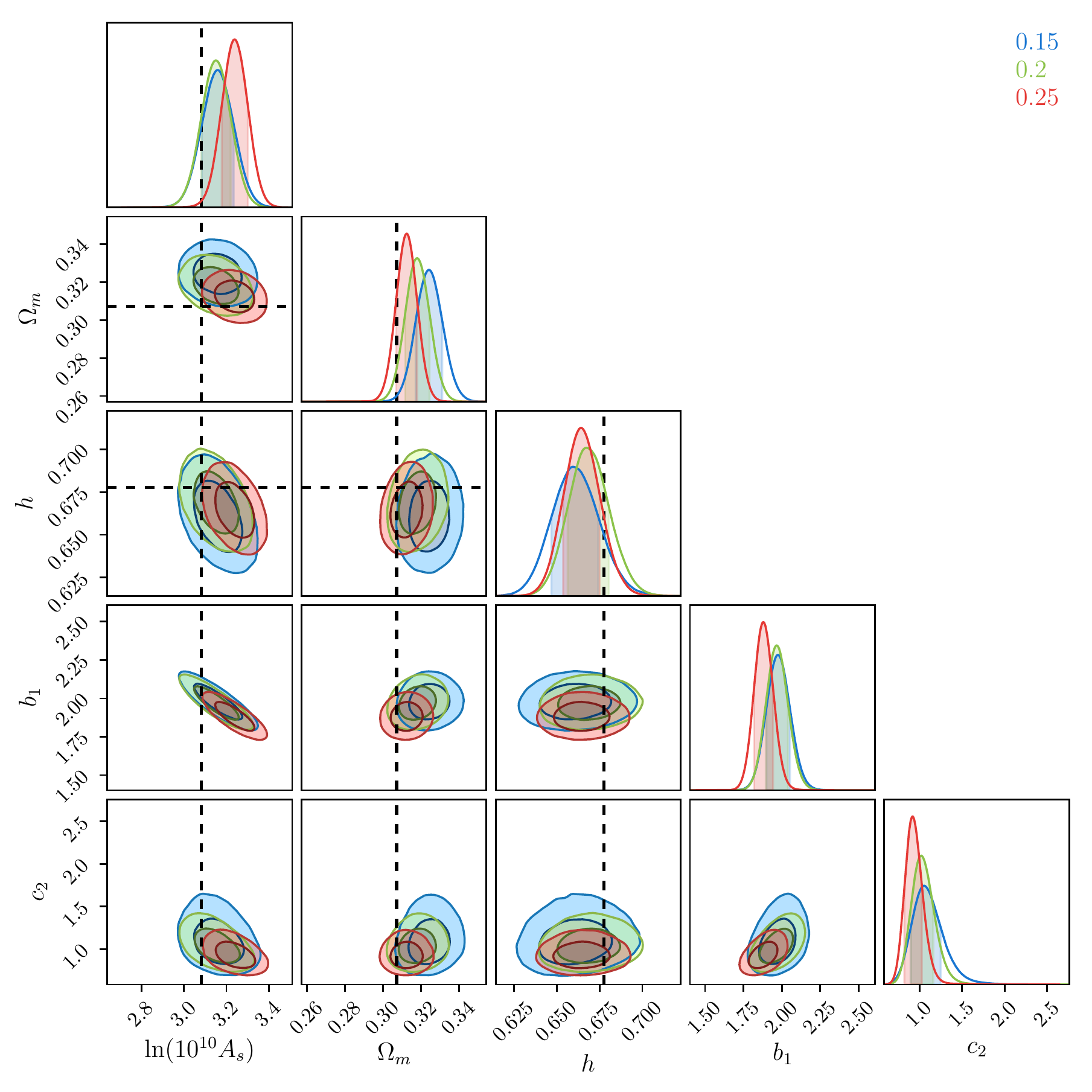}
\caption{\small 2D posterior distributions of the cosmological parameters $\ln (10^{10} A_s)$, $\Omega_m$, and $h$, and the EFT parameters $b_1$ and $c_2$, obtained from fitting the power spectrum up to $\kmax = 0.2 \hinvMpc$ of the challenge boxes ABFG ({\emph{left}}). On the right, we plot the same quantities for box A, as a function of $\kmax$ (\emph{right}).
We can see that we can measure all the cosmological parameters as well as the EFT parameters. All degeneracies are broken, though some to a different extent than others.
We see that the four HOD models on the left differ mainly for their value of $b_1$. The results for the cosmological parameters, which should be equal for all of them, appear to be remarkably consistent.}
\label{fig: challenge_power_corner}
\end{figure}

On the left of Fig.~\ref{fig:posteriorchallenge}, we show the marginalized posterior distributions of the cosmological parameters measured from the Challenge boxes, by fitting the individual power spectrum of each box with the partially marginalized likelihood of eq.~\eqref{eq:likelyhoodfour}. By power spectrum, here and in the rest of the paper, we mean the monopole and quadrupole of the power spectrum. On top of each posterior distribution, we give the expectation value of the cosmological parameter with its statistical error and theory-systematic error, to be read from top to bottom. In Fig.~\ref{fig: challenge_power_corner} we plot level contours of the partially marginalized likelihood, after marginalizing over all  but two EFT parameters.

We see that each box is fitted in very comparable terms. Concerning the cosmological parameters, we see that in each box we can measure $A_s,\Omega_m$ and $h$, breaking all degeneracies without any significant input from Planck or other CMB experiments~\footnote{This statement refers just to the inclusion of the sound horizon prior, which is a typical procedure to calibrate BAO data (see for example~\cite{Eisenstein:2005su,Percival:2007yw}). We remind the reader that we are fixing $\Omega_b/\Omega_c$ and $n_s$ to their Planck value, and that we discussed in footnote~\ref{gootnote:firstfisher} about the degradation of our measurements when this priors are relaxed, concluding that we would still have a significant measurement of the parameters. }. We will comment in the next subsection on the physical origin of this.

We call $\sigma_{\rm sys}$ an estimate of the theoretical error associated with the higher-order terms in the EFTofLSS prediction that we have not computed yet and so are not included in the model we fit to the data. Usually $\sigma_{\rm sys}$ is called a systematic bias, but we do not use this nomenclature in order to avoid confusion with the EFTofLSS parameters or with other forms of systematic errors. 
The challenge boxes A B F and G are highly correlated, as their initial conditions are the same (since they are extracted from the same dark-matter simulation, the Big MultiDark). Thus, $\sigma_{\rm sys}$ is measured by computing the distance between the true value and the $68\%$ confidence region of the average over the four boxes of the distribution of the posteriors of each box for a given parameter. In particular, $\sigma_{\rm sys}$ is taken to be zero when the true value is within the $68\%$ confidence region. We perform the same procedure for box D. Then, since box D and boxes A B F and G are constructed from independent simulations, we combine the two measurements of the distance of cosmological parameters from the fiducial ones as independent Gaussian distributions, and measure the theoretical systematic error as the distance of the one-sigma region of the resulting distribution from zero. 

Due to the finiteness of the statistical error of the boxes, this procedure implies that our  systematic error could be underestimated by an amount equal to $\sigma_{\rm stat}/{\sqrt{2}}$ for Challenge boxes. On the other hand, we are declaring the detection of a theoretical systematic error if the fiducial value of the simulation is just off the 68\% confidence region of our posteriors. This is clearly a conservative choice. Given this, and given how larger is the combined volume of these simulations compared to the one of BOSS, we consider the measurements we perform of the theoretical error accurate enough.

The challenge boxes are high precision N-body simulations with HOD modeling.
On box D, we find no evidence of theoretical systematic error in $\Omega_m$ and $h$ all the way to $\kmax=0.25\hinvMpc$, except a small $\sigma_{\rm sys}=0.007\simeq 0.3 \sigma_{\rm stat, \,CMASS}$ on $h$ at $\kmax=0.15\hinvMpc$. On $A_s$, we find a theoretical systematic error that starts very small at $\kmax = 0.2$, where we measure $\sigma_{\rm sys}=0.02\simeq\sigma_{{\rm stat,\, CMASS\, x\, low-}z}/7$, and becomes comparable to $\sigma_{\rm sys} \simeq0.4\sigma_{{\rm stat,\, CMASS\, x\, low-}z}$ and $0.5\sigma_{{\rm stat,\, CMASS\, x\, low-}z}$ at $\kmax=0.23\hinvMpc$ and $\kmax=0.25\hinvMpc$ respectively.

On the combination of boxes ABFG, we similarly find no evidence of theoretical systematic error in $\Omega_m$ and $h$ all the way to $\kmax=0.25\hinvMpc$, with the exception of a small $\sigma_{\rm sys}=0.006\simeq 0.4 \sigma_{\rm stat, \,CMASS}$ on $\Omega_m$ at $\kmax=0.15\hinvMpc$ {and a tiny $\sigma_{\rm sys}=0.002\simeq 0.1 \sigma_{\rm stat, \,CMASS}$ on $h$ at $\kmax=0.25\hinvMpc$} . Similarly, on $A_s$, we find a  theoretical systematic error that is {tiny} up to $\kmax=0.20\hinvMpc$, {and it is larger at $\kmax=0.23$, where we measure $\sigma_{\rm sys}=0.04\simeq\sigma_{{\rm stat,\, CMASS\, x\, low-}z}/3$, while it becomes a sizable $\sigma_{\rm sys}=0.09$ at $\kmax=0.25\hinvMpc$.} When we combine the measurements of the mean of ABFG and D, we can measure the theoretical systematic error more accurately by a factor of $\sim1/\sqrt{2}$.
We find no evidence of any systematic error on $h$ up to $\kmax=0.25\hinvMpc$, with a minimum detectable error of {$\sigma_{\rm sys,\; min}=0.008$}. For $\Omega_m$, we detect a systematic error $\sigma_{\rm sys}=0.0037$ at $\kmax=0.15\hinvMpc$ and $\sigma_{\rm sys}=0.0017$ at $\kmax=0.20\hinvMpc$, with a minimum detectable error $\sigma_{\rm sys,\; min}=0.0032$. In comparison to the error bars we will obtain on the observational data, all these systematic errors are negligibly small ($\lesssim \sigma_{{\rm stat,\, CMASS\, x\, low-}z}/6$), with the possible exception of $\kmax=0.15\hinvMpc$ where the error is about $\sigma_{\rm stat,\, CMASS}/4$, which we however assume to be due to a statistical fluctuation. Instead, on $\lnAs$, we detect a systematic error that becomes significant at $\kmax=0.20\hinvMpc$, {equal to 0.06, and at $\kmax=0.23\hinvMpc$, equal to 0.09}.  At $\kmax=0.20\hinvMpc$, we detect a systematic error on $\lnAs$ equal to 0.036, which is  is about $1/4\cot\sigma_{{\rm stat,\, CMASS\, x\, low-}z}$, which is sufficient for being negligible~\footnote{This interpretation is probably reinforced by the fact that our measurement of the theoretical systematic error is in this case potentially affected by a statistical fluctuation that is already visible at low $\kmax$.}. 

We conclude that we can safely perform the analysis of the data with negligible theoretical error up to {$\kmax=0.20\hinvMpc$}.
It is now worthwhile to remark  that a systematic error even as large as the statistical error can be accounted for by  shifting the posterior of the parameter by about this quantity. This has a very small effect on any statistically-significant statement that one can make on  the posterior of the parameters. Therefore, overall, the systematic errors appear to be small even in the highest wavenumbers that we consider. This in particular means that even though we will conservatively decide not to quote the result obtained $\kmax=0.23\hinvMpc$ as our main results, the systematic error in this case is still quite small, and one could decide to consider the ones obtained with this $\kmax$ to be the most significant results.

Though the challenge boxes are high precision N-body simulations with accurate HOD modeling, they lack the window function, redshift sampling and fiber collisions, that characterize a survey such as BOSS. While the challenge boxes are therefore a good tool to test for the theoretical systematic, they do not test for those. However, as we have discussed in Sec.~\ref{sec:window}, \ref{sec:redshiftselection}, and \ref{sec:fibercollisions}, all of these effects have been accurately implemented in our pipeline, and so we expect negligible systematic error from them. 

Overall, one should keep in mind that the window function does  not give a huge effect: even analyzing the data without accounting for it, the shift in the cosmological parameters is of the order of the statistical error. So, an implementation which is accurate just to about 10\% is enough to make any error negligible. However, to check the accuracy of the treatments of these survey systematics in our pipeline to an even higher level of safety, we analyze $16$ Patchy lightcone NGC boxes. The total volume is roughly equivalent to the one of one Challenge box, which allows us to measure survey-systematic errors to the same precision as we measured the theory-systematic errors using the Challenge boxes, though one should keep in mind that the level of accuracy of the mocks is less than the one of an N-body simulation. 
We  fit the $16$ boxes separately and do the product of the individual 1D marginalized posteriors. Even if this procedure is not fundamentally correct, as the correct procedure would be to multiply the full likelihoods, it is rather conservative and good enough to measure the theoretical systematic error. Another reason to fit the Patchy boxes individually is to measure the response of the EFT to our choice of priors on the EFT parameters for one volume equivalent to the one of the data. Indeed, for a volume as large as the one of Challenge box, all EFT parameters are well determined. However, when we turn to a volume like the one of Patchy or data, we might encounter unbroken degeneracies that can potentially drive the fit to an unphysical region. Therefore, it is important to test if our choice of priors, which follows solely from theoretical considerations, is enough to keep the fit in the physical region. 

\begin{figure}[h!]
\centering
\includegraphics[width=0.496\textwidth,draft=false]{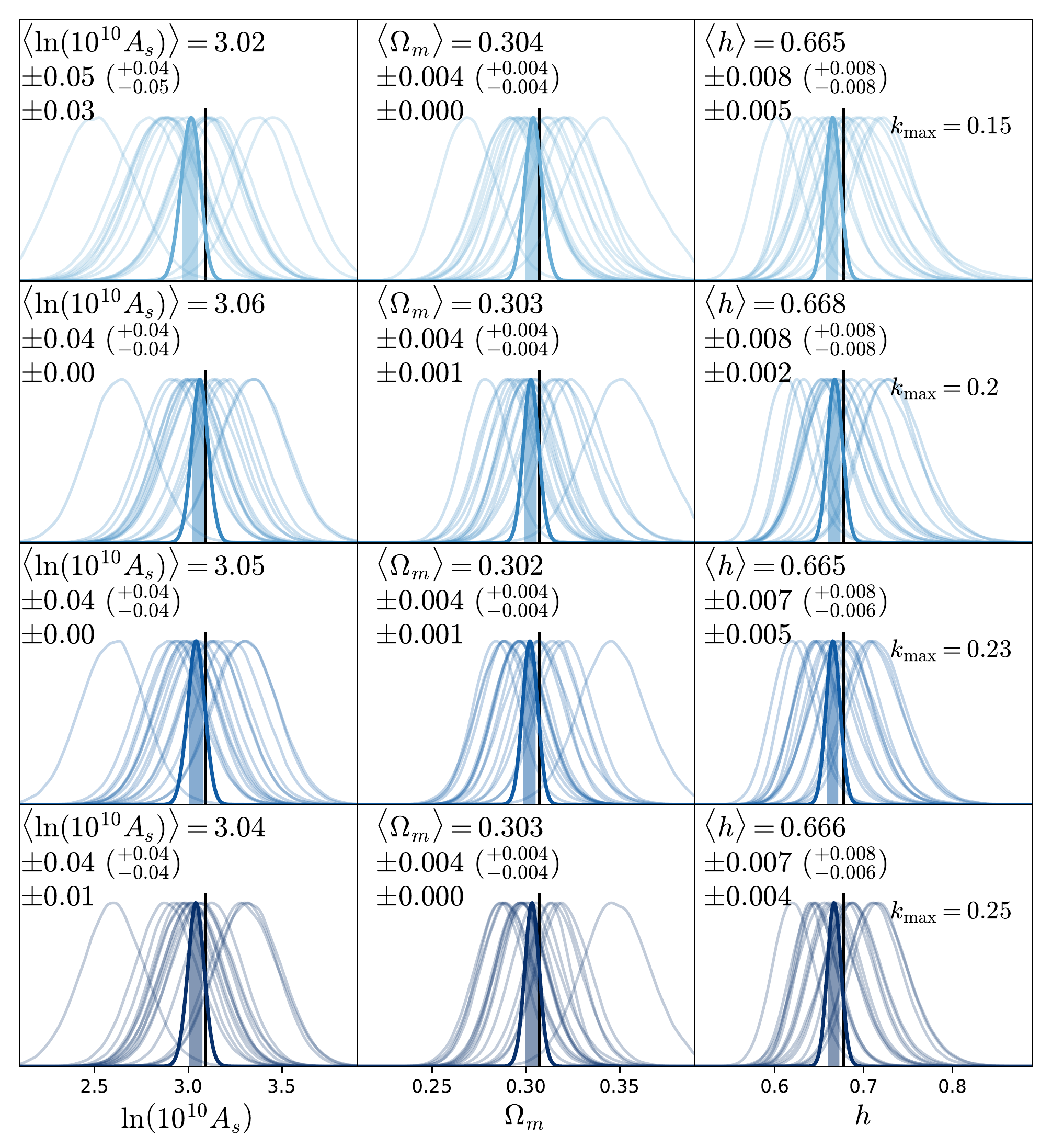}
\includegraphics[width=0.496\textwidth,draft=false]{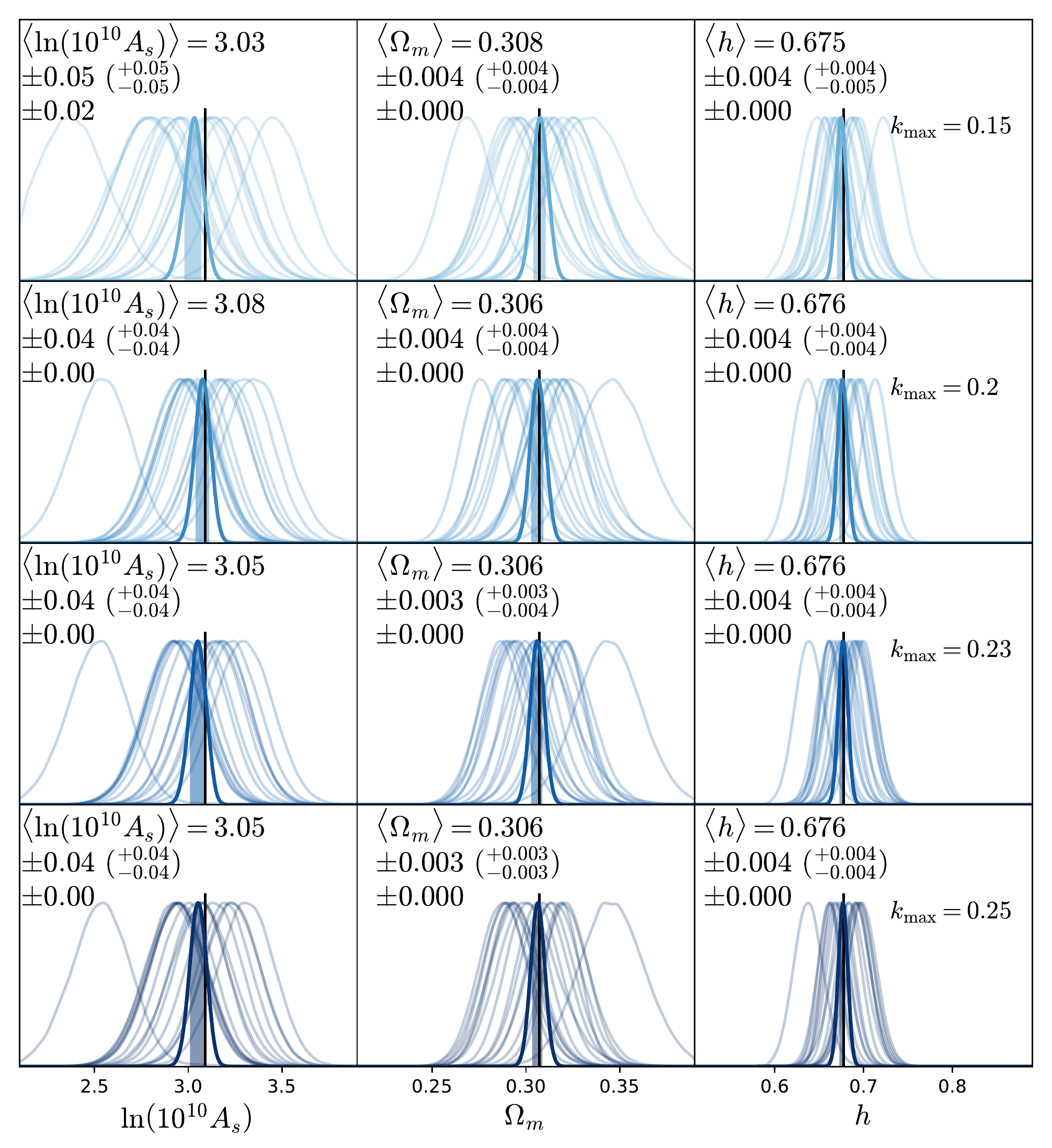}
\caption{\small Marginalized posterior distributions of the cosmological parameters obtained from fitting the power spectrum of $16$ Patchy lightcone NGC boxes, without ({\it left}), and with  ({\it right})  the inclusion of Planck sound horizon prior. For each cosmological parameter, we quote the expectation value together with the statistical error and the theory-systematic error of the product of the individual 1D posteriors (from top to bottom in each subplot). The vertical lines represent the fiducial cosmology of the simulation.
 \label{fig:posteriorpatchy} }
\end{figure}

On the left of Fig.~\ref{fig:posteriorpatchy}, we show the marginalized posterior distributions of the cosmological parameters measured from the Patchy mocks, by fitting the individual power spectrum of each box with the partially marginalized likelihood of eq.~\eqref{eq:likelyhoodfour}.  We find for each cosmological parameter at each $\kmax$ that the product of the 1D posteriors appear with negligible small systematics errors, with vanishing ones for $A_s$ and $\Omega_m$, and tolerably small $\sigma_{\rm sys} \lesssim 0.005 \simeq \sigma_{{\rm stat,\, CMASS\, x\, low-}z}/4$ on $h$. 

We conclude that, on one hand, all survey geometrical and selection effects are well accounted for in our pipeline, and, on the other hand, our theoretical priors are suitable to keep the parameters in the physical region for a volume like BOSS data.

Before moving on, we would like to make the following comment about the theory-systematic error as extracted from simulations. One can take the point of view that the simulations and the HOD modeling are exact, {\it i.e.} they represent the exact distribution of a population of galaxies generated according to some physical mechanism. Any deviation from the exact representation of the physical model (for example if the trajectory of the particles had a numerical error) would induce  a numerical systematic error, that we here assume to be negligible~\footnote{Of course, due to modeling issues, it is unlikely that this assumption is entirely true.}.  It is a well-recognized limitation that the HOD models do not actually simulate the formation of galaxies from first principles, but rather `model' them according to some rule guided by a mixture of physical processes and data. We point out that, from the EFTofLSS point of view, it is absolutely not important that the HOD model correctly represents a population of galaxies  that actually exists in the universe. In fact, as long as the galaxy-formation physics is modeled by the  HOD with a process that satisfies the principle of general relativity, {\it i.e.} it is controlled by locally observable quantities, then the EFTofLSS {\it is expected} to be able to describe the distribution of this population with arbitrary accuracy. Of course, for a given model, the non-linearities induced by the higher order biases can be larger than in other models. In this case, to reach a given accuracy, one might need to include higher-order terms in the EFTofLSS. For example, in~\cite{Perko:2016puo, Nadler:2017qto}, higher-order derivative terms were included in order to reproduce to a given accuracy the distribution of highly-biased tracers. This discussion makes it clear that, in the absence of numerical systematic error, the inferred cosmological parameters obtained from the four HOD models should not only be equal to the true ones, but also the same for each model. By looking at Fig.~\ref{fig: challenge_power_corner}, we notice that indeed all HOD models of the challenge boxes give very similar results for all the cosmological parameters. All the EFT-parameters that have not been already integrated out have almost identical values apart from the linear bias $b_1$, which seems to be the one over which the differences in the HOD models accumulate. It is therefore not a complete  surprise that the model with the smallest systematic error is the Challenge box~G, which has the smallest $b_1$, and so the neglected loops are the smallest.

\subsection{Physical considerations}\label{sec:physicalconsiderations}

We have seen above that we can measure the three cosmological parameters $A_s,\,\Omega_m,$ and~$h$, without any significant theory-systematic error. As we will see, this measurement can be done also on the SDSS/BOSS data. Given that the official analysis of the SDSS/BOSS data do not provide independent measurements of all of these quantities without the addition of external priors, here we briefly highlight why it is actually possible to measure them~\footnote{We stress again that we are fixing $n_s$ and  $\Omega_b/\Omega_c$ to the fiducial value of the simulations, or, for the data, to the central value from Planck2018. However, we fix these two cosmological parameters just for computational reasons, the analysis could be generalized to include them as well. We discussed earlier on the implication of this choice on the error bars of the parameters.}. 

First, we  remind the reader that we are working in the $\Lambda$CDM setting. Thanks to  the accurate modeling of LSS provided by the EFTofLSS, we are allowed to use, implicitly or explicitly, all the relations among quantities that are implied by this (as for example the dependence of $f$ on $\Omega_m$ from eq.~(\ref{eq:fdef}), or the dependence of the sound horizon at decoupling on $\Omega_m$ that is implicitly encoded in the transfer functions), without the need to introduce artificial free parameters beyond the ones given by the EFTofLSS. In this sense, our analysis is identical to the standard CMB analysis of Planck or WMAP. 

We  also remind the reader that we are analyzing all the simulation data up to {$\kmax=0.2 \hinvMpc$ and $0.23 \hinvMpc$}, without making any splitting between oscillating and broad-band signal. However, for the purpose of explaining the physical origin of the signal, we can separate the part of the signal that originates from the BAO and from the broad-band shape of the power spectrum. 

Focusing on the BAO signal, this is well approximated by an oscillating signal in the power spectrum, corresponding to a peak in the real space correlation function. The wavelength of this oscillation, equivalent to the location of the peak in real space, is the sound horizon at the end of the baryon-drag epoch. In the case of the CMB, the observed quantity is the angle at which this peak appears, $\theta_{\rm CMB}$, which scales as $\theta_{\rm CMB}\simeq r_d(z_{\rm CMB})/D_A(z_{CMB})$, where $r_d(z_{\rm CMB})$ is the  sound horizon at the time of decoupling, $z_{\rm CMB}\sim 1100$, and $D_A$ is the angular diameter distance. Ref.~\cite{Percival:2002gq} (see Sec. 4.4) carefully derives that such a measurement is mainly dependent on the combination $\Omega_m h^{3.4}$~(\footnote{This derivation applies if we keep $\Omega_b$ as an independent parameter. If the ratio $\Omega_b/\Omega_c$ is held fixed, then the scaling is rather $\Omega_m h^5$ (see Eq.~(11) of \cite{Aghanim:2018eyx}). However, for the purpose of our general discussion, we are only interested in describing how the degeneracy between $\Omega_m$ and $h$ can be broken by single-redshift LSS observations alone, and we therefore focus on the more general case where $\Omega_b$ is left as a free parameter.}). In the case of LSS, the angle (or ratio of length scales) under which the BAO oscillation is observed, $\theta_{\rm LSS}$, scales as the geometric mean of what is observed parallel and perpendicular to the line of sight (see for example~\cite{Eisenstein:2005su,Percival:2007yw})
\be\label{eq:BAOtotal}
\theta_{\rm LSS}\simeq \frac{ r_d(z_{\rm CMB})}{\left(D_A(z_{\rm LSS})^2\cdot c\, z_{\rm LSS}/H(z_{\rm LSS})\right)^{1/3}}\ .
\ee
Straightforwardly repeating the analysis of~\cite{Percival:2002gq}, we find that, 
\be
\left.\frac{\partial\log  \theta_{\rm LSS}}{\partial\log \Omega_m}\right|_{ \Omega_{m}=0.3,\; h=0.7}=-0.12\,\ ,  \qquad \left.\frac{\partial\log  \theta_{\rm LSS}}{\partial\log h}\right|_{ \Omega_{m}=0.3,\; h=0.7}=0.48\ ,
\ee
for $z_{\rm LSS}=0.5$. $\theta_{\rm LSS}$ therefore mainly depends on the combination $\Omega_m h^{-4}$ (\footnote{Eq.~(\ref{eq:BAOtotal}) summarizes the information that we acquire if we only observe the monopole. In reality the parallel and perpendicular directions to the line of sight can be disentangled by observing also the quadrupole. In this case, there are two angles or ratios of length scales that are observed: 
\be
\theta_{\rm LSS,\; \parallel }\simeq \frac{ r_d(z_{\rm CMB})}{D_A(z_{\rm LSS})}\ ,\qquad
\theta_{\rm LSS,\; \perp }\simeq \frac{ r_d(z_{\rm CMB})}{ c\, z_{\rm LSS}/H(z_{\rm LSS})}\ .
\ee 
$\theta_{\rm LSS,\; \parallel }$ mainly depends on the combination $\Omega_m h^{-3}$, while $\theta_{\rm LSS,\; \perp }$ mainly depends on the combination $\Omega_m h^{-10}$. This also allows to break the degeneracy between $\Omega_m$ and $h$.}). 

The relative amplitude of the BAO peak with respect to the smooth part instead gives a measurement of $\sim \Omega_m \, h^2$. This is quite intuitive as the amplitude of the oscillating part, unlike its wavelength, is not affected by projection effects, and therefore simply scales as the density of baryons and dark matter at the time of recombination, which, in the case of fixed $\Omega_b/\Omega_c$, simply scales as $\Omega_m h^2$ (see~\cite{Mukhanov:2003xr} for a very pedagogical review and derivation).  

We now pass to include the broad-band signal. The linear power spectrum monopole and quadrupole give a measurement of $b_1^2 A_s^{(\kmax)}$ and $b_1 f A_s^{(\kmax)}\sim b_1\Omega_m^{1/3} A_s^{(\kmax)}$. Here $A_s^{(\kmax)}$ represents the amplitude of the linearly evolved power spectrum at the maximum wavenumber of our analysis, as this is where the signal peaks.    $A_s^{(\kmax)}$ is not quite the same as the primordial amplitude, $A_s$, because the evolution of modes that enter the horizon before or after matter-radiation  equality is different. In fact, a scale-invariant primordial power spectrum decays, roughly, as $k^{-2}$ around the scales of interest (see for example~\cite{Carrasco:2013sva}). We therefore have:
\be\label{eq:Askmax}
A_s^{(\kmax)}\equiv A_s \left(\frac{k_{\rm eq}}{\kmax}\right)^2\ ,
\ee
where $k_{\rm eq}$ is the wavenumber that enters the horizon at matter-radiation equality, and scales as $k_{\rm eq}[\hinvMpc]\propto (\Omega_m h^2)^{1/2}$. We also approximated $f$ with the well known approximate relation $f(z)\sim \Omega_m(z)^{4/7}$ (see, for example,~\cite{Hamilton:2000tk}), which, for $z\simeq 0.5$ and accounting for the $z$-dependence of the matter density, gives $f(z\simeq 0.5)\sim \Omega_m(z=0)^{1/3}$.

  These scalings allow us to solve for $b_1\propto \Omega_m^{1/3}$, $A_s^{(\kmax)}\propto b_1^{-2}\propto \Omega_m^{-2/3}$. The constraint on the sound horizon from the BAO gives $h\sim \Omega_m^{1/4}$. Having expressed $b_1, h,$ and $A_s^{(\kmax)}$ in terms of~$\Omega_m$, we can finally use the amplitude of the BAO peak, that, after substituting the scaling of  $h$, goes as $\propto \Omega_m^{1/2}$, to determines $\Omega_m$ and, in consequence, $b_1, h,$ and $A_s^{(\kmax)}$ and, finally, from eq.~(\ref{eq:Askmax}), $A_s$.  

On top of this information, the power spectrum has a broad shape which also carries information. For example, the power spectrum changes slope for $k$'s around $k_{\rm eq}$. This carries further information about the combination $\Omega_m h^2$. Furthermore, the nonlinear power spectrum additionally contributes to lower the error bars on the cosmological parameters, as, though there are many new EFT parameters that are introduced at this stage, each contribution carries a different functional shape dependence, and so can be measured, at least in principle, independently. Since the number of shapes is larger than the number of parameters, this adds information. 

The way the cosmological and EFT parameters are determined can indeed be recognized by the contours of Fig.~\ref{fig: challenge_power_corner}. The constraint on the linear power spectrum from the monopole clearly must play a leading role, due to the smallness of the error bars. This amplitude scales as $b_1^2A_s^{(\kmax)}\sim b_1^2 A_s k_{\rm eq}^2\propto b_1^2 A_s \Omega_m h^2$. This explains the anticorrelated degeneracies between $A_s\,\&\, b_1$, between $A_s\, \&\,h$. As just described, all the degeneracies are broken by the information from the quadrupole and the BAO, but in general in a subleading way so that these (anti)correlated degeneracies are still measurable.  This might explain why we do not find a strong anticorrelation in $\Omega_m\,\&\,h$, but maybe, at least in the Challenge boxes, a slight correlation between $\Omega_m$ and $h$. The scaling of the observed BAO angle in LSS, $\theta_{LSS}$, in fact might explain this.

The fact that we are able to independently determine the cosmological parameters allows for another interesting application of our results to the realm of astrophysics, beyond the study of primordial cosmology. In fact, the breaking of the degeneracy between $A_s$ and $b_1$ allows for a measurement of $b_1$. As it is evident from the left panel of Fig.~\ref{fig: challenge_power_corner}, we can also measure the other EFT parameters. But Fig.~\ref{fig: challenge_power_corner} tells us that the bias $b_1$  correlates with the specific HOD model, and, in general, all the EFT parameters correlate with it. This means that by measuring the power spectrum, we can infer the EFT parameters, and, from them, the galaxy formation mechanism that is at play.

 \subsection{Inclusion of Planck sound horizon prior\label{sec:planckprior}}
 
As we have just discussed in Sec.~\ref{sec:physicalconsiderations},  the measurement of the sound horizon at decoupling using the BAO is very important in order for us to measure the cosmological parameters, as it strongly constraints the linear combination $\Omega_m/ h^4$. While we have just shown that this can be done using only redshift-space clustering data, it is a fact that the CMB experiments, and the Planck satellite most lately, have provided exquisite measurement of  the sound horizon at decoupling, or rather of the angle $\theta_{\rm CMB}$ described earlier, to per-mille precision. It is therefore interesting to use this measurement as a prior for the LSS analysis. In fact, contrary to other CMB measurements, this one is particularly insensitive to physical processes that act in the late universe.
  
  The combination of the LSS data with the CMB-measured sound horizon has a long history of important applications in the community, dating back to the first measurements of this effect (see for example~\cite{Eisenstein:2005su,Percival:2007yw,Lemos:2018smw} and references therein), and it is normally considered essential to obtain constraints on the  Hubble parameter from LSS data. However, contrary to the usual cases, here we will use it simply to {\it improve} our constraints, as we already have interesting constraints, and in general to explore the effect of such a CMB-motivated prior.

\begin{figure}[h]
\centering
\includegraphics[width=0.49\textwidth,draft=false]{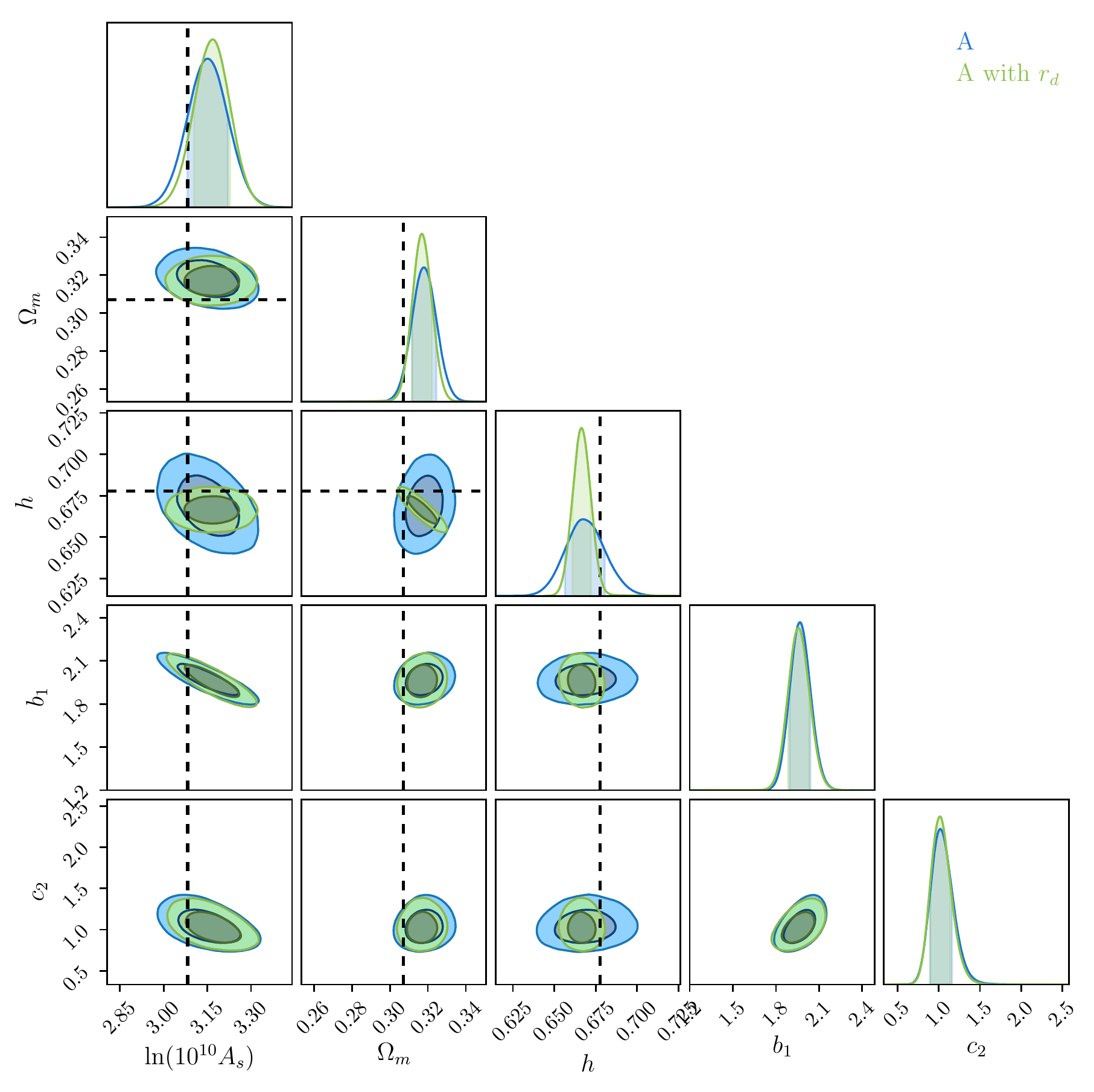}
\includegraphics[width=0.49\textwidth,draft=false]{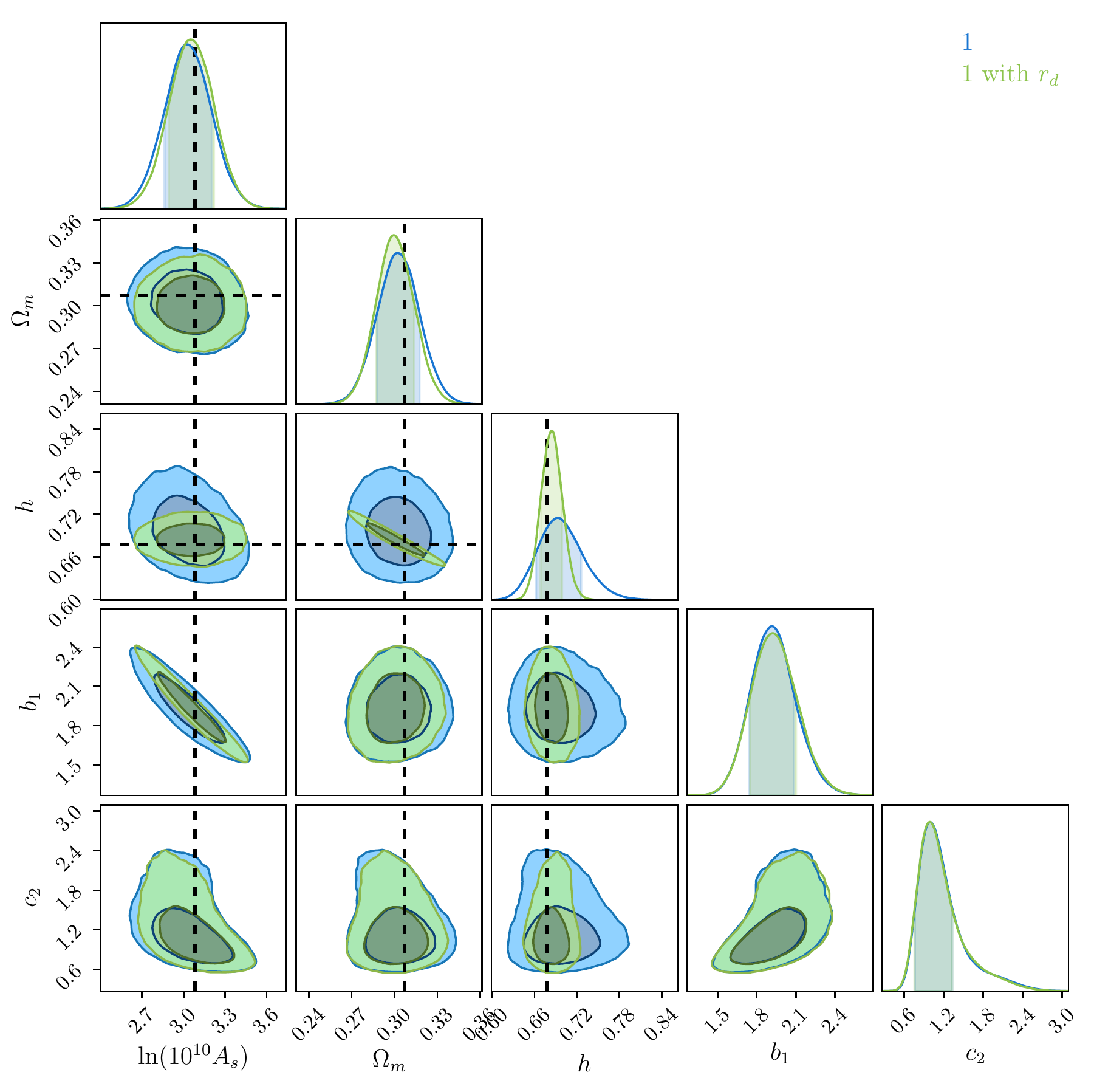}
\caption{\small 2D posterior distributions of the cosmological parameters $\ln (10^{10} A_s)$, $\Omega_m$, and $h$, and the EFT-parameters $b_1$ and $c_2$, with and without Planck sound horizon prior, obtained from fitting the power spectrum up to $\kmax = 0.2 \hinvMpc$ of the challenge box A ({\emph{left}}) and the Patchy box~1~(\emph{right}). In comparison to Fig.~\ref{fig: challenge_power_corner}, we can see that the Planck constraint on the sound horizon has restricted the allowed region for $\Omega_m$ and $h$, and left us with a very strong, anticorrelated, potential degeneracy between $\Omega_m$ and $h$,  We remind that this potential degeneracy had already been broken by the SDSS data, without the Planck sound horizon prior. }.
 \label{fig:Planck_challenge_power_corner} 

\end{figure} 

As we saw in the former section, the theory-systematic error that we measured is much smaller than the statistical error of BOSS DR12. However, once we apply the Planck prior on the sound horizon, one might worry that our statistical errors will drop and our theoretical systematics, which naively will not change, will dominate the error budget. This motivates us to test the EFTofLSS against the simulations again, using now a prior on the sound horizon. We refer to App.~\ref{app:sound_horizon} for details on the formulas for the sound horizon at decoupling and  for the numerical value of the prior we use.

The reason why the result of this test is non trivial is because it can happen that the Planck prior will reduce not only the statistical error, but also the theory-systematic error. On the right of Fig.~\ref{fig:posteriorchallenge}, we plot the marginalized likelihood for the cosmological parameters, after the inclusion of the Planck prior on the sound horizon. In Fig.~\ref{fig:Planck_challenge_power_corner} we plot the contours of the posterior for the partially marginalized likelihood.  By focusing on Challenge ABFG and D for definiteness, we see that, typically, the statistical error on $\lnAs$ has been reduced in a marginal way, the one on $\Omega_m$ by about around 25\% on average, while the one on $h$ is reduced by about a factor of two. In terms of theoretical systematic error, on ABFG we find overall tolerably small systematic errors for $\kmax\leq 0.23\hinvMpc$: 
\begin{itemize}
\item[-] {On $\lnAs$, $\sigma_{\rm sys}=0.02 \lesssim \sigma_{{\rm stat,\, CMASS}}/7$ at $\kmax=0.15 \hinvMpc$ and $\kmax=0.2\hinvMpc$, $\sigma_{\rm sys}=0.04 \simeq \sigma_{{\rm stat,\, CMASS\, x\, low-}z}/3$ at $\kmax=0.23\hinvMpc$, and $\sigma_{\rm sys}=0.08 \simeq 0.6\cdot \sigma_{{\rm stat,\, CMASS\, x\, low-}z}$ at $\kmax=0.25\hinvMpc$; }
\item[-] {On $\Omega_m$, $\sigma_{\rm sys} = 0.006 \simeq 0.4 \sigma_{\rm stat\,, CMASS} $ at $\kmax=0.15\hinvMpc$, $\sigma_{\rm sys} = 0.002 \simeq \sigma_{\rm stat\,, CMASS}/5$ at $\kmax=0.2\hinvMpc$ and $\kmax=0.25\hinvMpc$, $\sigma_{\rm sys} = 0.001 \simeq \sigma_{{\rm stat,\, CMASS\, x\, low-}z}/9$ at $\kmax=0.23\hinvMpc$;}
\item[-] {On $h$, $\sigma_{\rm sys} = 0.006 \simeq 0.4 \sigma_{\rm stat\,, CMASS}$ at $\kmax=0.15\hinvMpc$, $\sigma_{\rm sys} = 0.002 \simeq \sigma_{\rm stat\,, CMASS}/6$ at $\kmax=0.2\hinvMpc$ and $\kmax=0.25\hinvMpc$, and $\sigma_{\rm sys} = 0.001 \simeq \sigma_{{\rm stat,\, CMASS\, x\, low-}z}/10$ at $\kmax=0.23\hinvMpc$.}
\end{itemize}
{The potentially dangerous systematic errors are at $\kmax=0.15 \hinvMpc$ on $\Omega_m$ with $\sigma_{\rm sys} = 0.006 \simeq 0.4 \sigma_{\rm stat,\, CMASS}$ and on $h$ with $\sigma_{\rm sys} = 0.006 \simeq 0.4\sigma_{\rm stat,\, CMASS}$, but we assume that at such a low $\kmax$ they are due to a statistical fluctuation. At $\kmax=0.25 \hinvMpc$ we find a large systematic error on $\lnAs$ of $\sigma_{\rm sys} = 0.08$, and a marginally important one at $\kmax=0.23\hinvMpc$.}

{On box D, we detect vanishing errors for $\Omega_m$ and $h$ at all $\kmax$'s.
On $\lnAs$, we find a theoretical systematic error that grows with $\kmax$. It starts with $\sigma_{\rm sys}=0.03\simeq\sigma_{\rm stat,\, CMASS}/5$ at $\kmax=0.15\hinvMpc$, then we find $\sigma_{\rm sys}=0.05\simeq\sigma_{\rm stat,\, CMASS}/3$ at $\kmax=0.2\hinvMpc$
 and becomes $\sim \sigma_{\rm stat,\, CMASS}/2$ at $\kmax=0.23\hinvMpc$ and $\sigma_{\rm sys}=0.1 \simeq 0.8 \sigma_{{\rm stat,\, CMASS}}$ at $\kmax=0.25\hinvMpc$.}

 When we combine the measurements of ABFG \& D, we find {on $\Omega_m$ and $h$ neglibly small systematic errors with respect to what we find in the data ($\sigma_{\rm sys} \leq \sigma_{\rm stat,\, CMASS}/5$). On $\lnAs$,  we find a large systematic error of $\sigma_{\rm sys} = 0.11$ at $\kmax=0.25 \hinvMpc$, while we have $\sigma_{\rm sys} = 0.07$ at $\kmax=0.23 \hinvMpc$ and $\sigma_{\rm sys} = 0.05\simeq0.4 \sigma_{\rm stat,\, CMASS}$ at $\kmax=0.15 \hinvMpc$ and $\kmax=0.2 \hinvMpc$. Though the systematic error so detected in $A_s$ at $\kmax=0.20\hinvMpc$ is not completely negligible, we notice that it is the same already at $\kmax=0.15\hinvMpc$, so we assume that its detected magnitude is affected by a statistical fluctuation, and so we neglect it and proceed with the analysis at $\kmax=0.20$ by assuming negligible systematic error.}

\begin{figure}[t!]
\centering
\includegraphics[width=0.30\textwidth,draft=false]{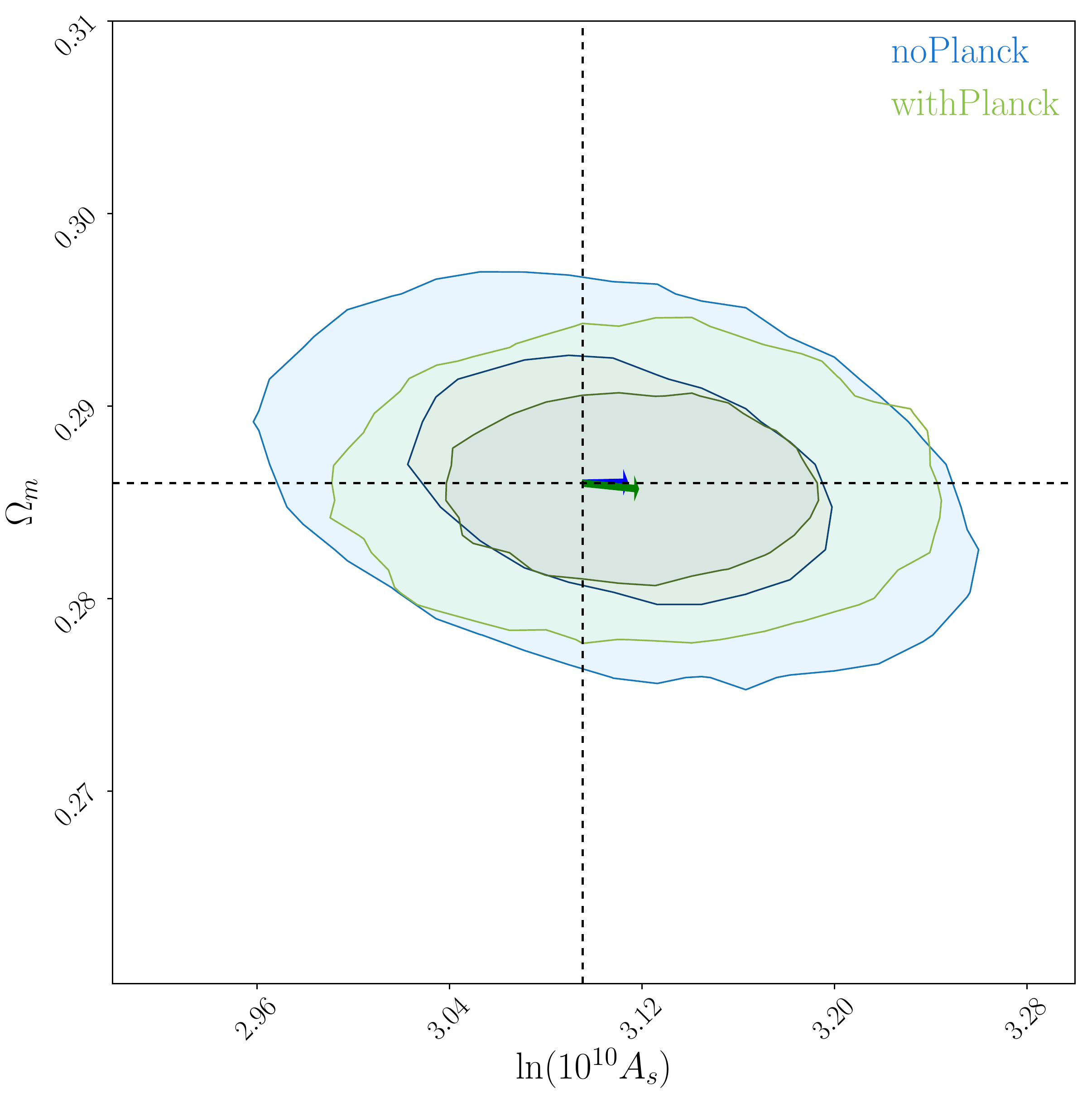}
\includegraphics[width=0.30\textwidth,draft=false]{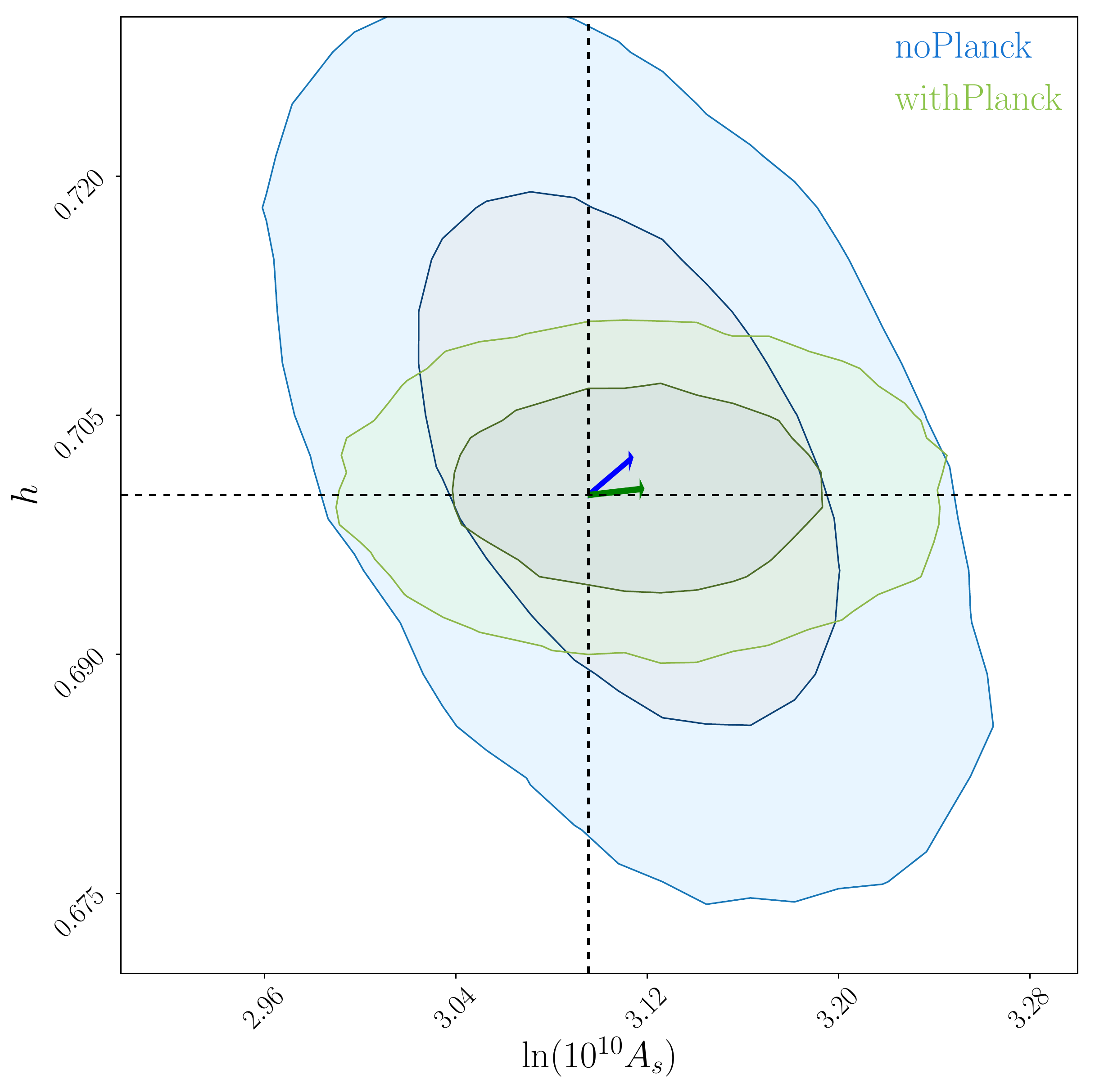}
\includegraphics[width=0.30\textwidth,draft=false]{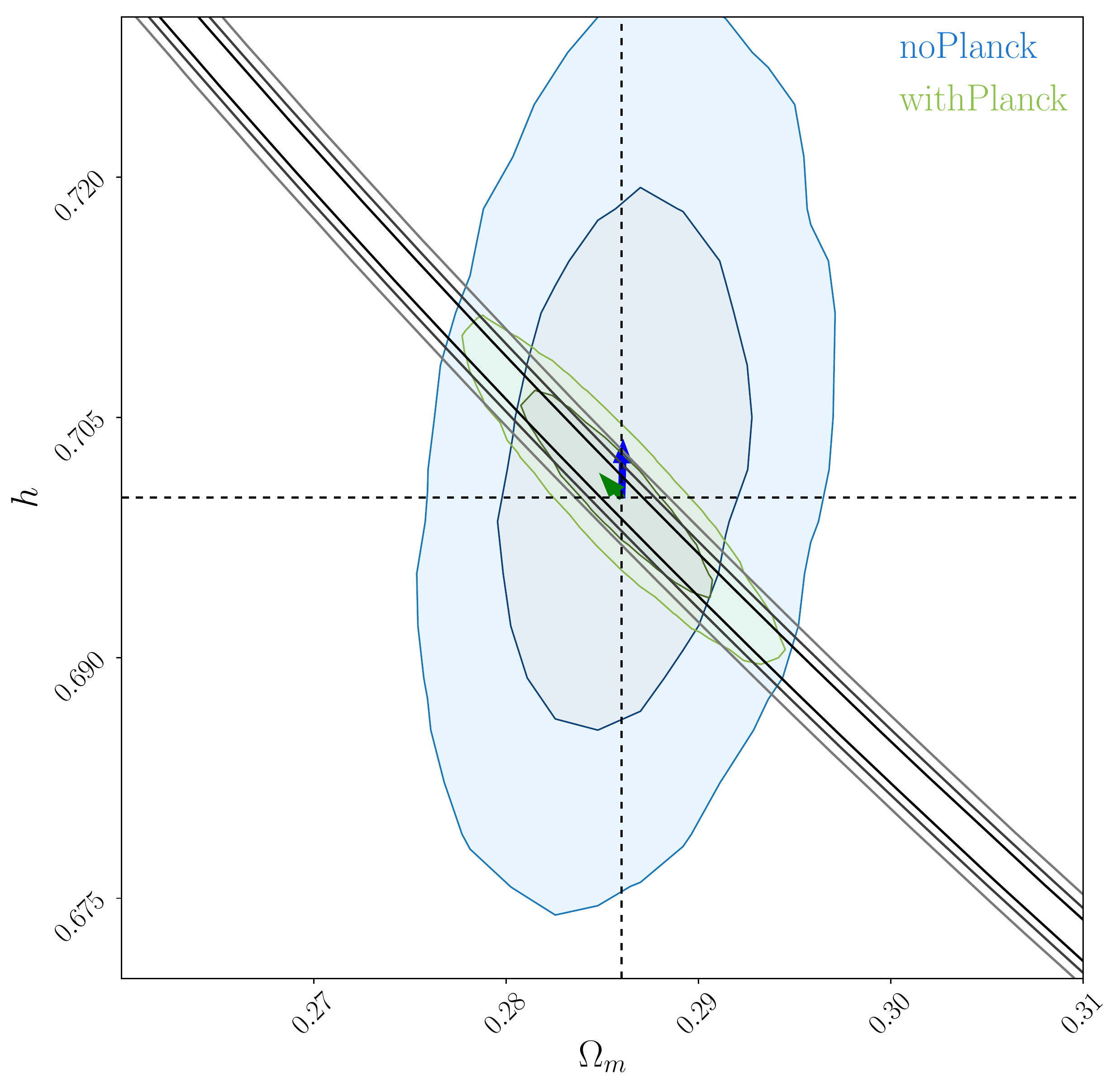}
\caption{\small In the absence of any prior, the theoretical systematic error can be represented as a vector living in the space of the cosmological parameters that goes from the true value to the maximum likelihood point as obtained from the MCMC's. $\sigmasys$ for each parameter is the component of this vector in the corresponding direction. If we impose a prior, the likelihood is forced to have support only on a region that is narrower or equal to the one without prior. Therefore, roughly, the resulting theoretical-systematic error obtained after imposing a prior is equal to the one we have without any prior, projected on the region supported by the prior. This tends to reduce the systematic error. On the $\Omega_m$-$h$ plane, we plot the level contour of the Planck sound horizon at decoupling $r_d$. \label{fig: planck_challenge_power_arrow} }
\end{figure}

It is worthwhile to explain why the systematic error has in some cases decreased, after the inclusion of the Planck sound horizon prior. This is naively surprising given the reduction in the statistical error, which also affects the size of the smallest systematic error we can detect. The reason is explained by Fig.~\ref{fig: planck_challenge_power_arrow}. In the absence of any prior, the theoretical systematic error can be represented as a vector living in the space of the cosmological parameters that goes from the true value to the average maximum likelihood point as obtained from the MCMC's. $\sigmasys$ for each parameter is nothing but the component of this vector in the corresponding direction (taking also into account the uncertainty associated with the finite number of simulations at our disposal). If we impose a prior, for example on the sound horizon at decoupling, the likelihood is forced to have support only on a region that is narrower or equal to the one without prior. Therefore, the resulting theoretical-systematic error that is obtained after imposing a prior is equal to the one we have without any prior, projected to the region where the prior has support. This reduces the systematic error, and, in general, it suggests that great care should be used when estimating how the the systematic error changes after the addition of priors.

{{These considerations suggest an effective way to include the theoretical error for a given~$\kmax$ without the use of simulations but taking into account the effect induced by the priors. One could add to the EFTofLSS prediction evaluated at a given order, {\it i.e.} the model, a reasonable expression for the functional form of a higher-order term, for example $ c_{\rm higher} (k/\knl)^4 P_{11}(k)$, whose amplitude is parametrized by a parameter, here $ c_{\rm higher}$. Then one can run the MCMC including the new contribution, with a prior on this parameter, here $c_{\rm higher}$, to be a number of order one. One can then measure the resulting shift induced on the maximum likelihood point for a cosmological parameter, and call that number the theory-systematic error. We leave the development of this potential method to estimate the theoretical systematic error to future work. }}

\subsection{Inclusion of bispectrum\label{sec:bispectrum}}

\begin{figure}[h!]
\begin{center}
\includegraphics[width=0.49\textwidth,draft=false]{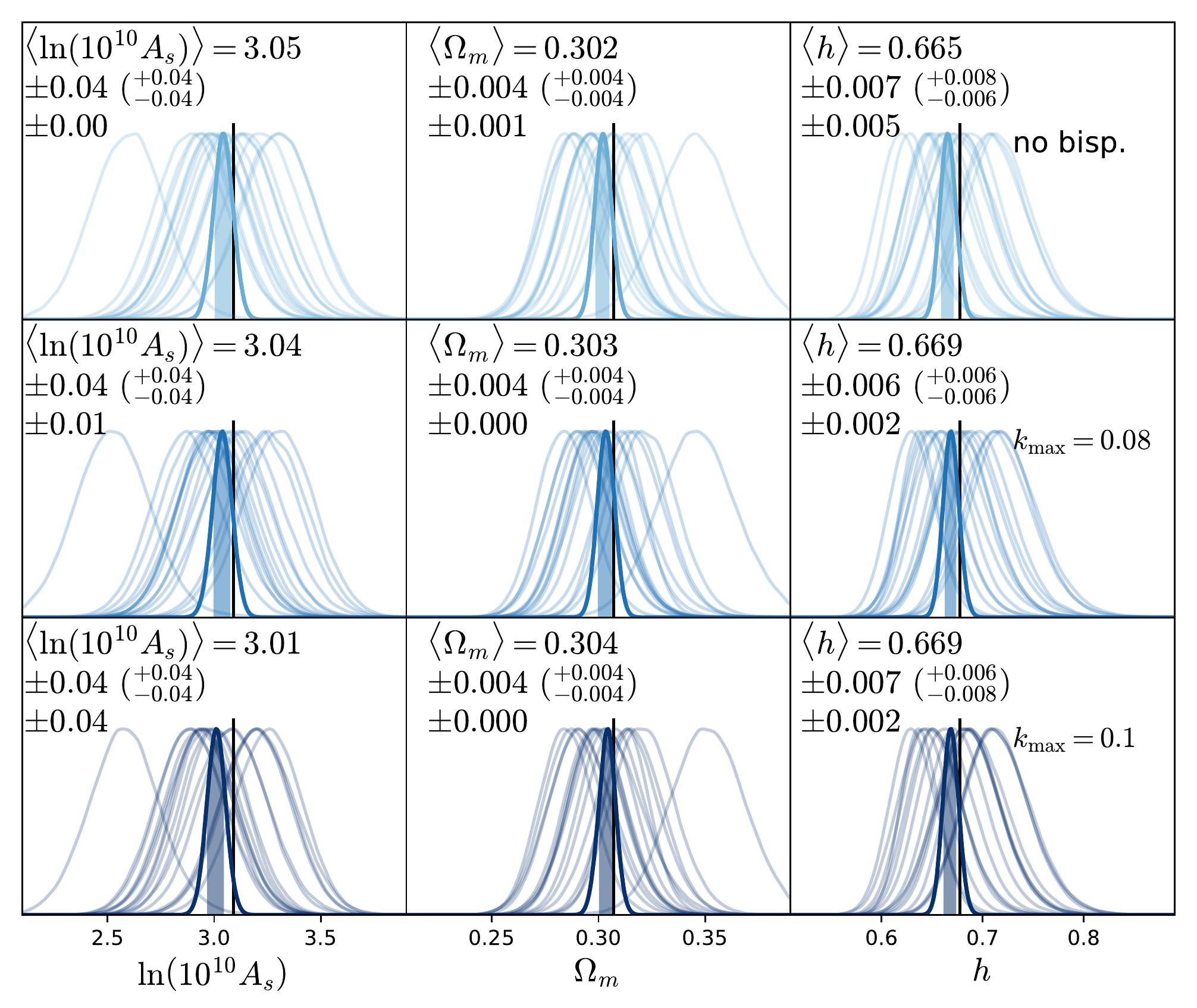}
\includegraphics[width=0.49\textwidth,draft=false]{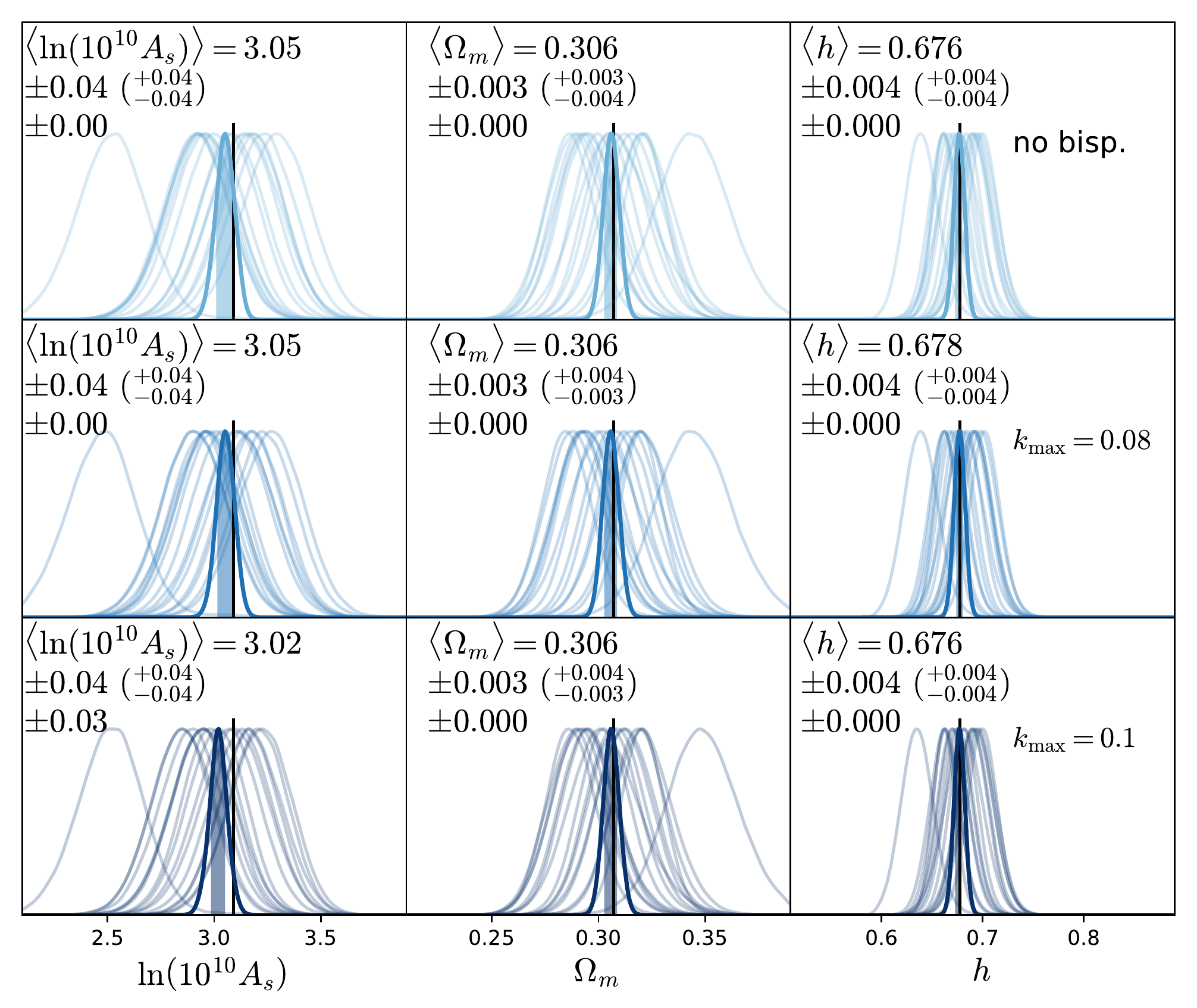}
\caption{\textcolor{blue}
\small Marginalized posterior distributions of the cosmological parameters obtained from fitting the power spectrum and the bispectrum monopole of 16 Patchy lightcones NGC ({\it left, thin lines}) and their average~({\it left, thick lines}), and with the inclusion of Planck sound horizon prior~({\it right}). We analyze the bispectrum either with $\kmax=0.08\hinvMpc$ or with $0.1\hinvMpc$. For each cosmological parameter, the expectation value of the product of the posteriors is given together with the statistical error and the theory-systematic error (from top to bottom in each subplot).
 \label{fig:posteriorpatchybisp} }
\end{center}
\end{figure}

We now discuss the effect of the inclusion of the bispectrum monopole in the likelihood, obtaining the likelihood given by eq.~(\ref{eq;marginalizedbispLikli}) in App.~\ref{sec:appendix_bispectrum_lik}. Here we use the Patchy mock catalogues to test our analysis pipeline, as we have measurements of the bispectrum monopole only for this kind of simulations.

We point out that, as discussed at the end of App.~\ref{sec:appendix_bispectrum_lik}, the application of the window function to the bispectrum is done approximately, and this might induce additional systematic error bars. In order to better limit the systematic error induced by the window function, we analyze the bispectrum monopole including  only triangles whose minimum wavenumber is $k_{\rm min}=0.04\hinvMpc$, and whose maximum one is either $\kmax=0.08\hinvMpc$ or $\kmax=0.1\hinvMpc$, adding $10$ and $34$ triangles, respectively. {For the power spectrum, we use $\kmax=0.23\hinvMpc$, when analyzed in combination with the bispectrum, though we noticed earlier that in this case the systematic error might be marginally important~\footnote{Since the theoretical error grows with $\kmax$, and since we will have negligible systematic error on simulation, this is justified.}.}

{When analyzing the Patchy mocks using the partially marginalized likelihood of eq.~(\ref{eq;marginalizedbispLikli}), we obtain the results presented in Fig.~\ref{fig:posteriorpatchybisp}. In the same figure, as we did in Sec.~\ref{sec:theorysys}, we also plot the product of the 1-D posteriors, which is a non-rigorous procedure, given that we should multiply the full posteriors, but we checked that, given our error bars, it is quite a conservative procedure.  We see that the error bars are reduced in a marginal way by the inclusion of the bispectrum monopole both with and without the inclusion of the Planck prior. We detect negligible systematic error upon the inclusion of the bispectrum monopole, though it is barely so when the bispectrum is analyzed up to $\kmax=0.1\hinvMpc$.}
  
We also point out that the improvement in the determination of the EFT parameters is limited. 
Comparing the statistical errors between the cases with and without the bispectrum in Table~\ref{tab:summarysims}, we see that we get a marginal improvement on $A_s$ of about $10\%$ and no improvement on $\Omega_m$ and $h$.
 Furthermore, we have checked that the improvement in the measurement of the EFT-parameters is much more significant if one were to use larger priors for the same parameters. 
 
Our conclusions about the inclusion of the bispectrum are clearly limited by several factors: the fact that we do not perform a high order calculation does not allow us to go to high wavenumbers; our approximate implementation of the window function forces us to exclude low wavenumbers. 
It would be interesting to improve in all of these aspects in future work.

\subsection{Summary of tests on simulations}\label{sec:simulsummary}

\begin{figure}
\centering
\includegraphics[width=1\textwidth,draft=false]{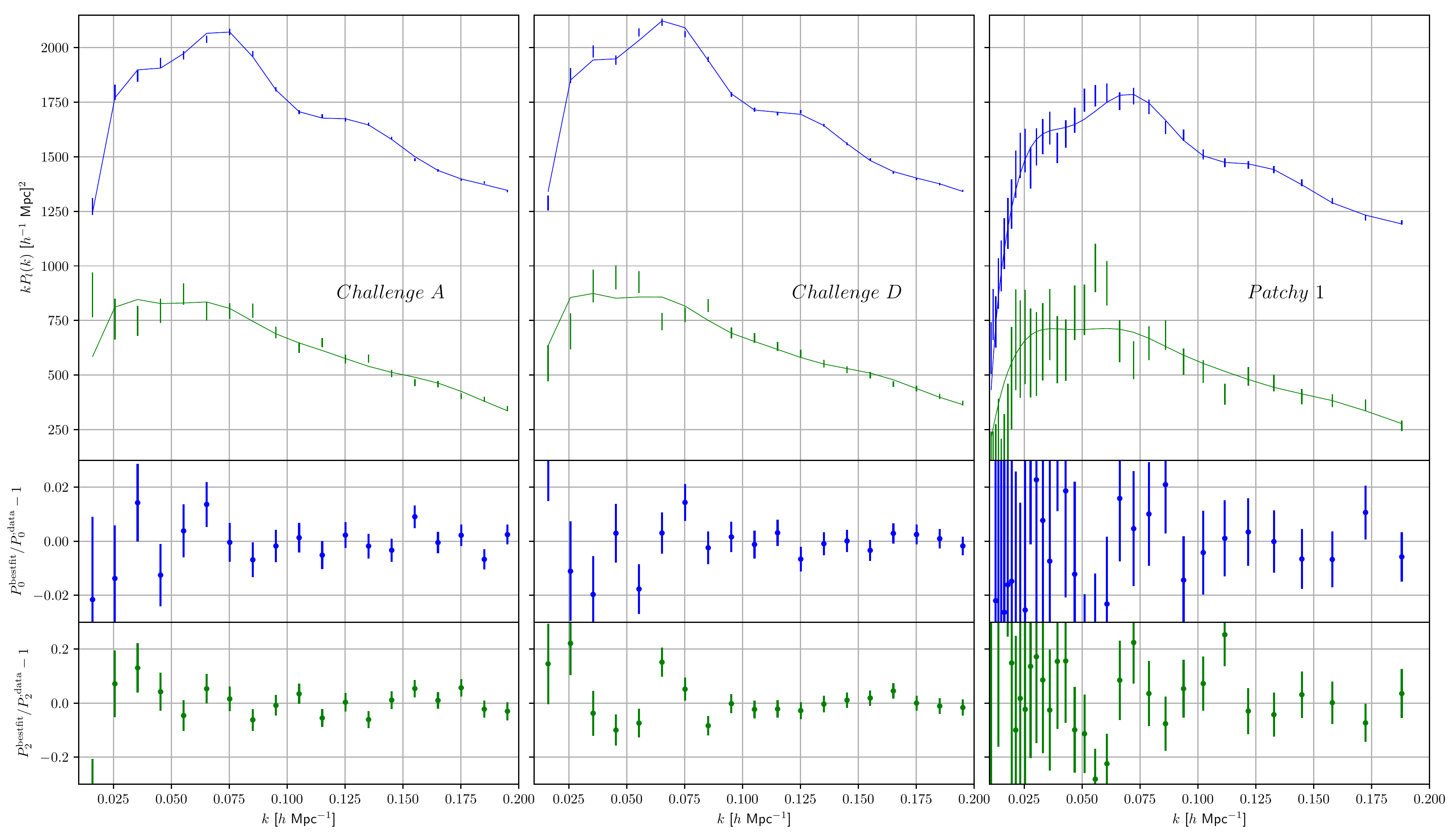}
\includegraphics[width=1\textwidth,draft=false]{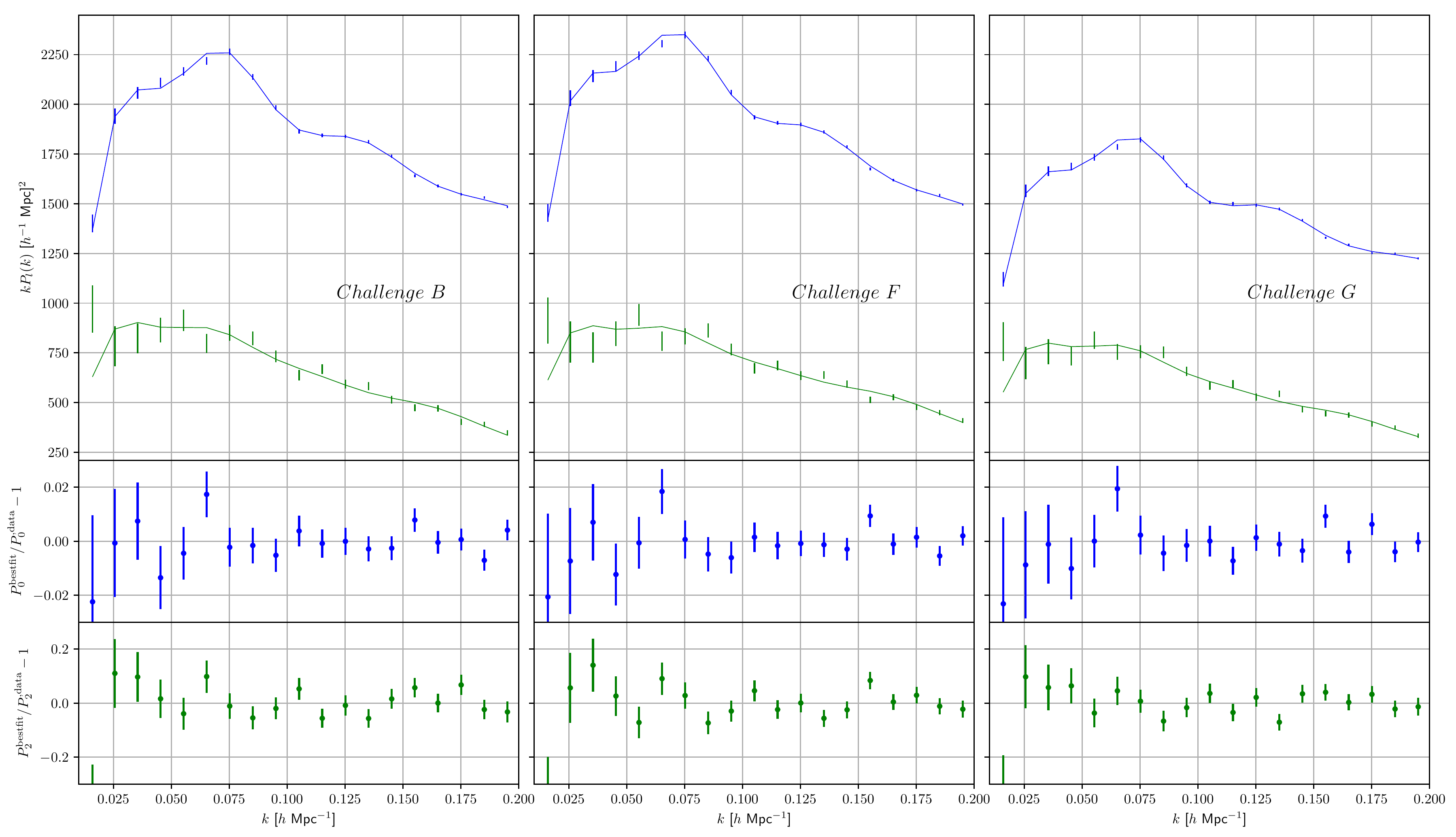}
\caption{\small  Best-fit power spectrum on measurements on (from top to bottom and left to right): Challenge box $A$, Challenge box D and  Patchy box 1; Challenge box B, F, G; and residuals. The error bars represent the diagonal terms of the covariance. The $\chi^2/{\rm d.o.f}$ is respectively $1.72$, $1.75$ and $1.17$ for the top row, and 1.72, 1.86, and 1.65 for the bottom row. We remind the reader that the error bars of Challenge are about a factor of four smaller than the ones of Patchy or the observational data we analyze.}
\label{fig:bestfit-simulation}
\end{figure}

We summarize the  results of our tests on numerical simulations in Table~\ref{tab:summarysims}.  We find that we can determine $A_s$, $\Omega_m$ and $h$ from the power spectrum data, with no significant role played by other observational data. We also can measure several of the EFT parameters, allowing, in principle, to learn about the galaxy formation mechanism. 

We give in Table~\ref{tab:bestfit} the best-fit results from the fit of the power spectrum up to {$\kmax = 0.2 \hinvMpc$}, from both the simulations and the BOSS DR12 data we analyze next. {All minimum $\chi^2$ per degree of freedom (min $\chi^2/$d.o.f.) give an acceptable $p$-value}, attesting for the goodness of the fit. The values of the EFT parameters are all of order of the linear bias $b_1$ ($\sim \mathcal{O}(2)$), all lying well within the chosen priors for this analysis. We remind that the Challenge boxes were run without shot noise, as this has been subtracted exactly from the measurements. We provide in App.~\ref{app:covariance} the correlation of all parameters as well as their 2D posterior distributions from the fits of the power spectrum of Challenge A and D. In Fig.~\ref{fig:bestfit-simulation} we plot the best-fit model against the power spectrum measured from all simulations: Challenge A, B, F, G and D and Patchy box 1. 
We do not notice any warning feature in the residuals.

\begin{table}[h!]
\centering
\footnotesize
\begin{tabular}{l|ll|ll|ll}\hline
                                 & \multicolumn{2}{l|}{$\lnAs$}                & \multicolumn{2}{c|}{$\Omega_m$}     & \multicolumn{2}{c}{$h$}      \\ \hline
                                 & $\sigma_{\rm stat}$ & $\sigmasys$ & $\sigma_{\rm stat}$ & $\sigmasys$ & $\sigma_{\rm stat}$ & $\sigmasys$ \\ \hline
Challenge ABFG                        & 0.07                & 0.04        & 0.005               & 0.000       & 0.011              & 0.000       \\
Challenge ABFG with $r_d$             & 0.07                & 0.04        & 0.004               & 0.001       & 0.005               & 0.001       \\
 Challenge D                     & 0.06                & 0.05        & 0.004               & 0.000       & 0.012                & 0.000       \\
 Challenge D with $r_d$             & 0.05                & 0.07        & 0.003               & 0.000       & 0.004               & 0.000       \\
 Challenge ABFG\;\&\;D                        & 0.05                & 0.036        & 0.004               & 0.002       & 0.008               & 0.000       \\
Challenge ABFG\;\&\;D with $r_d$             & 0.05                & 0.05        & 0.003               & 0.001       & 0.004               & 0.001       \\

Patchy NGC                       & 0.18                & 0.00        & 0.014               & 0.001       & 0.030               & 0.002       \\
Patchy NGC with $r_d$            & 0.16                & 0.00        & 0.013               & 0.000       & 0.014               & 0.000       \\
Patchy NGC with Bisp.            & 0.15               & 0.04        & 0.014               & 0.000       & 0.029               & 0.002       \\
Patchy NGC with Bisp. with $r_d$ & 0.14                & 0.03        & 0.013               & 0.000       & 0.015               & 0.000   \\ 
\hline
\end{tabular}
\caption{\small Summary of the individual analysis over simulations up to  $\kmax = 0.20 \hinvMpc$ for the power spectrum and up to $\kmax = 0.1 \hinvMpc$ for the bispectrum monopole (in which case we use the power spectrum up to $\kmax=0.23\hinvMpc$). As explained in the text, we use a conservative way to asses the systematic error, which is however limited to about $\sigma_{\rm stat}$ for Challenge boxes, to  $\sigma_{\rm stat}/\sqrt{2}$ for the combination of all the Challenge boxes, and to $\sigma_{\rm stat}/4$ for Patchy mocks.}
\label{tab:summarysims}
\end{table}

\begin{table}\scriptsize
\centering
{\bf{Best fits}}\\
\begin{tabular}{c|c|c|c|c|c|c}
\hline\hline
& $\ln (10^{10}A_s)$ & $\Omega_m$ & $h$ & $b_1$ & min $\chi^2/$d.o.f. & $p$-value   \\ \hline

Challenge A & 3.18 & 0.317 & 0.668 & 1.94 & 50/(38-9) & 0.007 \\ \hline

Challenge B & 3.24 & 0.312 & 0.671 & 1.96 & 50/(38-9) & 0.008 \\ \hline

Challenge F & 3.17 & 0.314 & 0.671 & 2.09 & 54/(38-9) & 0.003 \\ \hline

Challenge G & 3.17 & 0.312 & 0.673 & 1.80 & 48/(38-9) & 0.014 \\ \hline

Challenge D & 3.19 & 0.290 & 0.708 & 1.83 & 51/(38-9) & 0.006 \\ \hline

Patchy NGC box 1 & 3.12 & 0.304 & 0.698 & 1.82 & 68/(68-10) & 0.18 \\ \hline

data CMASS NGC & 2.70 & 0.311 & 0.725 & 2.22 & 65/(68-10) & 0.27 \\ \hline

data  CMASS SGC & 2.98 & 0.306 & 0.670 & 2.06 & 60/(68-10) & 0.41 \\ \hline

data LOWZ NGC & 3.24 & 0.312 & 0.638 & 1.73 & 69/(76-10) & 0.40 \\ \hline

data CMASS NGC with bisp. & 2.73 & 0.303 & 0.686 & 2.31 & 103/(68+34-11) & 0.18 \\ \hline \hline

& $c_2$/$c_4$ & $b_3$ & $ c_\text{ct}$ & $c_{r,1}$ ($+c_{r,2}$) & $c_{\epsilon,1}/\bar{n}_g$ & $c_{\epsilon,\rm quad}$ \\ \hline

Challenge A & 1.0 & -1.6 & -0.7 & -6.9 & - & -4.1 \\ \hline

Challenge B & 1.1 & -2.0 & -1.1 & -8.6 & - & -3.9 \\ \hline

Challenge F & 1.1 & -2.3 & -2.2 & -7.0 & - & -4.2 \\ \hline

Challenge G & 0.9 & -1.8 & 1.1 & -6.7 & - & -3.3 \\ \hline

Challenge D & 1.0 & -1.5 & -0.8 & -7.3 & - & -3.6 \\ \hline

Patchy NGC box 1 & 1.0 & -1.2 & 0.2 & -9.9 & 0 & -1.5 \\ \hline

data CMASS NGC & 1.2 & 0.1 & 0.4 & -7.7  & 0 & -3.7 \\ \hline

data CMASS SGC & 1.2 & -0.6 & -0.1 & -9.0 & 0 & -1.1 \\ \hline

data LOWZ NGC & 1.0 & -1.0 & 0.2 & -10.3 & 0 & -2.1 \\ \hline

data CMASS NGC with bisp. & 1.8/-0.5 & 0.2 & -2.0 & -9.5 & 0 & -1.1 \\ \hline
\end{tabular}
\caption{\small Best-fit values and goodness of fit of the power spectrum up to $\kmax = 0.2 \hinvMpc$, with $k_\textsc{m}= 0.7 \hinvMpc$ and $\bar{n}_g = \{4.5,\,4.0\} \cdot 10^{-4} \hinvMpc$, respectively.  The values of $\bar n_g$ for Challenge boxes is given in footnote~\ref{footnote:shotnoisechallenge}. For $b_1$ of challenge A, the true value of $b_1$ is $2.04$. For the observational data, the bins below $k\lesssim 0.07$ are correlated, therefore the value of the  $\chi^2/{\rm d.o.f}$ should be taken with care. Performing a similar calculation for the data with the minimum wavenumber $k=0.05\hinvMpc$ still gives good values  of the fit. Regarding the $p$-value of the Challenge boxes, we remind that they have error bars roughly a factor of four smaller than the data we analyze in the observational data. Fitting without imposing priors results into a $p$-value of $\sim 0.03$ both for Challenge A and D, while the EFT-parameters best-fit values are still of physical size. 
As we do not fit the hexadecapole in this analysis, we set $c_{r,2} = 0$. Thus, its contribution (mainly in the quadrupole) is absorbed by $c_{r,1}$, which takes overall higher values about two times its natural size.  \label{tab:bestfit} }
\end{table}

The relative statistical error on the cosmological parameters on BOSS-NGC-like Patchy mocks is about {{18\%, 4\% and 4.4\%} for $A_s,\;\Omega_m$ and~$h$ respectively at $\kmax=0.2\hinvMpc$}. The inclusion of the bispectrum monopole reduces the error bars marginally. Upon inclusion of the Planck prior on the sound horizon, the error bars are decreased by very roughy about {10\%, 20\% and 50\%} for $A_s,\Omega_m$ and~$h$ respectively.  We have argued that the systematic error is always comfortably smaller than the statistical errors we will find in the observational data as long as {$\kmax\leq 0.2\hinvMpc$}: $\lesssim 1/4\cdot \sigma_{{\rm stat,\, CMASS\, x\, low-}z}$ from the Challenge boxes analyses for all cosmological parameters, and on $h$ $\lesssim 1/4\cdot \sigma_{{\rm stat,\, CMASS\, x\, low-}z}$ from the analyses of the Patchy mocks (otherwise vanishing for the other cosmological parameters). These are negligibly small as long as {$\kmax\leq 0.20\hinvMpc$}, which we therefore will take as the maximum wavenumber for our CMASS analysis.  

We also observe how the statistical error decreases with $\kmax$. Ideally, the errors should decrease as $\kmax^{-3/2}$, but this is affected by the fact that some parameters are dominated by the information in the BAO peak, which is already well  measured at low $\kmax$, as well as by the fact that there are degeneracies with some cosmological and EFT parameters. Taking for definiteness the power spectrum-only analysis of the Challenge ABFG boxes, we find that the error decreases roughly as~$\kmax^{-0.4}$, $\kmax^{-0.7}$ and $\kmax^{-0.6}$, for $A_s$, $\Omega_m$ and $h$, respectively~\footnote{In order to check, at least partially, the correctness of our pipeline, we have performed a Fisher-matrix study of the expected error bars on the parameters around the best fit point for the power spectrum analysis for Challenge box A. We find results similar, though somewhat smaller than expected, to the ones we find with the MCMC. We notice here that we also performed the same Fisher-matrix analysis around the best fit point for the data, again finding compatible results to what we find with the MCMC.  

We have also checked through the Fisher-matrix analysis that we obtain very similar error bars if we do not include the AP effect in the Fisher forecast. This is consistent with the fact that the AP effect is just a calculable geometric effect that relates how the Fourier-space power-spectra are obtained out of the sky map, and therefore should carry no cosmological information. Indeed, if we analyzed directly the three-dimensional information in the $z$-dependent angular correlation function, usually referred as $C_\ell(z,z')$, there would be no need of the AP parameters and no information would be lost. 

We use the same Fisher-matrix analysis to study the degradation of our error bars if we lower the $\kmax$ of the analysis. In particular, we find from the Fisher-matrix analysis of the CMASS NGC sample that the error bars on the parameters at $\kmax=0.1\hinvMpc$ are worse than the ones at $\kmax=0.25\hinvMpc$ by a factor of $133\%,\, 180\%,$ and $92\%$ for $A_s$, $\Omega_m$ and $h$ respectively. Maybe even more significantly, if we do not fix $n_s$ and, for $\Omega_b/\Omega_c$, we rather put a prior on $\Omega_b/\Omega_c$ of the order of four times the percentage error on $\Omega_b h^2$ obtained by BBN (which is about 20$\sigma$'s of Planck2018)~\cite{Tanabashi:2018oca}, the expected error bars at $\kmax=0.1\hinvMpc$ are increased by a factor of $3.7, 2.2, 4.2$ for $A_s$, $\Omega_m$ and $h$ respectively, to roughly about $60\%,\, 30\%$ and $50\%$  for $\ln A_s$, $\Omega_m$ and $h$ respectively, for CMASS NGC, making the overall measurement scarcely significant. On the other hand, as mentioned earlier, the relative degradation of the measurements when we loosen the prior on $n_s$ and $\Omega_b/\Omega_c$ is much smaller at the $\kmax$'s of interest here. }. It would be interesting to see how this scaling is modified if one adds higher $n$-point function in the analysis without the limitation we described earlier for our analysis.

It is worth to finally notice that the EFTofLSS at one-loop order is able to fit accurately, without introducing unacceptably large systematic errors, the power spectrum measured from the challenge boxes, whose volume is sixteen times larger than the effective one of BOSS NGC, up to $\kmax = 0.2\hinvMpc$.  This is promising in  light of the forthcoming large-size surveys.




\newpage
\section{Analysis of the SDSS/BOSS data}

\subsection{Main results}

\begin{figure}[h!]
  \begin{center}
  \textbf{ CMASS NGC}
    \end{center}
\begin{center}
\includegraphics[width=0.466\textwidth,draft=false]{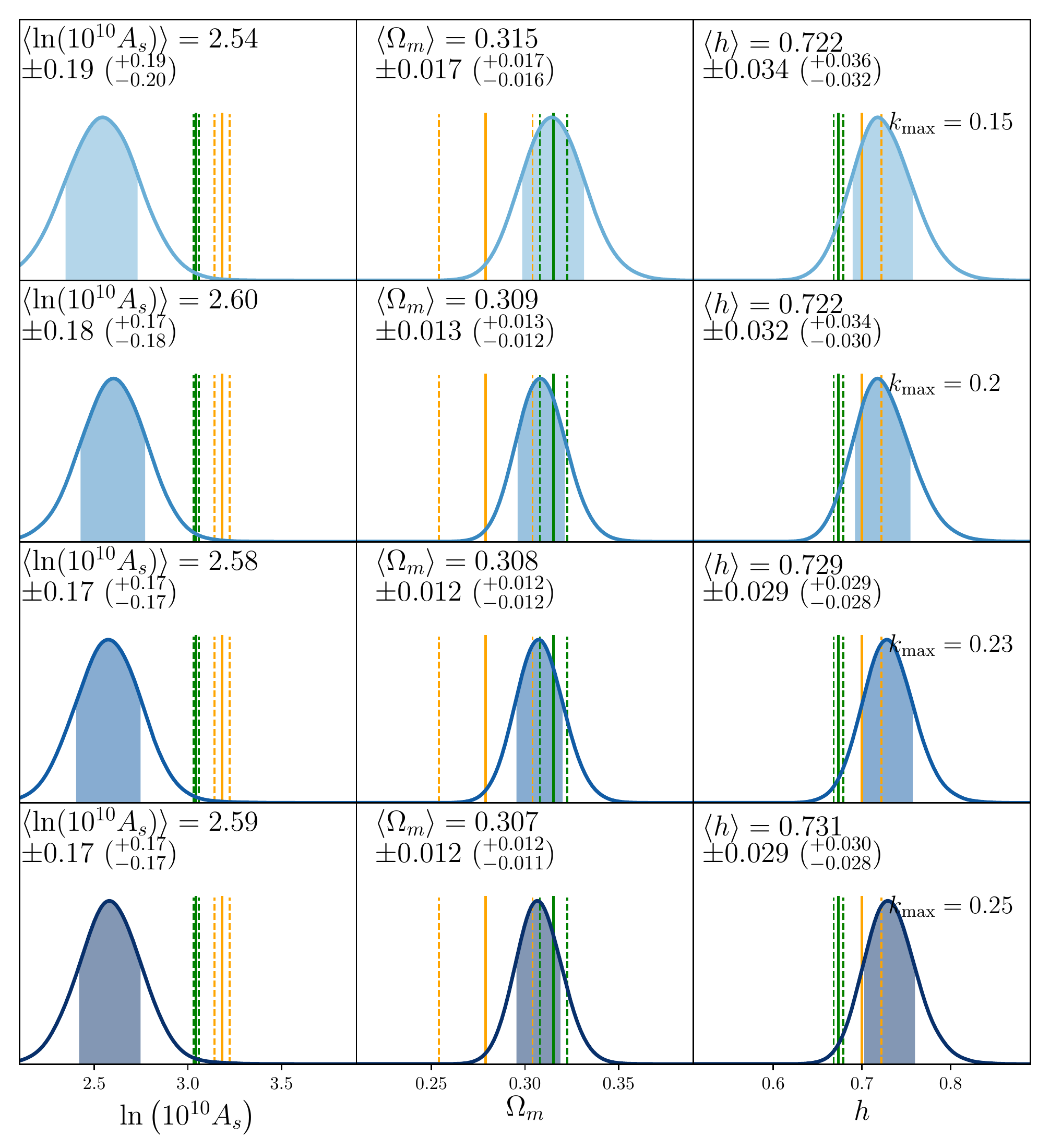}
\includegraphics[width=0.466\textwidth,draft=false]{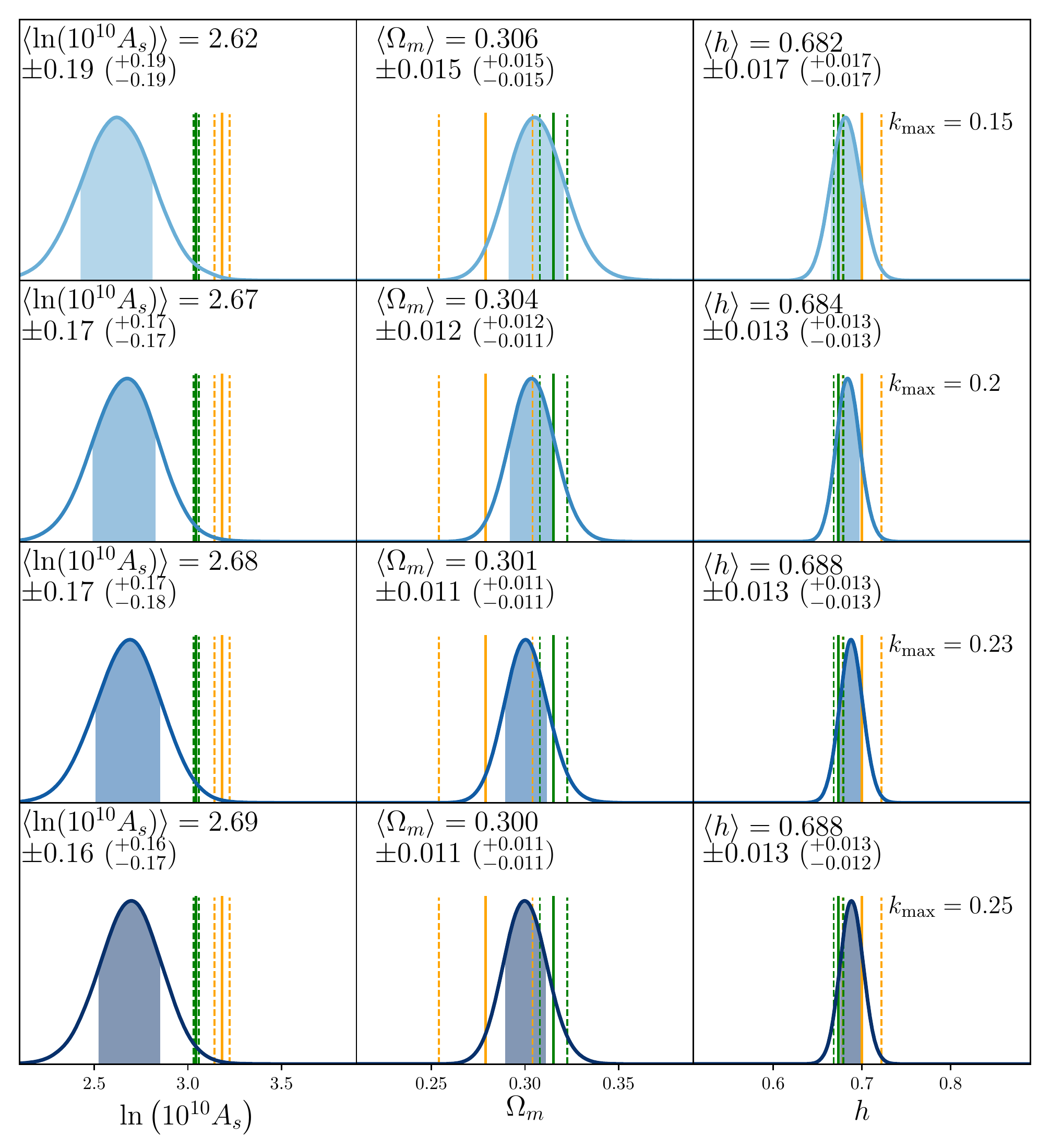}
\end{center}
 \begin{center}
  \textbf{CMASS SGC}
    \end{center}
\begin{center}
\includegraphics[width=0.466\textwidth,draft=false]{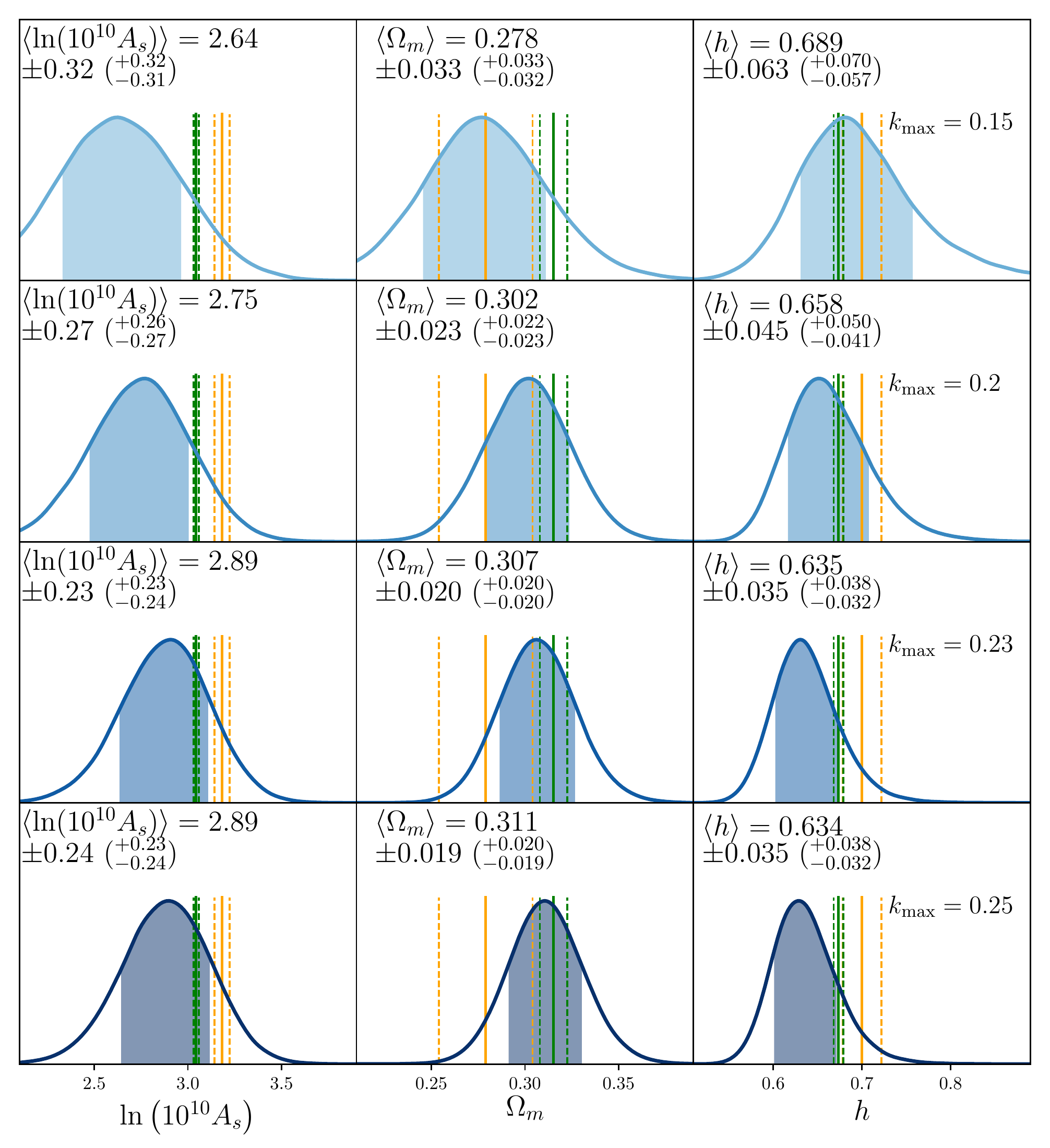}
\includegraphics[width=0.466\textwidth,draft=false]{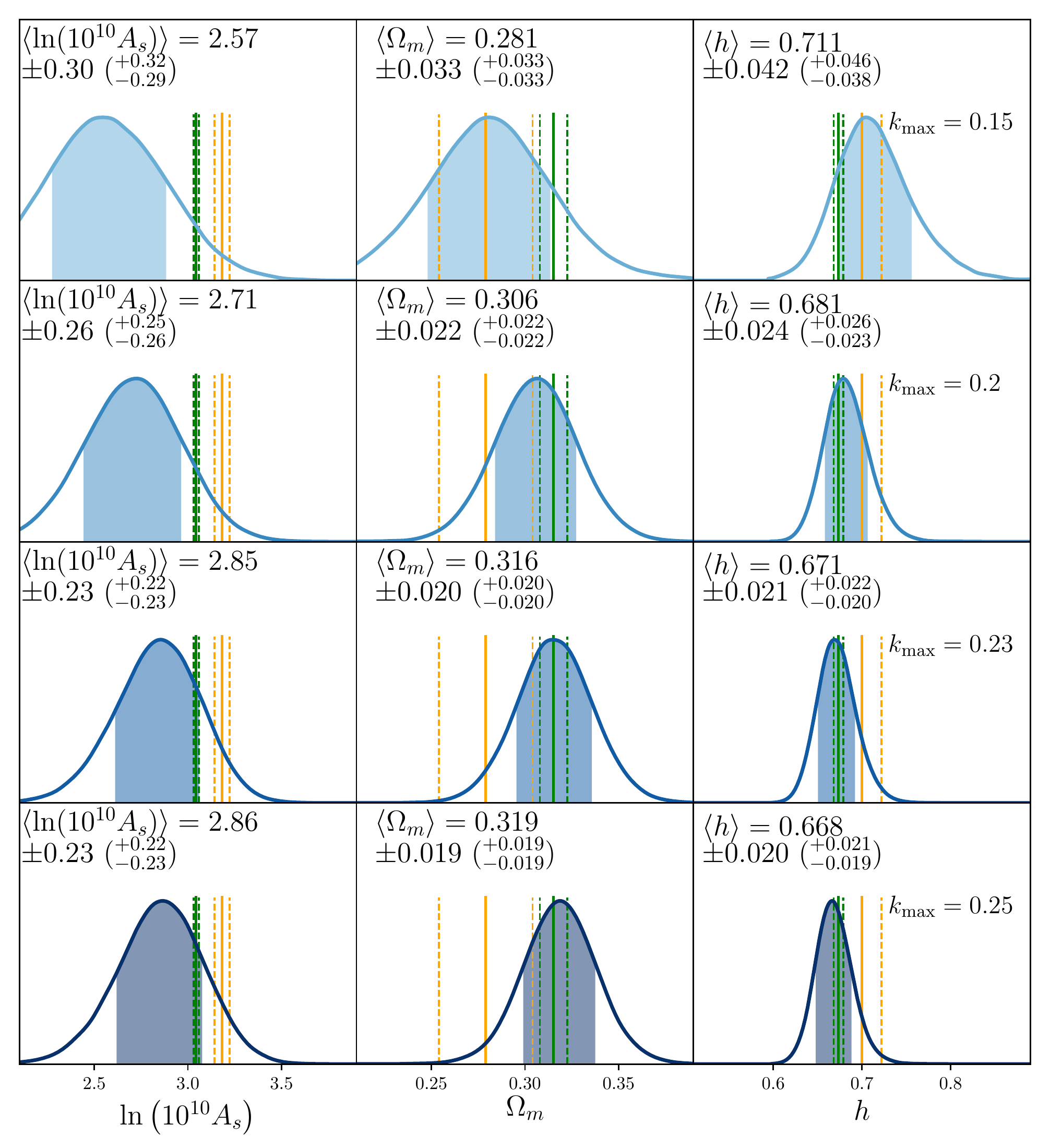}
\caption{\small Marginalized posterior distributions of the cosmological parameters obtained from fitting the power spectrum of the CMASS sample NGC ({\it top}) or SGC ({\it bottom)} at $z_{\rm eff} = 0.55$ without ({\it left}), and with ({\it right}) the inclusion of Planck sound horizon prior ({\it right}). The green lines represent the 1-$\sigma$ constraints from Planck2018 while the orange lines represent the constraints from WMAP9yrs. 
 \label{fig:posteriordataNGC} }
\end{center}
\end{figure}

\begin{figure}[h]
    \begin{center}
  \textbf{ CMASS NGC$\times$SGC}
    \end{center}
\begin{center}
\includegraphics[width=0.496\textwidth,draft=false]{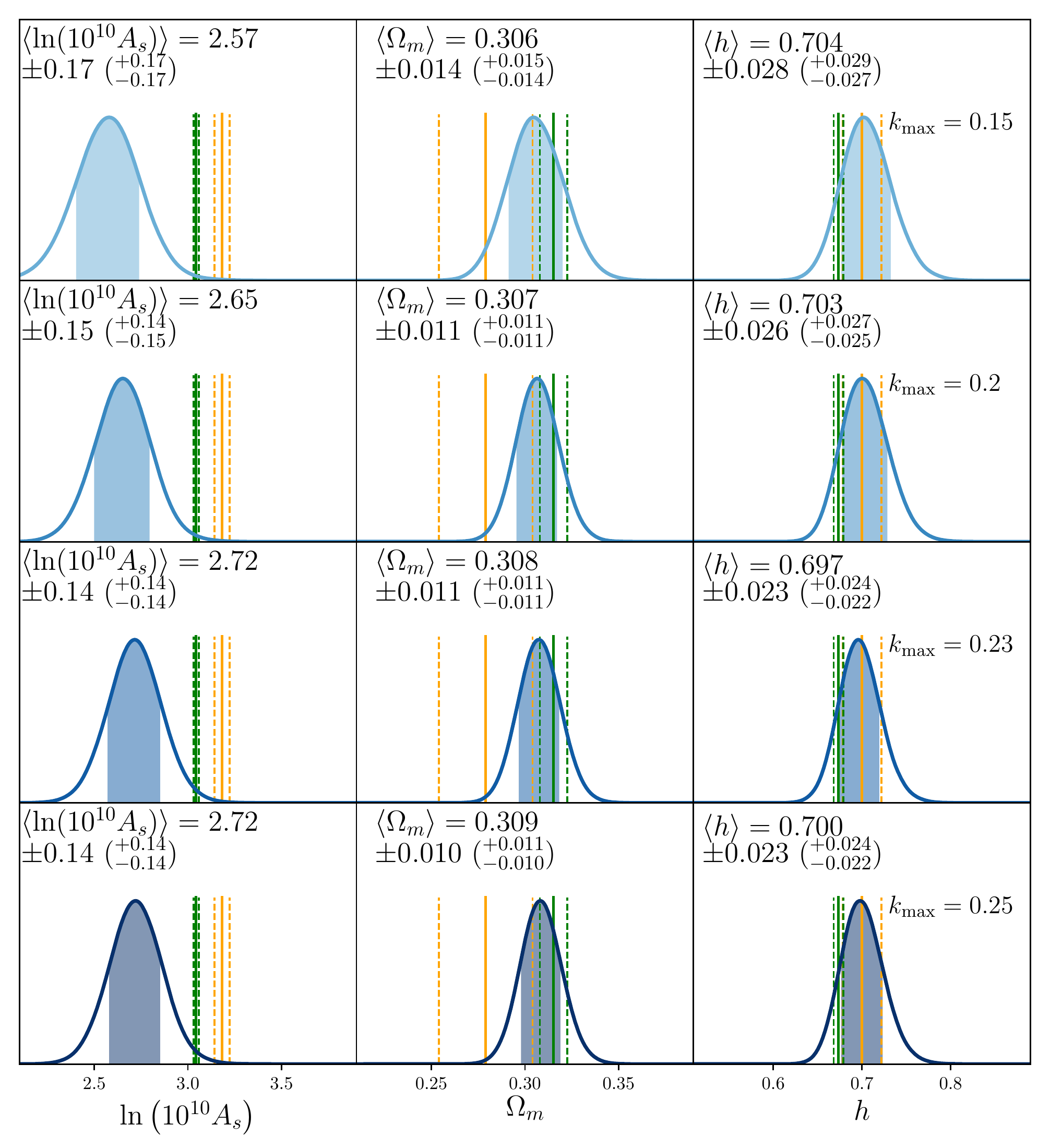}
\includegraphics[width=0.496\textwidth,draft=false]{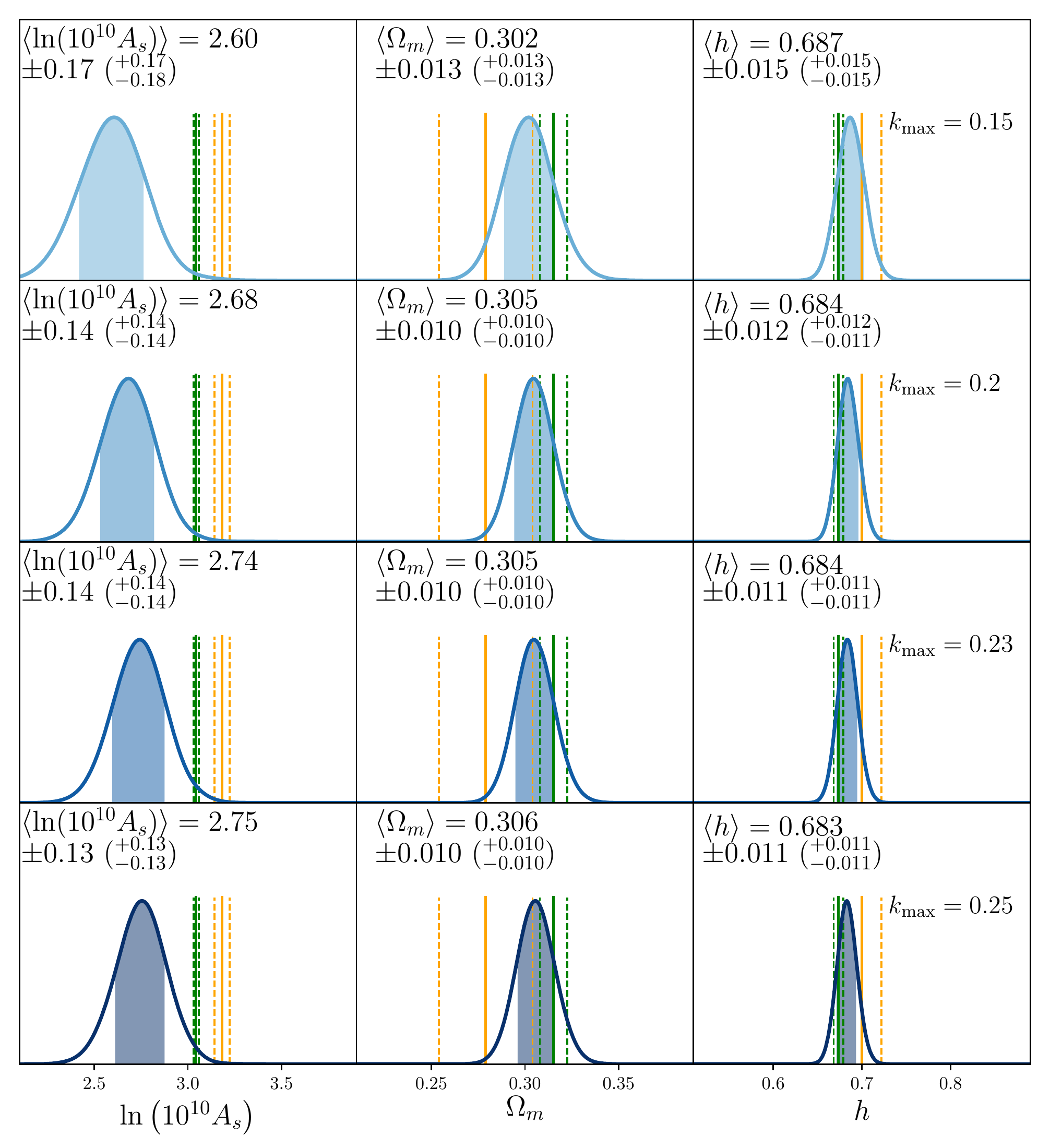}
\caption{\small Marginalized posterior distributions of the cosmological parameters obtained from fitting the power spectrum of the CMASS sample NGC$\times$SGC at $z_{\rm eff} = 0.55$  ({\it left}), and with the inclusion of Planck sound horizon prior. The green lines represent the 1-$\sigma$ constraints from Planck2018 while the orange lines represent the constraints from WMAP9yrs.
 \label{fig:posteriordataNGCxSGC} }
\end{center}
\end{figure}

\begin{figure}[h!]
    \begin{center}
  \textbf{LOWZ NGC}
    \end{center}
\begin{center}
\includegraphics[width=0.496\textwidth,draft=false]{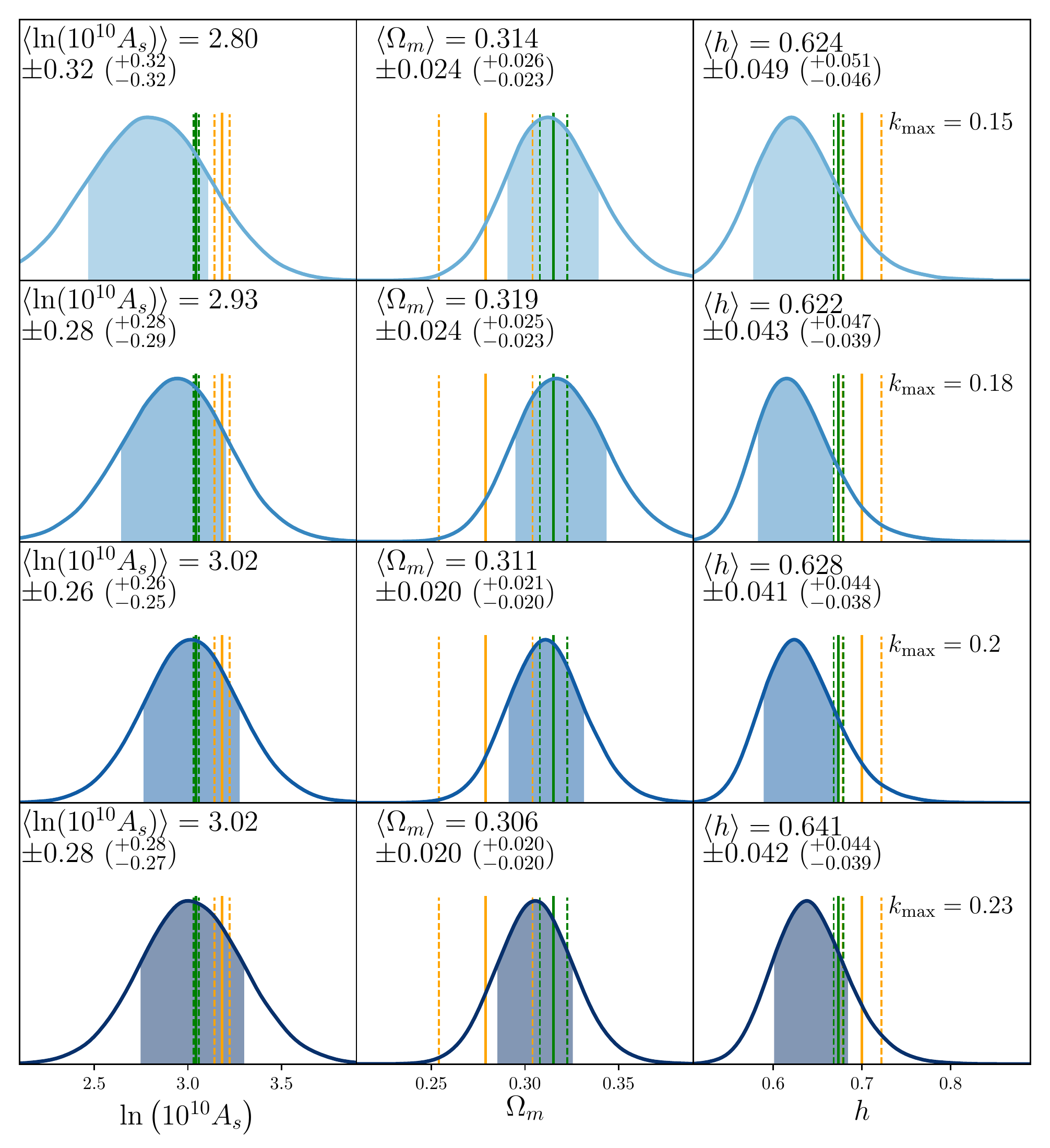}
\includegraphics[width=0.496\textwidth,draft=false]{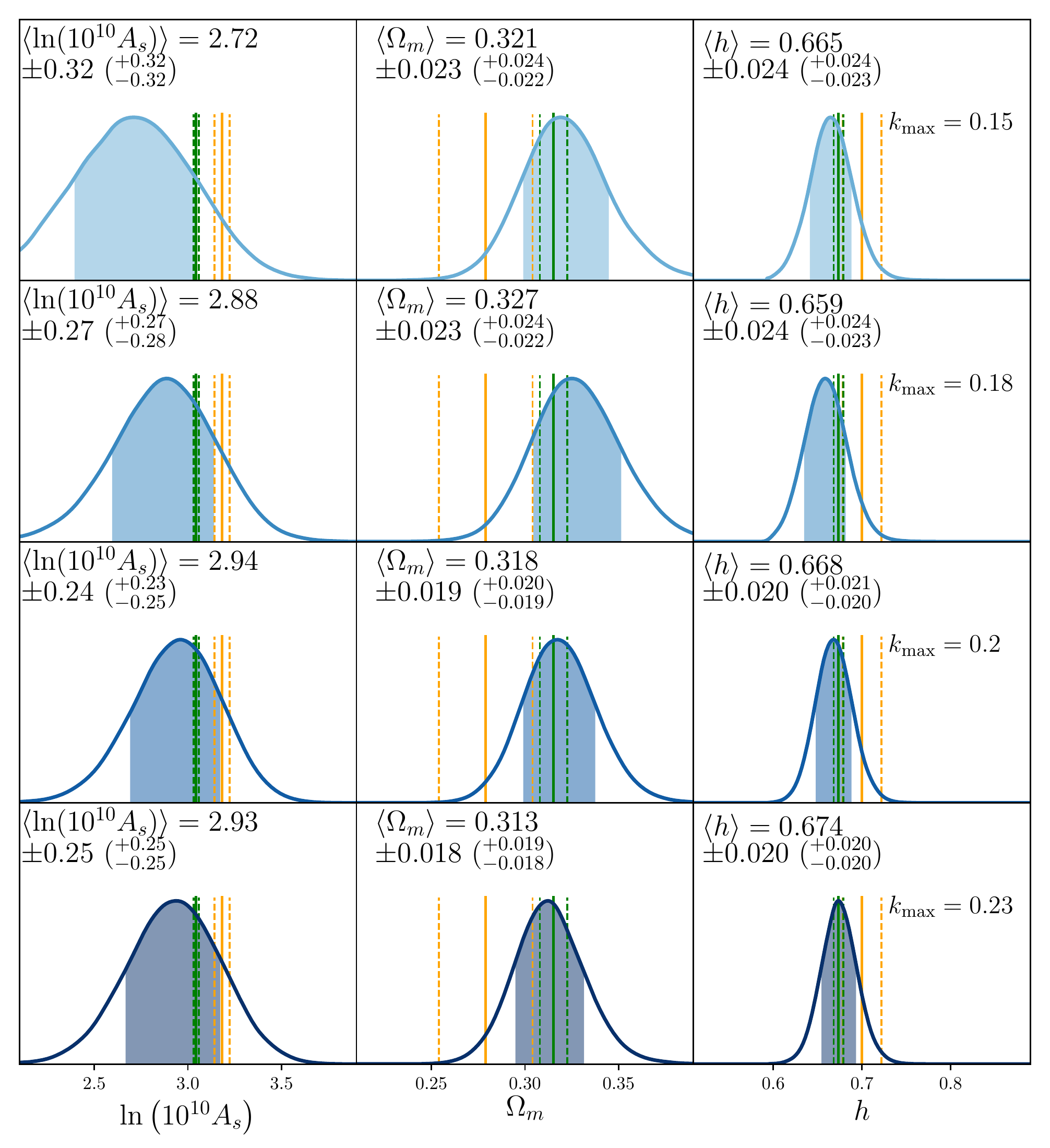}
\caption{\small Marginalized posterior distributions of the cosmological parameters obtained from fitting the power spectrum of LOWZ NGC at $z_{\rm eff} = 0.32$, without ({\it left}) and with ({\it right}) the inclusion of Planck sound horizon prior. The green lines represent the 1-$\sigma$ constraints from Planck2018 while the orange lines represent the constraints from WMAP9yr. 
 \label{fig:posteriordataNGClowz} }
\end{center}
\end{figure}

\begin{figure}[h]
   \begin{center}
  \textbf{CMASS NGC$\;\times$ CMASS\;SGC $\;\times$  LOWZ NGC}
    \end{center}
\begin{center}
\includegraphics[width=0.51\textwidth,draft=false]{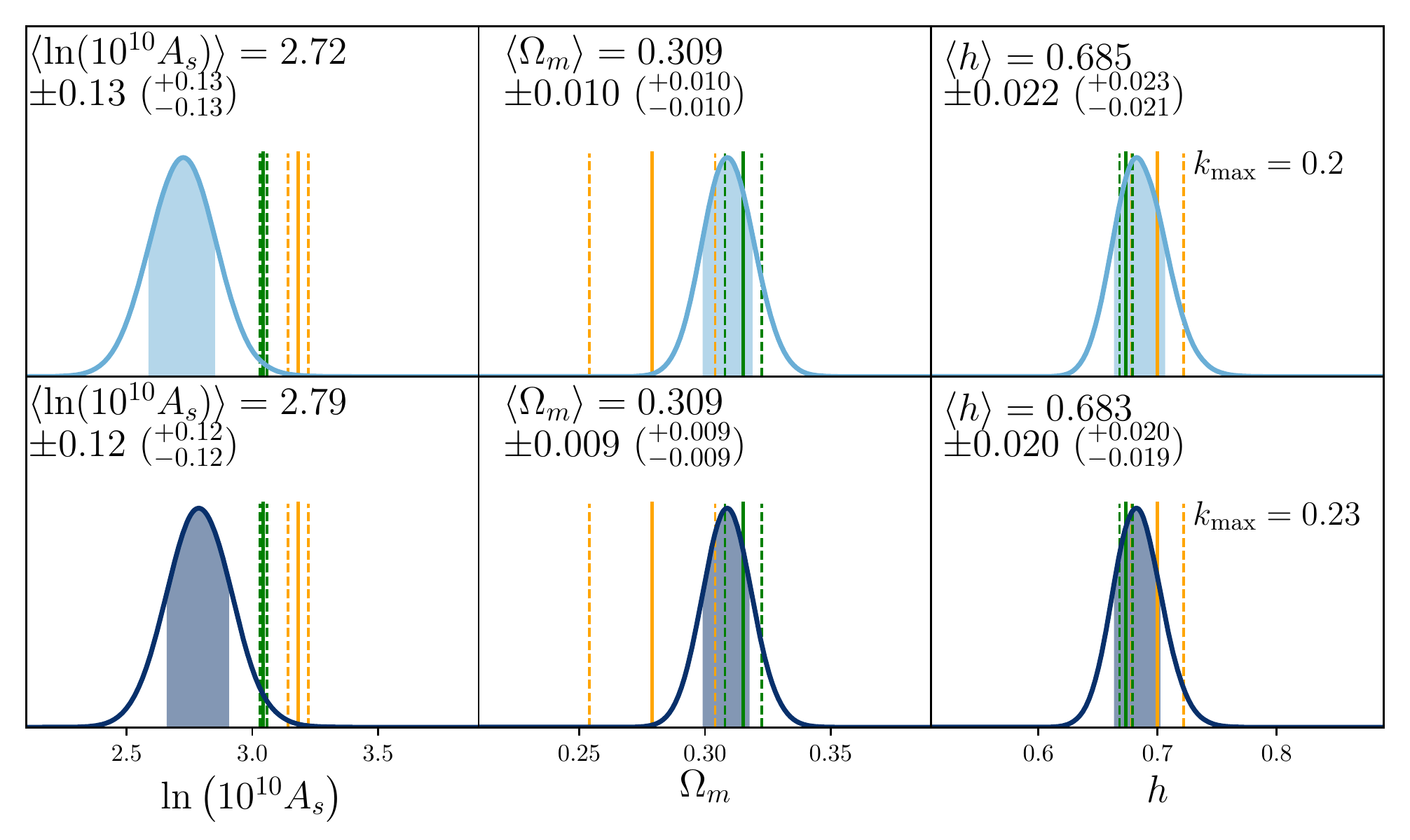}
\includegraphics[width=0.51\textwidth,draft=false]{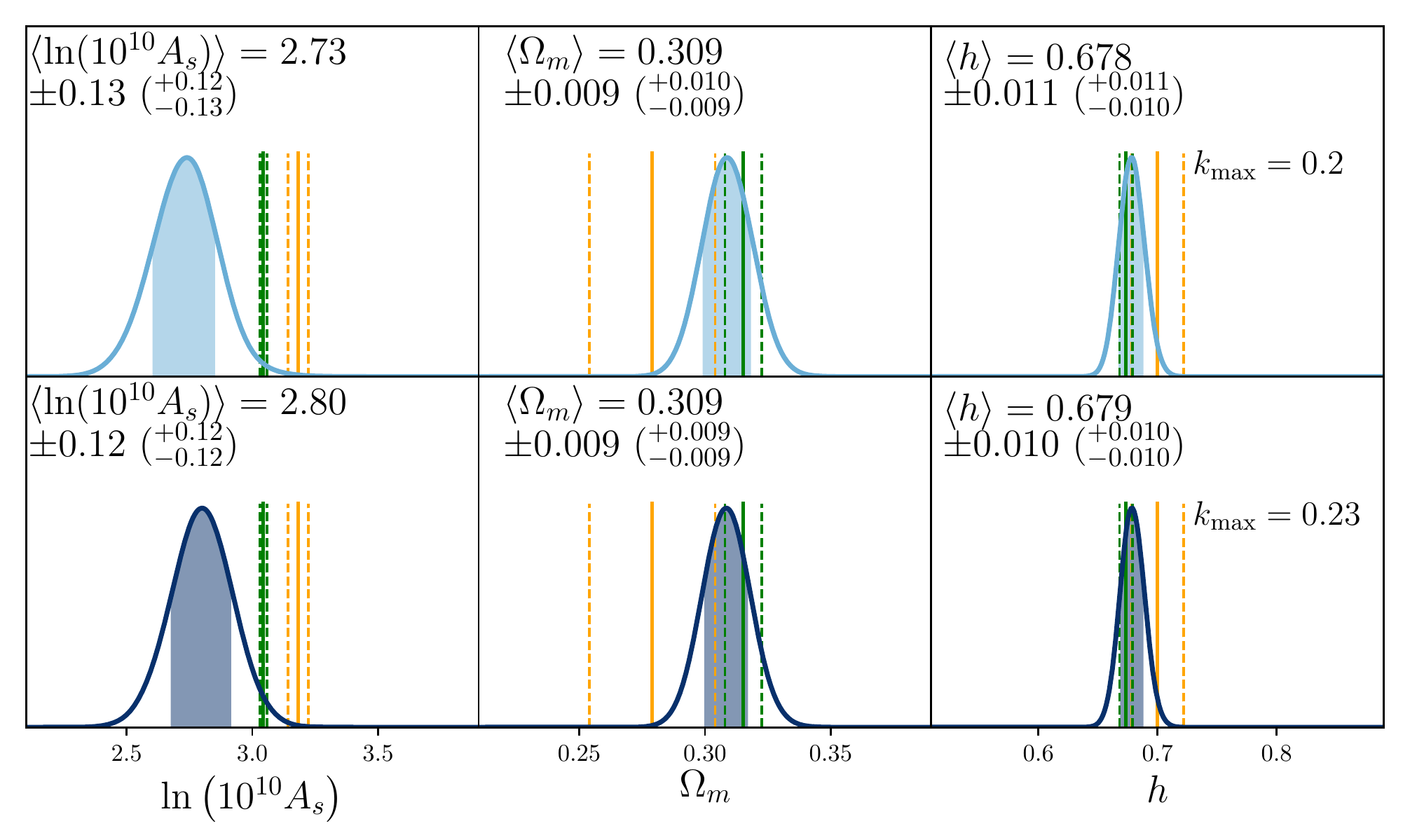}
\caption{\small Marginalized posterior distributions of the cosmological parameters obtained from fitting the power spectrum of the CMASS sample NGC and SGC at $z_{\rm eff} = 0.55$ and LOWZ NGC at $z=0.32$, with ({\it top}) and without ({\it bottom}) the inclusion of the Planck sound horizon prior. In each plot, on the top row, we use $\kmax=0.20\hinvMpc$ for CMASS and $\kmax=0.18\hinvMpc$ for LOWZ. On the bottom row, we use $\kmax=0.23\hinvMpc$ for CMASS and $\kmax=0.20\hinvMpc$ for LOWZ. The green lines represent the 1-$\sigma$ constraints from Planck2018 while the orange lines represent the constraints from WMAP9yrs. 
 \label{fig:posteriordatacombinedall} }
\end{center}
\end{figure}

\begin{figure}[h!]
\centering
\includegraphics[width=1\textwidth,draft=false]{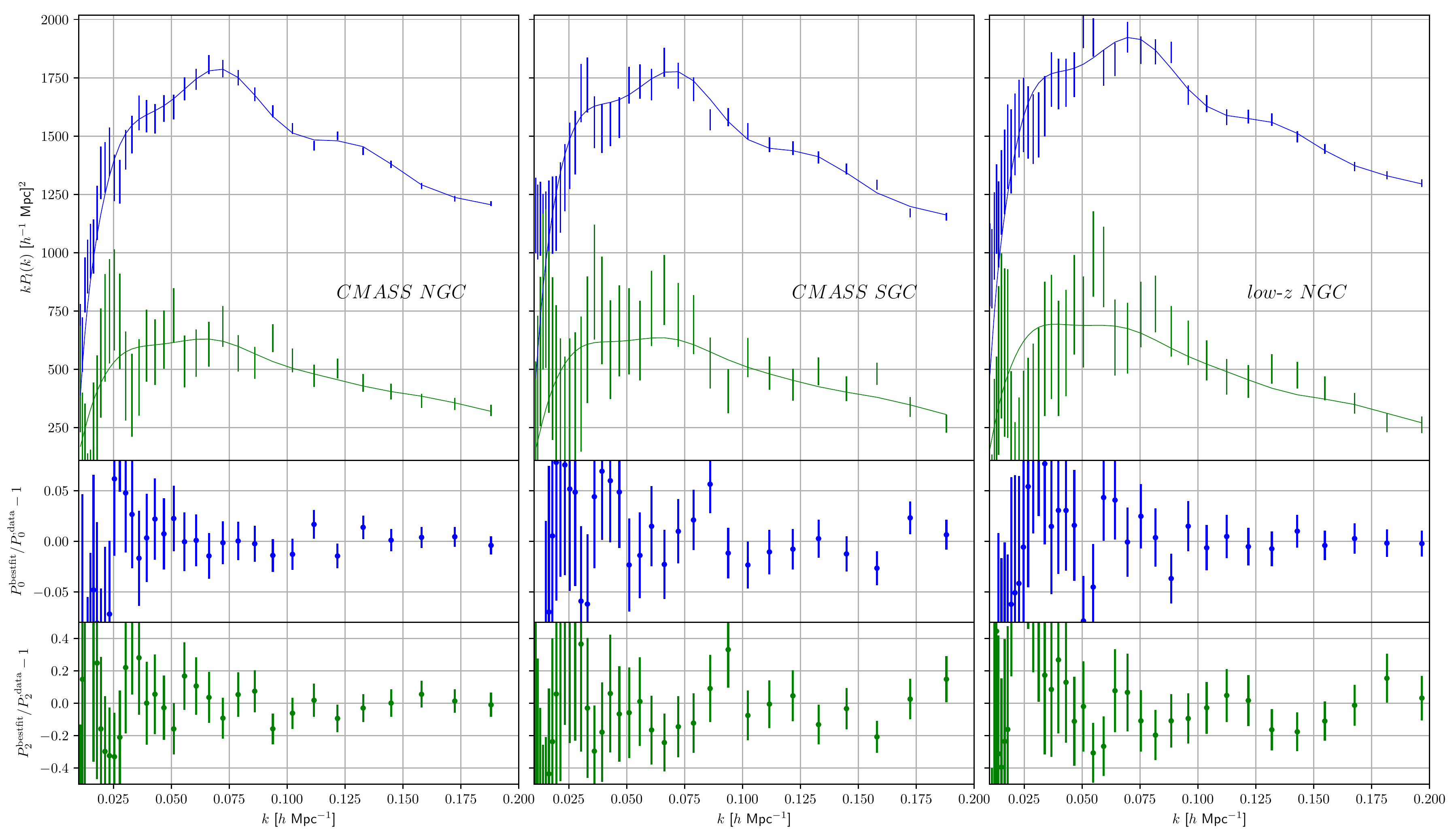}
\caption{\small Best-fit CMASS NGC ({\it left}) and CMASS SGC ({\it center}) and LOWZ  NGC ({\it right}) power spectrum up to $ \kmax= 0.2 \hinvMpc$ for CMASS and $\kmax=0.20\hinvMpc$ for LOWZ  and residuals. The error bars represent the diagonal terms of the covariance. Notice that the bins below $k\simeq 0.07$ are highly correlated as the binning is smaller than the fundamental mode of the survey. The $\chi^2/{\rm d.o.f}$ is $1.10$ for NGC, $1.02$ for SGC and $0.85$ ($p\text{-value}\simeq 0.88$) when combined, while $1.10$ for LOWZ NGC. When we combine CMASS and LOWZ NGC, the $\chi^2/{\rm d.o.f}$ is $0.85$, corresponding to a $p$-value of $\simeq0.92$. We stress that given the correlation of the low-$k$ points, the interpretation of the $\chi^2$ and the $p$-value should be taken with care.  \label{fig:bestfit-data} }
\end{figure}

\begin{figure}[h!]
\begin{center}
\includegraphics[width=0.485\textwidth,draft=false]{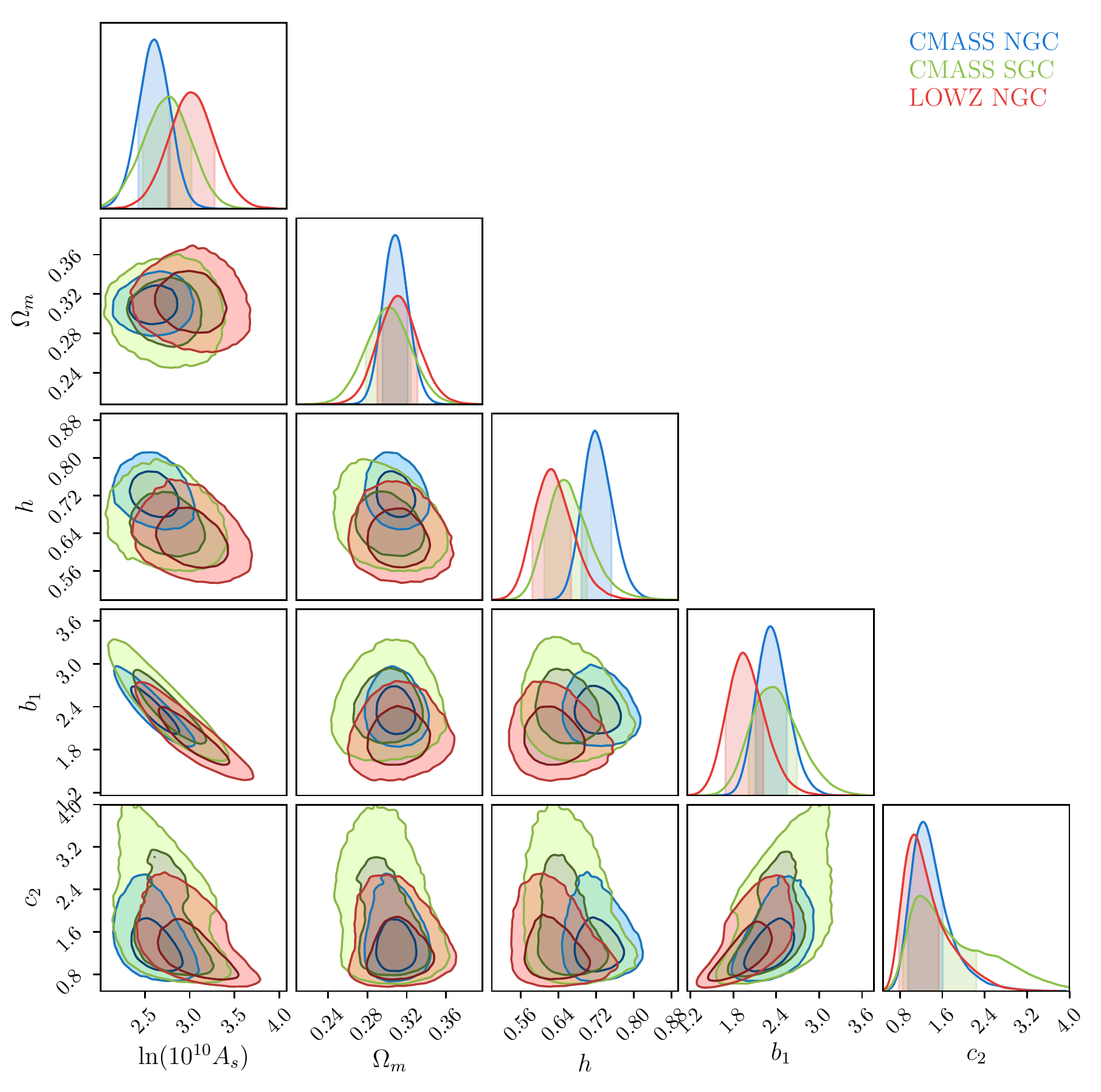}
\includegraphics[width=0.485\textwidth,draft=false]{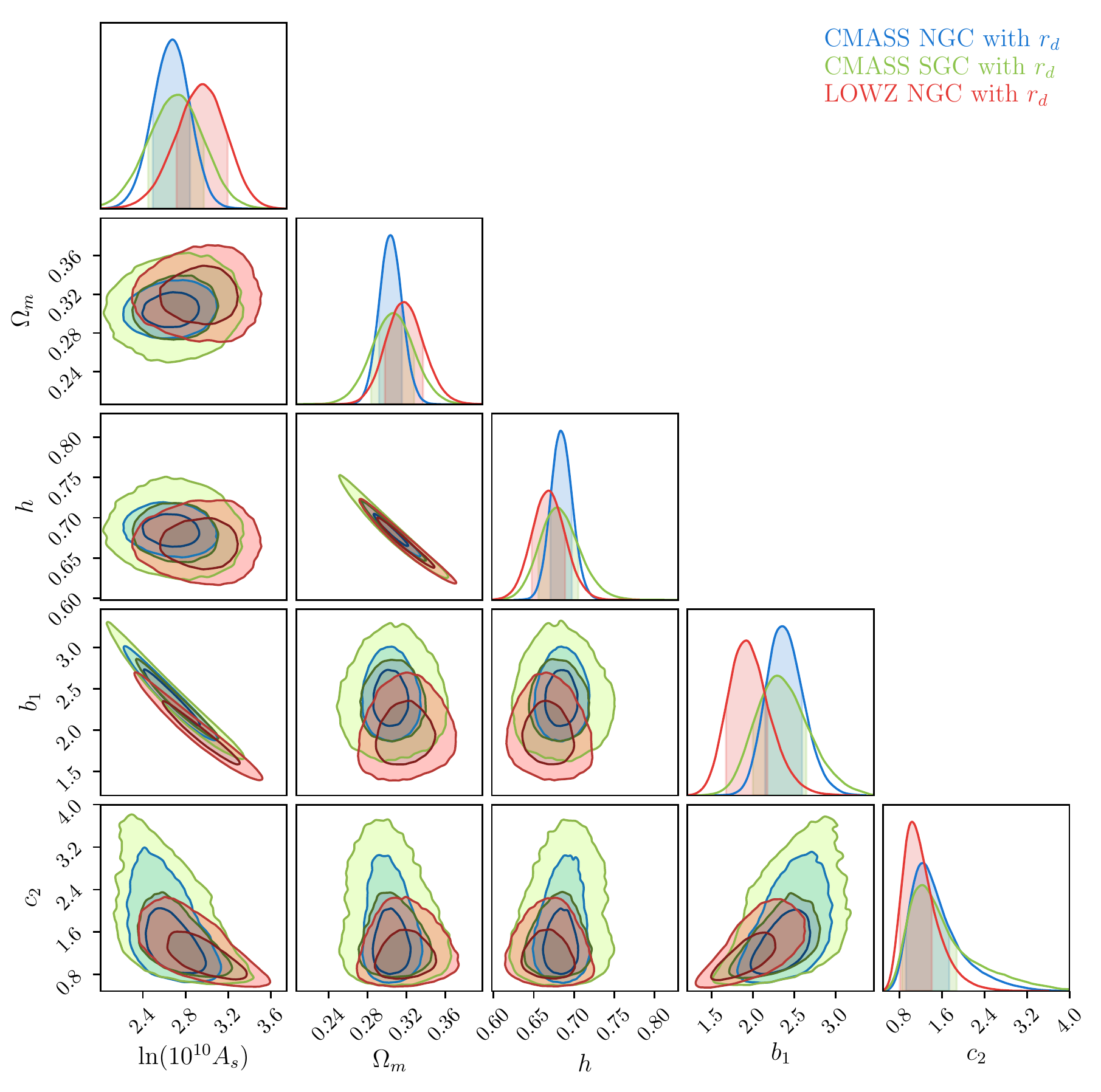}
\end{center}
\caption{\small 2D posterior distributions of the cosmological parameters $\ln (10^{10} A_s)$, $\Omega_m$, and $h$, and the EFT-parameters $b_1$ and $c_2$, obtained from fitting the CMASS NGC and SGC, and the LOWZ NGC power spectra up to $\kmax = 0.2 \hinvMpc$. We can see the determination of the cosmological and the EFT parameters, as well as the consistency of NGC and SGC data sets.} \label{fig:data_corner}
\end{figure}

\begin{figure}[h!]
\begin{center}
\includegraphics[width=0.496\textwidth,draft=false]{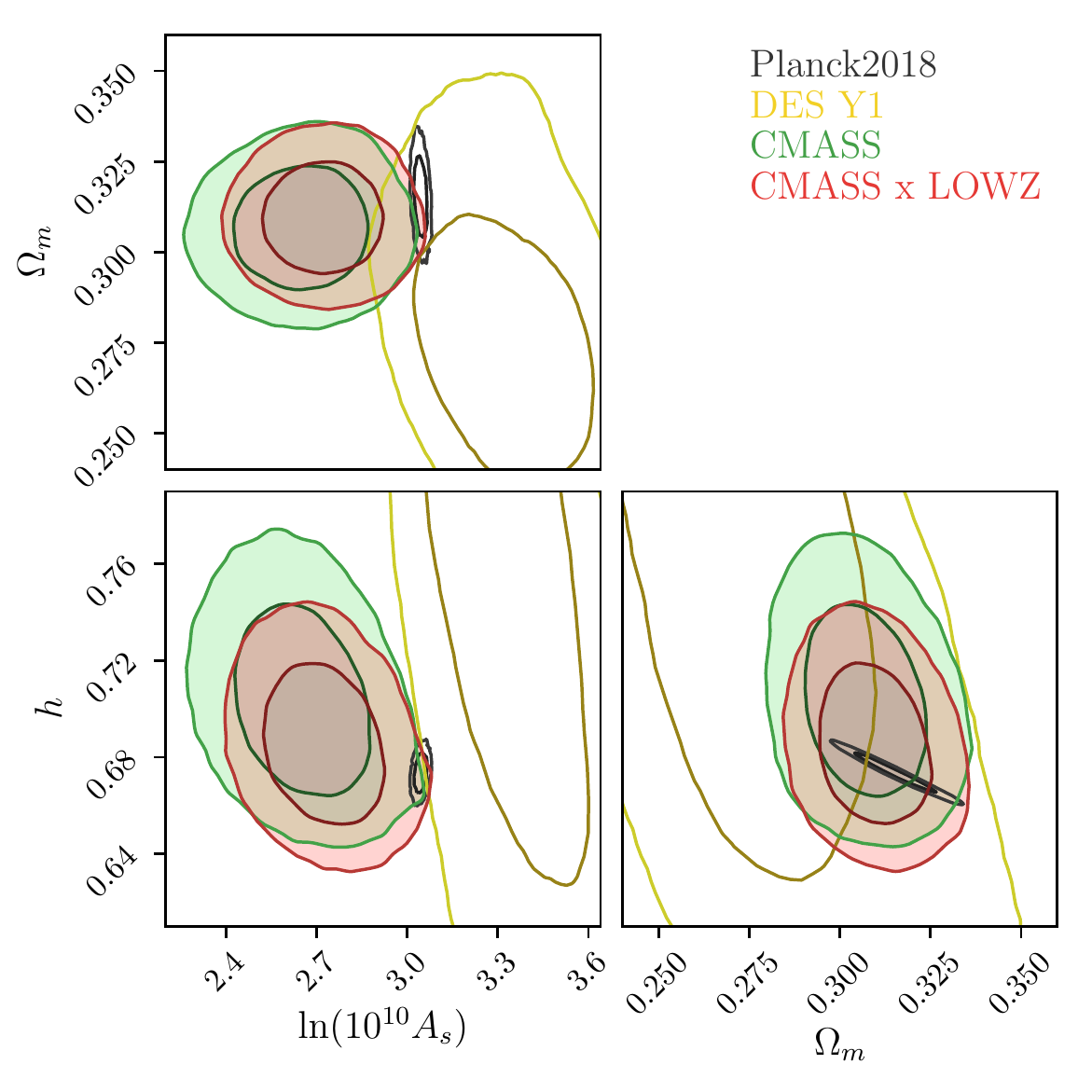}
\includegraphics[width=0.496\textwidth,draft=false]{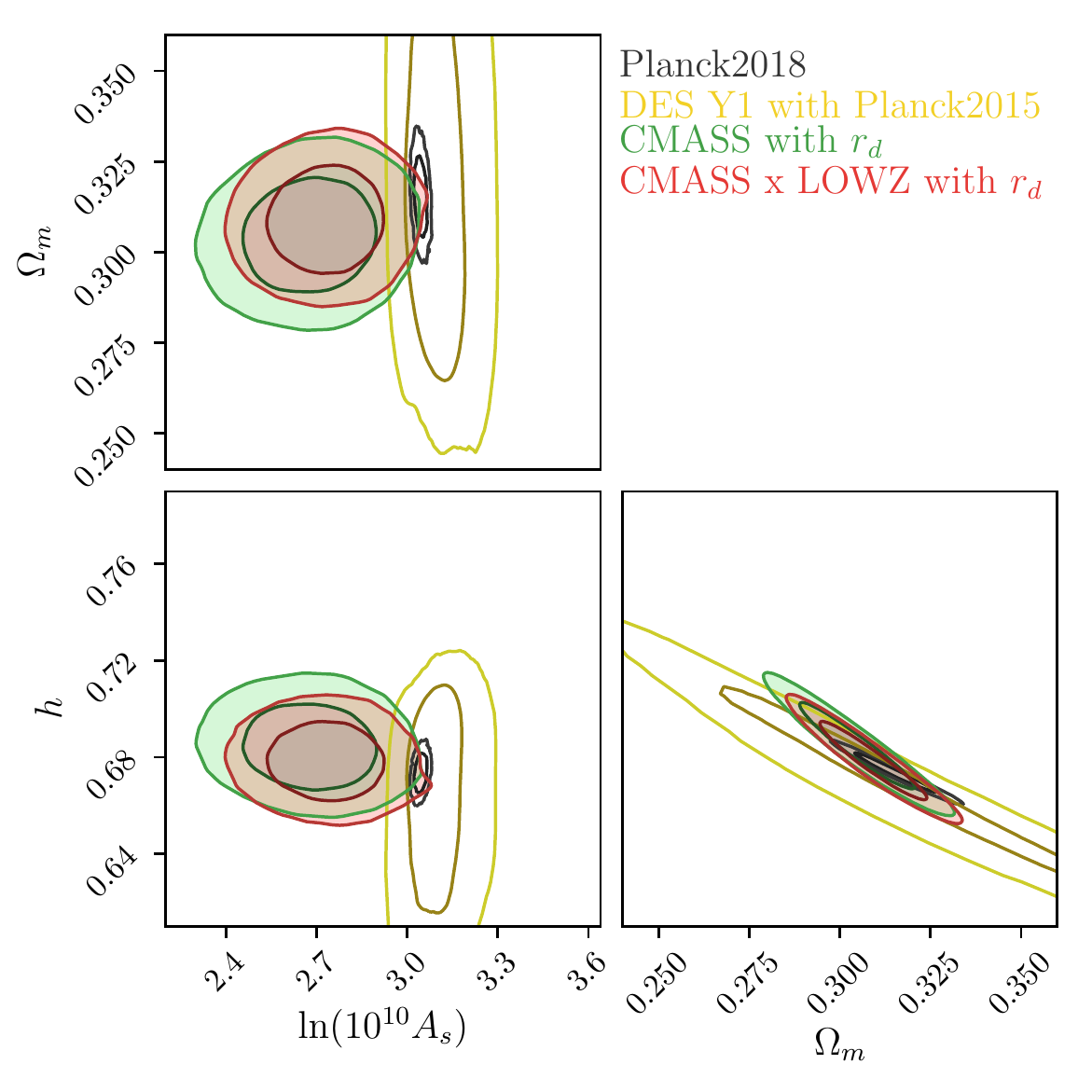}
\end{center}
\caption{\small 2D posterior distributions of the cosmological parameters $\ln (10^{10} A_s)$, $\Omega_m$, and $h$ obtained from fitting the CMASS and LOWZ NGC power spectra up to $\kmax = 0.2 \hinvMpc$ and $\kmax = 0.18 \hinvMpc$ respectively ({\it left}), and with the inclusion of the sound horizon prior ({\it right}), together with the same quantities from Planck2018~\cite{Aghanim:2018eyx} and DES1Yr~\cite{Abbott:2017wau}. We can see the consistency with Planck2018.} \label{fig:data_corner1}
\end{figure}

\begin{figure}[h!]
 \begin{center}
  \textbf{CMASS NGC power spectrum and bispectrum}
    \end{center}
\begin{center}
\includegraphics[width=0.49\textwidth,draft=false]{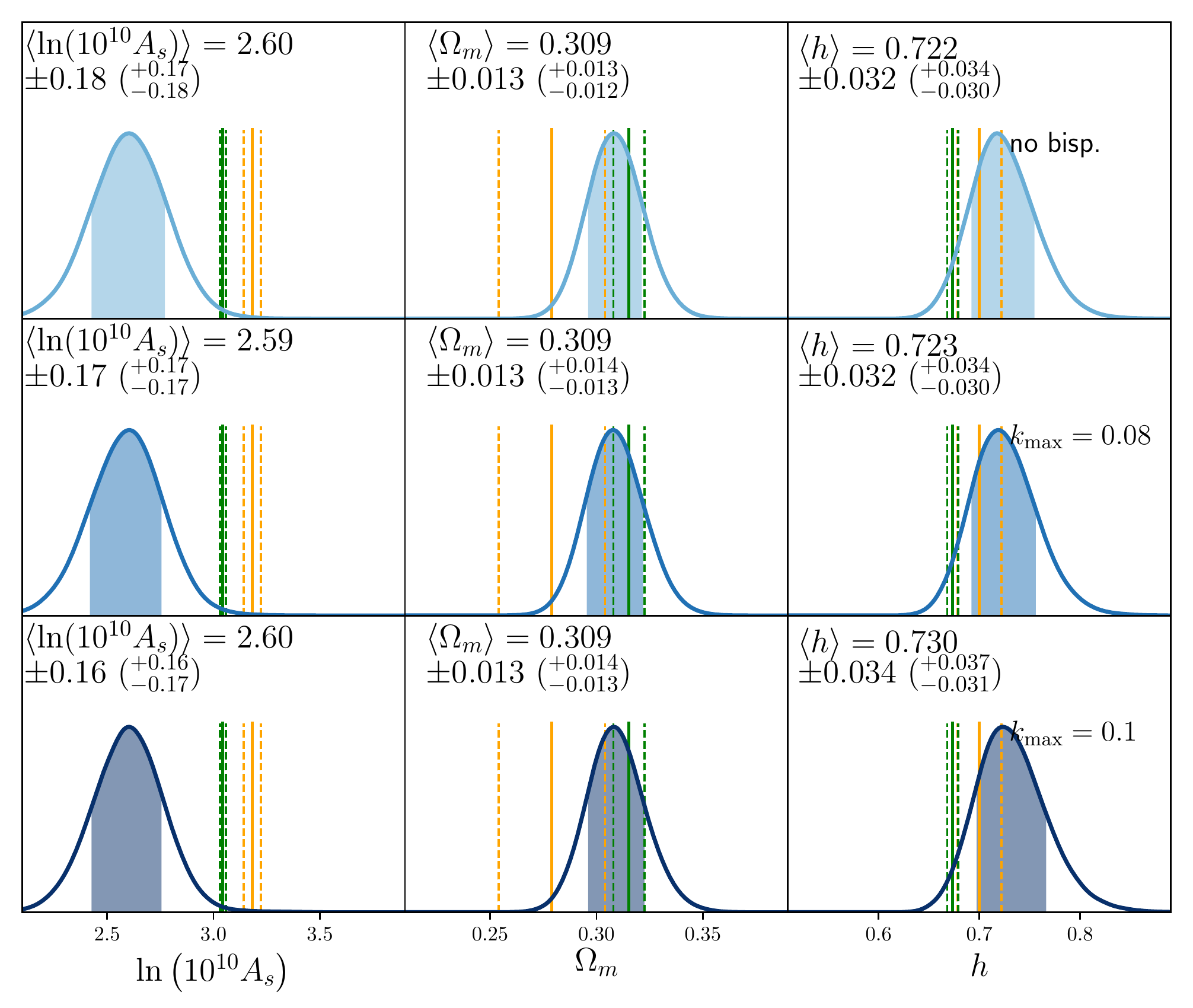}
\includegraphics[width=0.49\textwidth,draft=false]{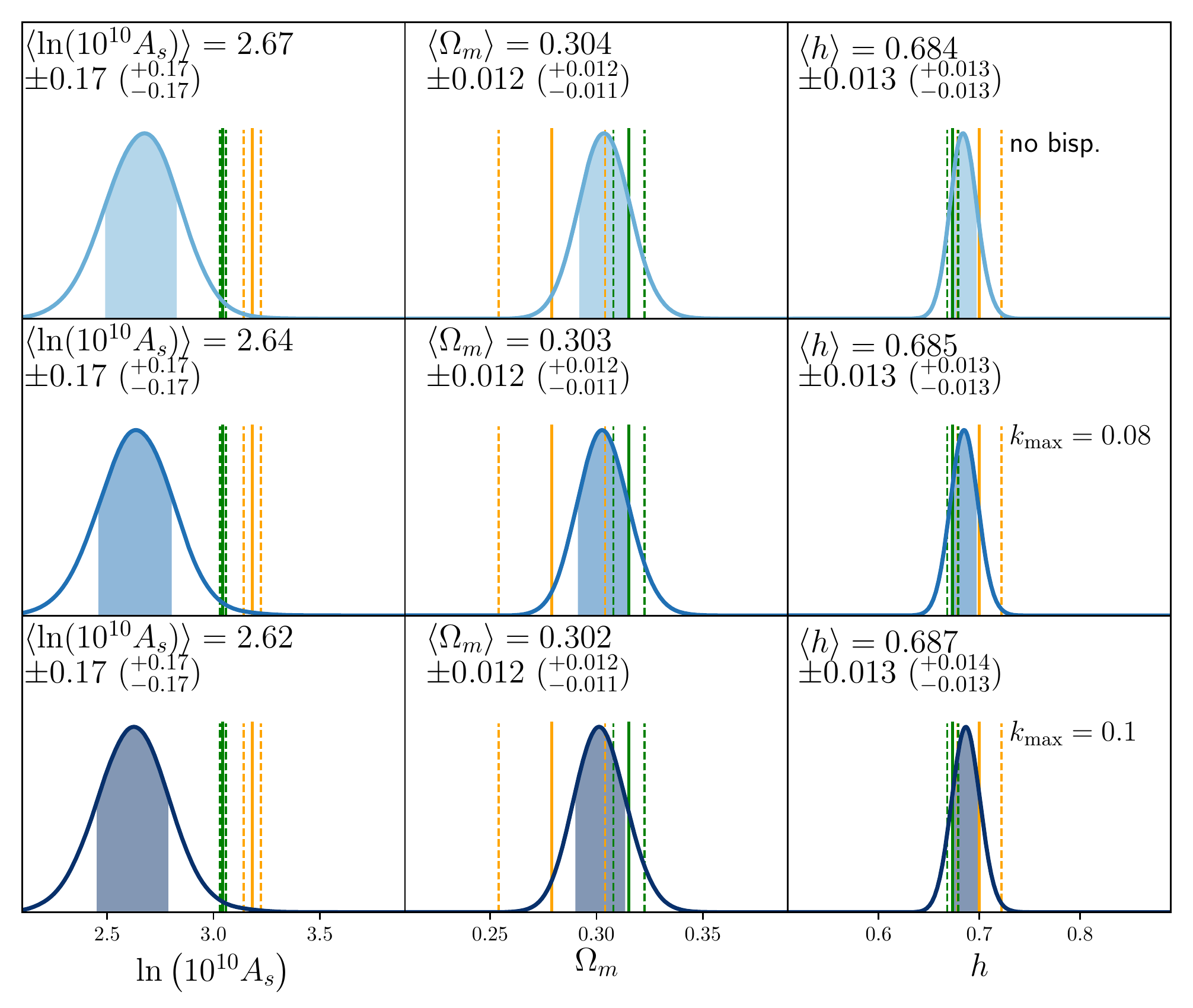}
\caption{
\small Marginalized posterior distributions of the cosmological parameters obtained from fitting the power spectrum and the bispectrum of the CMASS sample NGC ({\it left}), and with the inclusion of the Planck sound horizon prior ({\it right}) .
 \label{fig:posteriordatabisp} }
\end{center}
\end{figure}

\begin{figure}[h!]
\begin{center}
\includegraphics[width=0.69\textwidth,draft=false]{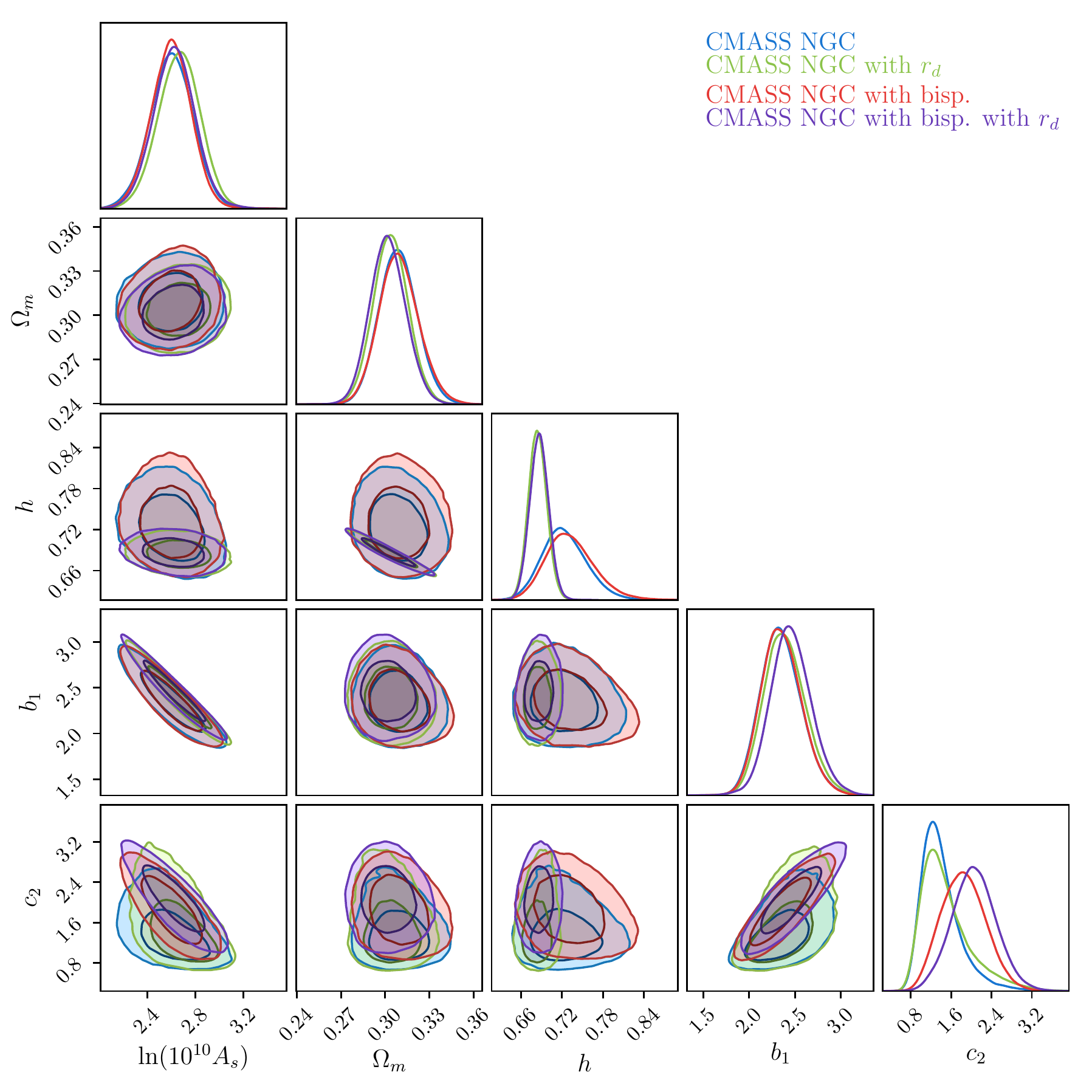}
\end{center}
\caption{
\small 2D posterior distributions of the cosmological parameters $\ln (10^{10} A_s)$, $\Omega_m$, and $h$, and the galaxies biases $b_1$ and $c_2$, obtained from fitting the CMASS NGC power spectrum up to $\kmax = 0.20 \hinvMpc$, with and without the bispectrum monopole up to $\kmax = 0.1 \hinvMpc$, and with and without the Planck sound horizon prior. This plot highlights the contribution of the various data sets.} \label{fig:data_corner2}
\end{figure}

 Having successfully tested our pipeline with simulations, we are ready to apply the same pipeline to the data. When we analyze the power spectrum only, we obtain the results shown in Fig.~\ref{fig:posteriordataNGC}, \ref{fig:posteriordataNGCxSGC}, \ref{fig:posteriordataNGClowz},~\ref{fig:posteriordatacombinedall},~\ref{fig:data_corner1} and~\ref{fig:data_corner2}. They correspond, respectively, to the analysis of the CMASS NGC, CMASS SGC, CMASS NGC combined with SGC, LOWZ NGC, and finally to the combination of these three data sets.
  When we apply the bispectrum, we obtain the results presented in Fig.~\ref{fig:posteriordatabisp} and~\ref{fig:data_corner}. The best-fit values for all parameters are given in Table~\ref{tab:bestfit} and the 68\% confidence intervals for the cosmological parameters are given in Table~\ref{tab:summarydata}. In Fig.~\ref{fig:bestfit-data}, we plot the best-fit power spectrum for CMASS NGC, CMASS SGC and LOWZ NGC, against the BOSS DR12 measurement and the residuals. We provide the correlation among all parameters and the full 2D posterior surfaces in App.~\ref{app:covariance}.

 Let us give some details of the analysis that are specific to the observational data. For the AP effect, we use  $\Omega_m=0.310$ and $z=0.57$ for CMASS and $z=0.32$ for LOWZ. Since we know neutrinos have a mass, we analyze the data as done in Planck2018 by fixing one massive neutrino at $0.06$eV. When we fit the different data sets, we keep all EFT parameters as independent. While this is completely justified as we go from the CMASS to the LOWZ sample, as the $z$-dependence of the EFT parameters is unknown, in principle CMASS NGC and SGC should have the same EFT parameters if they observe the same population of galaxies. However, following BOSS collaboration~\cite{Beutler:2016arn}, we conservatively keep the bias parameters separate because we wish to account for different selection criteria among the samples~\footnote{We observe that, however, given our priors and the size of the error bars of CMASS, all EFT parameters are statistically consistent between the CMASS NGC and SGC samples, as it is can be seen from the full 2D posterior surfaces in Fig.~\ref{fig:contournonmargData}. We show in an App.~\ref{app:CMASSNGCSGCsameEFT} the results on CMASS if the NGC and SGC samples are analyzed using the same set of EFT parameters with the exception of the shot noise. We find that results are quite consistent, with a very marginal reduction of the error bars by about 10\%, and, as we will see, most significantly a slightly larger value of $A_s$ by about half a statistical error.}.

 Our comparisons with numerical simulations have shown that we do not have any significant systematic error until {$\kmax=0.2\hinvMpc$ for $z\simeq0.55$ (in reality, due to $k$-binning, we will use $\kmax=0.188\hinvMpc$ at $z=0.55$, but we will refer to it as $\kmax=0.2\hinvMpc$ for simplicity)}. We will analyze also the LOWZ NGC sample, which is centered at $z\simeq0.32$.   In order to establish until which wavenumber we can trust our LOWZ prediction, we rescale the theoretical error as measured at $z=0.55$ to $z=0.32$, and compare it to the statistical error of the data finding that we can fit the LOWZ data up to {$\kmax=0.18\hinvMpc$}~(\footnote{In detail, we use the following procedure. The limitation in the $k$-reach of the EFTofLSS is given by the first order in the perturbative expansion that we  do not compute. In our case, this is the two-loop contribution. As a function of $z$, this contribution scales with good approximation as $D(a)^6 P_{11}(k)\left(k/\knl\right)^{2.4}$, where we used the fact that, very approximately, the linear power spectrum goes as $k^{-1.8}$ in this range of $k$'s. At $z=0.55$, our comparison with simulations establishes that the theoretical error is a significant fraction of the error bar of the data, $\sigma_{\rm data}(k,z)$, at $\kmax=0.2\hinvMpc$, which allows us to set the normalization of the two-loop term. We therefore rescale accordingly the $\kmax$ by imposing the following equality to hold:
 \be
 \sigma_{\rm data}(\kmax (z_{\rm low}),z_{\rm low})=\sigma_{\rm data}(\kmax (z_{\rm high}),z_{\rm high}) \frac{D(z_{\rm low})^6}{D(z_{\rm high})^6} \frac{P_{11}(\kmax (z_{\rm low}))}{P_{11}(\kmax (z_{\rm high}))} \left(\frac{\kmax (z_{\rm low})}{\kmax (z_{\rm high})}\right)^{2.4}\ .
 \ee 
 We find $\kmax (z_{\rm low})\simeq0.18\hinvMpc$.}). The results at {$\kmax=0.20\hinvMpc$ for CMASS and at $\kmax=0.18\hinvMpc$} for LOWZ should be considered as our main results.\\

\begin{table}[!]
\centering
\scriptsize
\begin{tabular}{l|lll} \hline
                                & $\lnAs$                              & $\Omega_m$                               & $h$                                      \\  \hline 
CMASS NGC                       & $2.60\pm 0.18 \, (^{+0.17}_{-0.18})$ & $0.309\pm 0.013 \, (^{+0.013}_{-0.012})$ & $0.722\pm 0.032 \, (^{+0.034}_{-0.030})$ \\
CMASS NGC with $r_d$            & $2.67\pm 0.17 \, (^{+0.17}_{-0.17})$ & $0.304\pm 0.012 \, (^{+0.012}_{-0.011})$ & $0.684\pm 0.013 \, (^{+0.013}_{-0.013})$ \\
CMASS SGC						 & $2.75\pm 0.27 \, (^{+0.26}_{-0.27})$ & $0.302\pm 0.023 \, (^{+0.022}_{-0.023})$ & $0.658\pm 0.045 \, (^{+0.050}_{-0.041})$ \\
CMASS SGC with $r_d$             & $2.71\pm 0.26 \, (^{+0.25}_{-0.26})$ & $0.306\pm 0.022 \, (^{+0.022}_{-0.022})$ & $0.681\pm 0.024 \, (^{+0.026}_{-0.023})$ \\
CMASS NGC x SGC           		 & $2.65\pm 0.15 \, (^{+0.14}_{-0.15})$ & $0.307\pm 0.011 \, (^{+0.011}_{-0.011})$ & $0.703\pm 0.026 \, (^{+0.027}_{-0.025})$ \\
CMASS NGC x SGC with $r_d$ 		 & $2.68\pm 0.14 \, (^{+0.14}_{-0.14})$ & $0.305\pm 0.010 \, (^{+0.010}_{-0.010})$ & $0.684\pm 0.012 \, (^{+0.012}_{-0.011})$ \\
LOWZ NGC                       & $2.93\pm 0.28 \, (^{+0.28}_{-0.29})$ & $0.319\pm 0.024 \, (^{+0.025}_{-0.023})$ & $0.622\pm 0.043 \, (^{+0.047}_{-0.039})$ \\
LOWZ NGC with $r_d$            & $2.88\pm 0.27 \, (^{+0.27}_{-0.28})$ & $0.327\pm 0.023 \, (^{+0.024}_{-0.022})$ & $0.659\pm 0.024 \, (^{+0.024}_{-0.023})$ \\ \hline
CMASS NGC x SGC x LOWZ NGC    & $2.72\pm 0.13 \, (^{+0.13}_{-0.13})$ & $0.309\pm 0.010 \, (^{+0.010}_{-0.010})$ & $0.685\pm 0.022 \, (^{+0.023}_{-0.021})$ \\
CMASS NGC x SGC  x LOWZ NGC with $r_d$ 		 & $2.73\pm 0.13 \, (^{+0.12}_{-0.13})$ & $0.309\pm 0.009 \, (^{+0.010}_{-0.009})$ & $0.678\pm 0.011 \, (^{+0.011}_{-0.010})$ \\
\hline
\end{tabular}
\caption{\small 68\% confidence interval for the cosmological parameters from the  individual analyses over the CMASS sample and LOWZ NGC sample of the BOSS data up to $\kmax = 0.20 \hinvMpc$ for the power spectrum of CMASS, up to $\kmax = 0.18 \hinvMpc$ for the power spectrum of LOWZ. \label{tab:summarydata}}
\end{table}

\begin{table}[!]
\centering
\footnotesize
    \begin{tabular}{l |ccc| ccc} \hline
   & \multicolumn{3}{c}{$p$-value of Planck $1\sigma$ value}   &  \multicolumn{3}{c}{effective-$\sigma$'s from Planck}      \\ \hline
      &       $\ln\left(10^{10} A_s\right)$ & $\Omega_m$ & $h$  &       $\ln\left(10^{10} A_s\right)$ & $\Omega_m$ & $h$ \\ \hline
CMASS NGC                       &  0.02 & 1.00 & 0.18 & 2.3 &  0.0 & 1.3 \\  
CMASS NGC with $r_d$            & 0.04 & 0.72 & 0.69 & 2.1 & 0.4 & 0.4 \\  
CMASS SGC                       & 0.29 & 0.77 & 0.82 & 1.1 &  0.3 & 0.2 \\  
CMASS SGC with $r_d$            & 0.21 & 0.93 & 0.92 & 1.3 & 0.1 & 0.1 \\  
CMASS NGC x SGC            		 & 0.014 & 0.91 & 0.37 & 2.46 & 0.1 & 0.9 \\ 
CMASS NGC x SGC with $r_d$      & 0.02 & 0.74 & 0.64 & 2.3 & 0.3 & 0.5 \\ 
LOWZ NGC                       & 0.96 & 1.00 & 0.33 & 0.1 &  0.0 & 1.0 \\  
LOWZ NGC with $r_d$            & 0.73 & 1.00 & 1.00 & 0.3 & 0.0 & 0.0 \\  
CMASS NGC x SGC x LOWZ NGC & 0.02 & 1.00 & 0.78 & 2.3 & 0.0 & 0.3 \\ 
CMASS NGC x SGC x LOWZ NGC  with $r_d$  & 0.03 & 1.00 & 1.00 & 2.2 & 0.0 & 0.0 \\
CMASS NGC with bisp.                      & 0.014 & 1.00 & 0.14 & 2.46 &  0.0 & 1.5 \\ 
CMASS NGC with bisp. with $r_d$             & 0.02 & 0.60 & 0.58 & 2.3 &  0.5 & 0.5 \\ 
 \hline
\end{tabular}%
\caption{\small Value of the probability~({\it i.e.~}$p$-value) for the closest $1\sigma$ value of the cosmological parameters as measured by Planck2018, with respect to the individual analyses of the BOSS data up to $\kmax = 0.2 \hinvMpc$ for the CMASS and LOWZ power spectrum, respectively, and $\kmax = 0.1 \hinvMpc$ for the bispectrum monopole. We also include the effective number of $\sigma$'s such a $p$-value would correspond to in a putative Gaussian distribution. We conclude that our results are consistent with Planck2018, with the largest, very-mild, discrepancy present in $\lnAs$ only in the case we include the bispectrum in the analysis.}
\label{tab:p-values}%
\end{table}%

\begin{figure}[h!]
\begin{center}
\includegraphics[width=0.496\textwidth,draft=false]{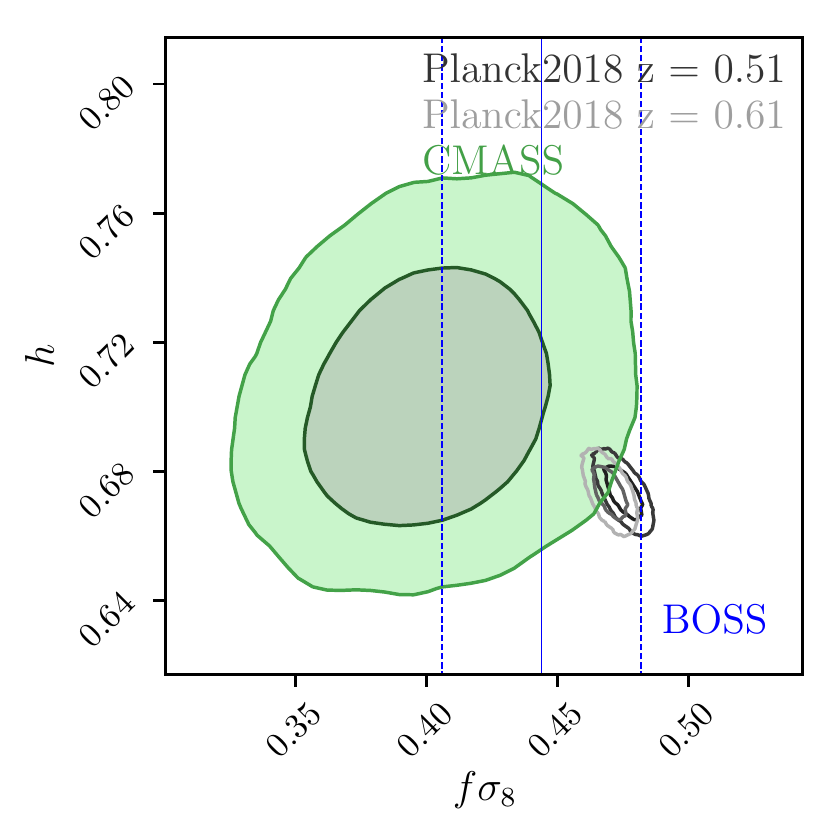}
\includegraphics[width=0.496\textwidth,draft=false]{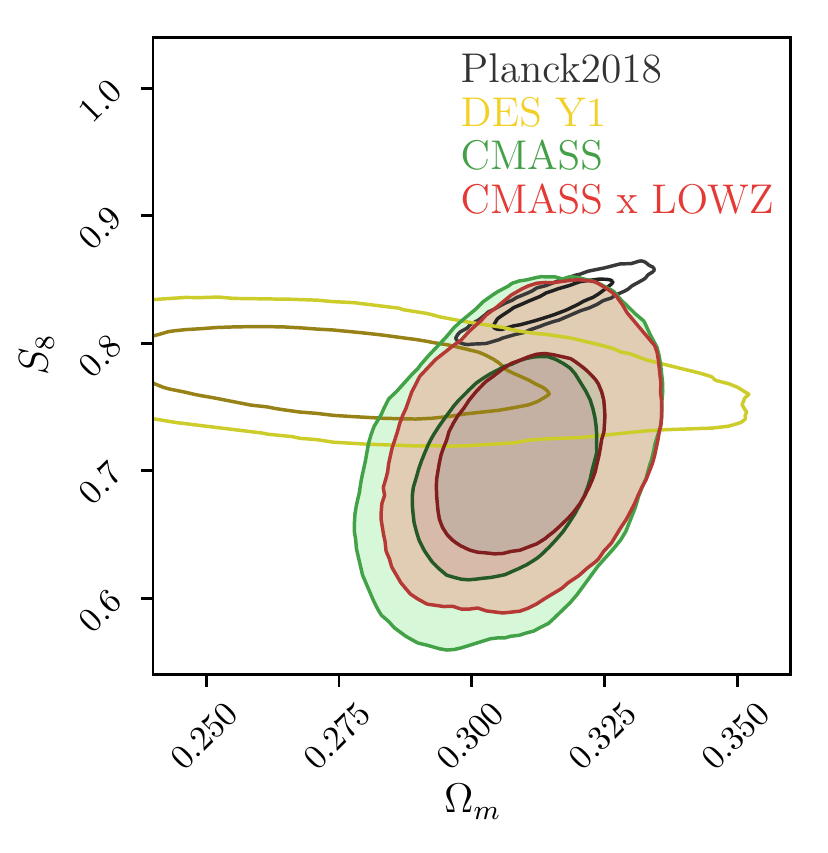}
\end{center}
\caption{\small  2D posterior distributions of the cosmological parameters $f(z=0.55)\sigma_8(z=0.55)$ and $h$ obtained from fitting the CMASS power spectrum sample ({\it left}), and of $S_8=\sigma_8(z=0) \left(\Omega_m/0.3\right)^{0.5}$ and $\Omega_m$ obtained from fitting the CMASS and the LOWZ NGC sample ({\it right}). We also plot the 2D posteriors for the same quantities from Planck2018~\cite{Aghanim:2018eyx} and DES1Yr~\cite{Abbott:2017wau} and the 1D posterior for $f\sigma_8$ from BOSS CMASS sample~\cite{Gil-Marin:2015sqa}. We find that the results are consistent. } \label{fig:data_corner3}
\end{figure}

 There are several features to discuss. 
\begin{itemize} 
\item Fitting the power spectrum up to {$\kmax=0.20\hinvMpc$ for the CMASS sample and $\kmax=0.18\hinvMpc$} for the LOWZ sample, determines $A_s, \Omega_m$ and $h$ to about {$13\%$, $3.2\%$ and $3.2\%$ respectively, as seen from Fig.~\ref{fig:posteriordatacombinedall}.
If we add the Planck prior on the sound horizon, $A_s, \Omega_m$ and $h$ are determined to about $13\%, 2.9\%$ and $1.9\%$}, respectively. This is quite consistent with what we see in simulations. Especially, the statistical errors on CMASS NGC are quite similar to the ones we find in Patchy mocks, see Fig.~\ref{fig:posteriordataNGC}.

\item The best fit model has a {good $p$-value} and the EFT parameters are well within what is theoretically expected, see Table~\ref{tab:bestfit}, and comparable to what we see in simulations. From Fig.~\ref{fig:bestfit-data} we do not see any surprising feature in the comparison of the data to the model. 

\item We obtain consistent constraints on the cosmological parameters from the power spectrum-only analysis as we change the data set, with and without the inclusion of the Planck sound horizon prior. The change in the statistical error as we include the Planck sound-horizon prior is quite consistent with what we saw in simulations. {More in detail, if we look at  Fig.~\ref{fig:data_corner}, we can see that the 1-$\sigma$ contours in the 2D posterior distributions for both the cosmological parameters and the EFT-parameters intersect or barely miss to do so, with CMASS SGC and LOWZ NGC having a very strong overlap.} This consistency remains true at the level of the 1D posterior distributions.

\item In Fig.~\ref{fig:posteriordatacombinedall}, we plot our results for the posterior distribution of the cosmological parameters together with the 68\% confidence interval from Planck2018~\cite{Aghanim:2018eyx} and WMAP9yr~\cite{Hinshaw:2012aka}. We see that while our measurements are clearly less precise than the ones of Planck2018 for $A_s$ and $h$, while the error bar is very similar for $\Omega_m$. Once we compare to WMAP9yr, we find that the size of our error bars is similar for $h$ and smaller for $\Omega_m$. Furthermore, our error bar on $h$ is only about {$20\%$ worse than the one obtained with strong lensing~\cite{Wong:2019kwg} and with the  latest Supernovae results~\cite{Yuan:2019npk}}. Given that, roughly speaking, BOSS is currently not the latest generation LSS experiment (in comparison with DES, DESI, etc.) similarly to how WMAP is not currently the latest generation CMB experiment, this comparison with WMAP9yr (and also with Planck2018) is highly suggestive and bears lots of hopeful expectations for the potentialities of LSS science in the next decade or two.

We remind the  reader that the accurate modeling of LSS provided by the EFTofLSS has allowed us to use all the relations among parameters implied by the $\Lambda$CDM  cosmology (such as for example the use of the transfer functions or the dependence of $f$ on $\Omega_m$), without the need to add some artificial free parametrization beyond the one provided by the EFTofLSS. In this sense, our analysis is identical to the standard CMB ones, such as those of WMAP and Planck.

\item  As can be seen from Fig.~\ref{fig:posteriordatacombinedall} and  Table~\ref{tab:p-values}, without the inclusion of the Planck prior on the sound horizon, the preferred values of the cosmological parameters are in statistical agreement with the ones measured from Planck2018. 

{Our value of $\lnAs$ is slightly, 2.3$\sigma$, lower than the one measured in Planck2018,} while the values of $\Omega_m$  and $h$ are extremely consistent with CMB measurements. Though statistically non-significant, the trend we see in our results is in accordance with trends found in other LSS measurements (see for example results from the DES collaboration~\cite{Abbott:2017wau}). For BOSS, this is in the same direction as the measurement of $f \sigma_8$ from of the high redshift bin~\cite{Alam:2016hwk}. 
By looking at Fig.~\ref{fig:data_corner}, we see that, in the 2D posteriors, the push to lower values of $A_s$ is driven by the CMASS NGC sample.

\item As expected, the central values of the posterior distribution for the cosmological parameters is slightly pushed towards Planck2018 values when we add the Planck prior on the sound horizon. {The compatibility with Planck2018 is similar to the case of no sound horizon prior}.

\item  {The derived quantity $f\sigma_8$ at $z=0$ from our analysis of the power spectrum up to $\kmax = 0.20\hinvMpc$ for CMASS and up to $\kmax=0.18$ for LOWZ NGC reads: {$f(z=0)\sigma_8(z=0) = 0.363\pm0.026$}. If use only the CMASS sample, we obtain the constraint on this quantity at $z=0.55$ to be {$f(z=0.55)\sigma_8(z=0.55) = 0.399\pm0.031$}}. We remind the reader that this is a derived quantity for us, as, since we assume $\Lambda$CDM and all the relations implied by that, our natural parameters as $A_s$, $\Omega_m$ and $h$. The measurement does not improve significantly if we add the Planck prior on the sound horizon. Our statistical error is roughly {20\% better} than the bound obtained by the final BOSS power spectrum analysis in~\cite{Gil-Marin:2015sqa,Beutler:2016arn}, $f(z=0.57)\sigma_8(z=0.57)=0.444\pm 0.038$, and our central value is a bit lower, though in a statistically compatible way. We plot the 2D posterior distribution for $h$ and $f\sigma_8$ in Fig.~\ref{fig:data_corner3}.  In the same figure, we also plot the 2D posterior distribution for $\Omega_m$ and the quantity $S_8=\sigma_8 \left(\Omega_m/0.3\right)^{0.5}$. The 1D posterior distribution reads, at $z=0$, {$S_8=0.704\pm 0.051$} and {is lower than the one of DES1Yr~\cite{Abbott:2017wau}, though, again, in a statistically very compatible way}. 

\item Finally, when we include the bispectrum monopole in the analysis of CMASS NGC, we find the results presented in Fig.~\ref{fig:posteriordatabisp} and~\ref{fig:data_corner2}. {As from Table~\ref{tab:bestfit}, the $p$-value of the best fit is $0.29$}. The values of the EFT-parameters and of the residuals, that we do not show to avoid clutter, do not show any particular feature.
The bispectrum monopole brings no significant statistical improvement on our results, quite consistent with what we saw in the Patchy mocks. The marginality of the improvement could be originating from several limitations in our bispectrum analysis that we mentioned in Sec.~\ref{sec:bispectrum}. We therefore do not use these data in our final constraints, though we here highlight its importance as a complementary statistical probe.

\item We finally note that the unprecedented precision of these measurements, as well as the breaking of previously unbroken degeneracies, should warn us against possibly yet unmeasured observational systematics. Though we find no issues on the theory side and in the comparison with the simulations, we make no assessment of the potential observational systematics that could have become relevant. To fully trust our findings, a reassessment of the observational systematic error is quite required.

\end{itemize}

\subsection{Neutrino mass}

We conclude the section on data analysis by highlighting an important fact that is shown by our analysis. It is well known that the leading effect of the non-vanishing  of the neutrino masses is to induce a damping in the amplitude of the evolved power spectrum at distances shorter than the free-streaming scale of the neutrinos. Since it is in general believed to be hard to obtain a measurement of $A_s$ and $\Omega_m$ from the clustering power spectrum alone, in order to detect the value of the neutrino masses from the analysis of this observable, it is therefore generically assumed that one has either to detect the step in the power spectrum induced by neutrinos, or to measure $A_s$ (or equivalently $\Omega_m$) using some other observable (for example, CMB, lensing, cluster abundance, etc.). In our analysis, we have seen that we can measure $A_s,\,\Omega_m$ and $b_1$ directly using only the clustering power spectrum. This clearly adds additional constraining power, as we are going to show.

An analysis of the neutrino masses in BOSS data goes beyond the scope of the current paper, and we leave it to future work~\cite{leoprogressTaylorSDSS}. As a proof of principle and to show the potential power of using the EFTofLSS to detect neutrino masses, here we perform  an analysis in flat $\Lambda$CDM with non-vanishing neutrino masses: we fix all the cosmological parameters to Planck2018, and fit the data by varying the mass of neutrinos, assuming normal mass hierarchy for definiteness. 

At the level of the EFT model, we do not implement the rigorous derivation of the neutrino effects developed in~\cite{Senatore:2017hyk}, but rather simply perform the EFT computation described in Sec.~\ref{sec:modeltheory} using the linear power spectrum produced with such neutrino spectrum. Such an analysis has several shortcomings: the dark matter and neutrino clustering could be computed more accurately, we should allow the galaxy  clustering to independently depend on the neutrinos overdensity, we should properly account for the $k$-dependence of the growth factors in the EFT-parameters (see for example~\cite{Chiang:2018laa}), one should impose Planck priors on the cosmological parameters,  rather than fix them to the most likely ones from Planck2018.  However, the effect we include should be the leading one~\cite{Senatore:2017hyk}. We stress instead that it is acceptable to impose the Planck priors on all the cosmological parameters, as it is a fair assumption to study first the physics that we are sure is present in the `standard cosmological model', such as flat $\Lambda$CDM with neutrino masses, and not to limit the power of such a `standard model' analysis by looking at the same time for more exotic physics, such as dynamical dark energy.

\begin{figure}[h!]
\begin{center}
\includegraphics[width=0.44\textwidth,draft=false]{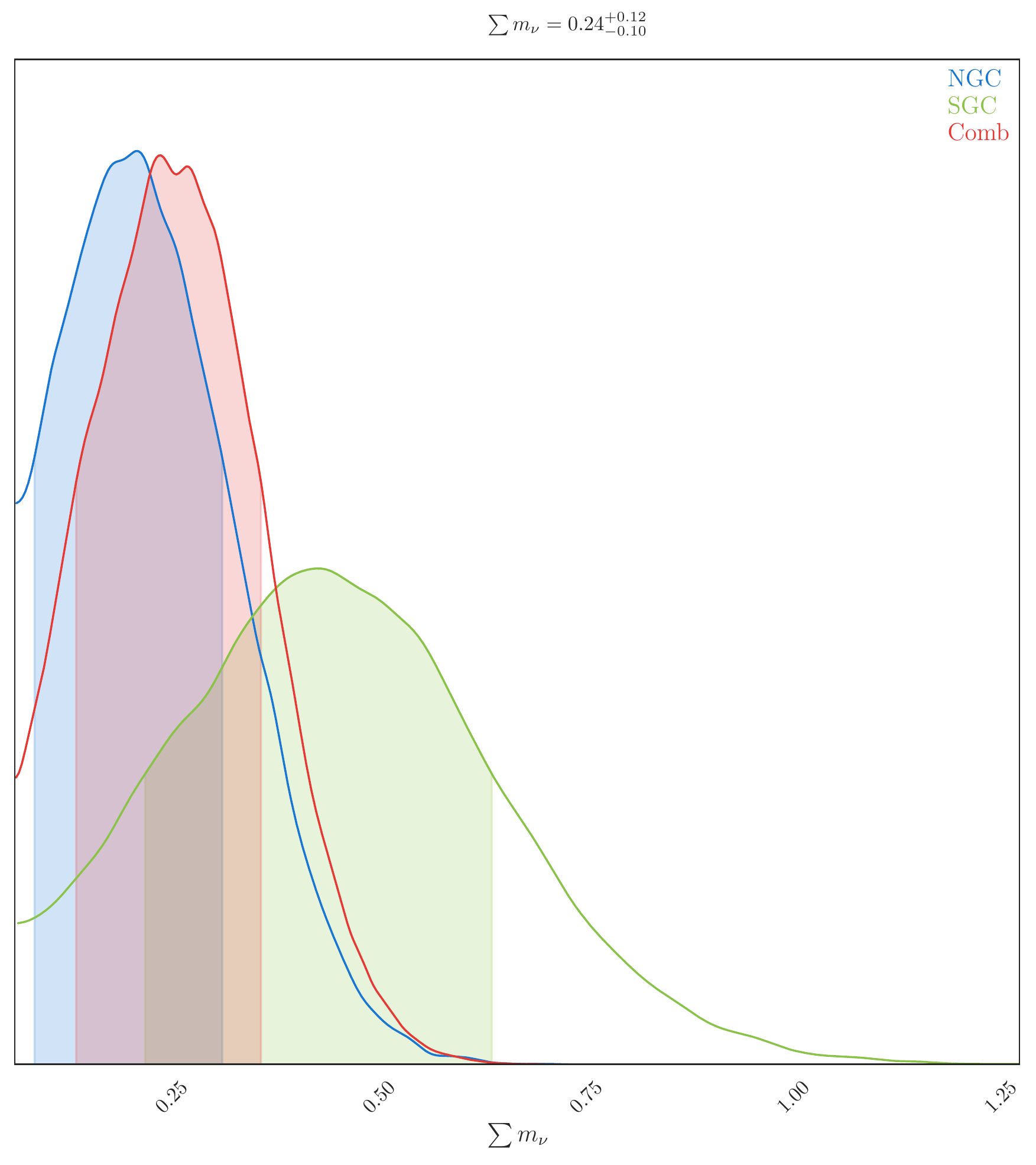}
\includegraphics[width=0.49\textwidth,draft=false]{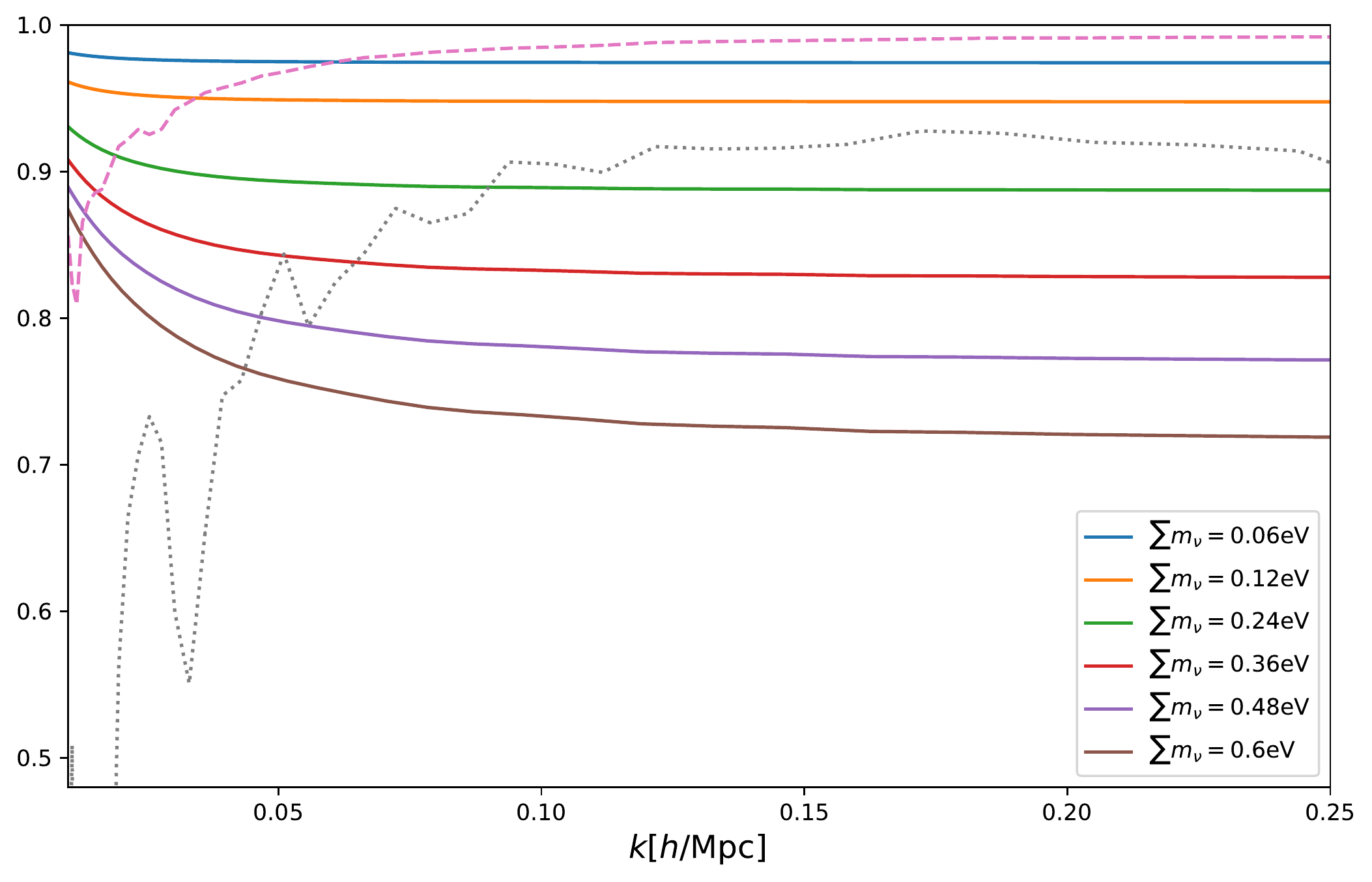}
\end{center}
\caption{\small {\it Left:} Posterior distribution for  the sum of the neutrinos in normal hierarchy, obtained from fitting the CMASS NGC, SGC and NGCxSGC power spectrum up to $\kmax = 0.2 \hinvMpc$, and by fixing all the remaining cosmological parameters to Planck2018. {\it Right:} Ratio of the linear power spectrum with and without neutrino masses, as a function of the sum of the neutrino masses in normal hierarchy. We see the step in the linear power spectrum. We also plot the monopole and quadrupole errors in red dashed and black dotted respectively. For fixed cosmological parameters, we expect a value of $\sum_i m_{\nu_i}$ to be detectable once the drop in the power spectrum is below the error from the quadrupole, suggesting $\sum_i m_{\nu_i}\sim 0.25$eV should be detectable.} \label{fig:data_neutrinos}
\end{figure}

With all these caveats in mind, we show on the left of Fig.~\ref{fig:data_neutrinos} the posterior distribution for the sum of the neutrino masses and for $A_s$ from the analysis of the power  spectrum of the CMASS sample of the BOSS DR12 data, for the analysis we mentioned earlier. We see that we have quite a tight constraint on the measurement of the sum of the neutrino masses, with error bars that are  roughly comparable to the ones from Planck2018. Of course, given  that  our preferred value  of $A_s$ is lower than Planck2018, we find a statistically-non-significant preference for non-vanishing masses once we fix $A_s$ to Planck2018. The origin of such a tight error bar can can be traced back to the plot on the right of Fig.~\ref{fig:data_neutrinos}, where we  plot the ratio of the linear power spectra with and without neutrino masses, together with the error bars of the monopole and the quadrupole. Given that $A_s,\; h$ and $\Omega_m$ are fixed, for given neutrino masses, the monopole can be fitted by lowering $b_1^2$. The amplitude of the quadrupole then drops by $b_1$, but once the drop is below the error allowed by the quadrupole, the chosen neutrino masses are excluded. We indeed can see that as the sum of the neutrino masses is above $\sim0.26$eV, the neutrino should begin to be  excluded. This is indeed comparable to the error bars we have in the data.

In summary, though we find no evidence of a non-vanishing neutrino mass, and with the caveats that we described earlier in this section, we see a slight statistical preference for a mass above the minimal one imposed by neutrino oscillation experiments once we fix $A_s$ to the Planck2018 value. Most importantly, the error bars are quite tight: with some optimism, a detection of the neutrino masses, using LSS, might be around the corner.

\newpage
 \section{Conclusions}

We have used the predictions from the Effective Field Theory of Large-Scale Structure (EFTofLSS) to analyze the data from the DR12 BOSS CMASS and LOWZ samples. In $\Lambda$CDM, by fixing $\Omega_b/\Omega_c$ and $n_s$ to the best fit values of Planck2018, we have analyzed first the power spectrum up to {$\kmax=0.2\hinvMpc$ for the CMASS sample and up to $\kmax=0.18\hinvMpc$} for the LOWZ sample, finding that we can measure {$A_s$, $\Omega_m$ and $H_0$ to about {$13\%$, $3.2\%$ and $3.2\%$} respectively. We have then added the Planck prior on the sound horizon, finding that this further improves the constraints to about {$13\%, 2.9\%$ and $1.6\%$} respectively}. We have also included the data on the bispectrum monopole, finding a limited further improvement on the measurement of $A_s$. We have performed extensive checks with numerical simulations, that include several different HOD models, finding that the theoretical systematic error of the EFTofLSS is negligibly small. We have also performed a careful implementation of the window function and fiber collision effects, as well as studied the redshift selection effects, concluding that they do not induce any residual sizable systematic error. We have confirmed this conclusion using simulations that include these observational effects.

The analysis of the BOSS data using the EFTofLSS has required us to develop  some interesting technical implementations. For example, we have developed a numerically-non-demanding way to apply the survey window function  directly in Fourier space (as the standard procedure of applying it in real space was numerically challenging due to peculiarities of the observables in the EFTofLSS)~\footnote{See also~\cite{Beutler:2013yhm} for a different numerical  implementation.}; to reduce the number of parameters to be explored in the MCMC, we have marginalized analytically over some of the EFT parameters that appear linearly in the power spectrum.\\ 

It appears to us that these measurements, that were enabled by the  EFTofLSS, represent a very substantial qualitative and quantitative improvement of  former analyses of redshift-clustering data. In fact, these parameters were never measured individually, and in such an accurate way. If confirmed, our results suggest that the Large-Scale Structure (LSS) of the Universe might contain much more exploitable cosmological information than previously believed. They therefore open up some research directions that we now briefly highlight.  \\

On the side of the EFTofLSS and of  the numerical simulations, we point out the following interesting  research avenues that could be pursued:
\begin{itemize}

\item Our analysis is limited by the order  at which we  have performed our calculation. Here we have used the one-loop power spectrum and the tree-level bispectrum. It would be useful to  go to the next perturbative order in both these observables, as well as to compute the trispectrum. This would allow us to additionally increase the accuracy of our predictions, as we mention briefly in the main text.

\item By the use of simulations, the theory systematic error has been measured to be small up to the $\kmax$ at which we have stopped the analysis, and we have therefore simply neglected it. It would be desirable to account for it in  a more systematic way, either within the EFTofLSS itself (as we have highlighted in the main text), or by using more accurate and precise simulations.

\item Notwithstanding the presence of  many EFT parameters, our result show the great constraining power provided by the EFTofLSS. In this paper we have made no use of simulations to determine any prior about the EFT parameters. Simulations have been used only to estimate the theoretical systematic error. However, our statistical errors would be much smaller if we had stronger priors on the EFT parameters. We have checked that if we were to put a 15\% prior on $b_1$, and more stringent priors on the  EFT-parameters,  our error bars would shrink. It would be very useful if simulations could provide  reliable priors of this order. For example, one could consider if it is possible to extend techniques such as the `separate universe approach' (see for example~\cite{Baldauf:2011bh,Baldauf:2015vio,Lazeyras:2015lgp,Chiang:2016vxa,Chiang:2017vuk}) to include galaxies in a reliable way. Similarly, one could consider to use the information from physically-motivated phenomenological models, such as the very-well-known `halo model'~(see for example~\cite{Cooray:2002dia}), to provide stringent priors on the EFT parameters. 

\item We have highlighted how the measurement of the galaxy-bias coefficients (and more generally of the EFT parameters) that we can perform, could in principle allow us to determine the galaxy formation mechanism in the LSS. It would be interesting if a mapping between galaxy formation models and EFT parameters were to be performed.

\item 
Our measurements are statistically compatible with the ones from Planck2018~\cite{Aghanim:2018eyx}, with a slight lower value for $A_s$. 
It would be interesting to see if this extremely mild tension can be ameliorated, for example, by including curvature or an equation of state for dark energy ({\it i.e.} $\Omega_K\neq 0$ or $w\neq -1$), or allowing the mass of the neutrinos is larger~(see~\cite{leoprogressTaylorSDSS}). Of course, we should not forget that these tensions are not statistically significant, but however go together with a general trend observed in analysis of other LSS data (see for example results from the DES collaboration~\cite{Abbott:2017wau}).
 
\end{itemize}

On the side of observations, we highlight a few interesting improvements and directions that could be pursued:
\begin{itemize}
\item[$\square$] In the EFTofLSS, several $n$-point functions, with $n>2$, are predicted using approximately the same parameters as the power spectrum. We have not analyzed any of them beyond  the bispectrum monopole, simply because we did not have such measurements at our disposal. Given the opportunity provided by the EFTofLSS, it would be desirable if measurements of multipoles of higher $n$-point functions, such as, for example, the bispectrum quadrupole or the trispectrum monopole, were available to us. The analysis of these data is rather straightforward from the EFTofLSS point of view. Similarly, we have not tried to optimize the analysis, for example studying the effect of the binning of the data or the use of better estimators. Indeed, the main objective of this paper has been to provide a first comparison of the EFTofLSS with data.

\item[$\square$] Similarly, it would be interesting to repeat our analysis using the configuration-space correlation function. In fact, the relatively-limited reach in $\kmax$ of the EFTofLSS at one-loop forces us to neglect the signal associated to the BAO oscillations at wavenumber higher than~$\kmax$. In fact, as already shown in~\cite{Senatore:2014vja}, the BAO features can be accurately reproduced in their entirety by the IR-resummed EFTofLSS. A configuration-space analysis, or a modification of the current analysis in Fourier space, should allow for this additional information to be used.  

\item[$\square$] Clearly, the fact that the EFTofLSS allows us to break degeneracies and measure several cosmological parameters with such an accuracy requires that we re-asses the measurement of the observational systematics. This is particularly true given the slight disagreement with CMB experiments for the measurement of $A_s$, in particular when we add the bispectrum measurements. We stress that this is very important to do to fully trust our findings.
 
 \item[$\square$] Of course, it would be nice to apply our analysis to larger LSS surveys, such as DES, DESI, eBOSS, Euclid, LSST, etc., as well as to intensity mapping experiments. Similarly, it would be interesting to include the most common extensions to $\Lambda$CDM, such as neutrinos (for which we presented a preliminary analysis), non-Gaussianities, $N_{\rm eff}$, curvature, {\it etc.}, {\it etc.}. Overall, we have performed this first analysis of the BOSS data using the EFTofLSS having in mind the current and next generation LSS surveys, and more general cosmological parameters.
 
\end{itemize}

We leave all of this, and more, to future work, some of which is already in progress~\cite{leoprogressTaylorSDSS}.

 \section*{Acknowledgements}

Preliminary, but already at an advanced stage of completion, results of this work were presented at the following avenues (with link to the publicly-available slides): conference `the non-linear universe', Slovenia, July 2018; conference `Analytic Methods in Cosmology', Institute Poincare, Paris, Sept. 2018; conference `Modern Inflation', Simons Center, NY, Sept. 2018; seminar, Harvard University, Nov. 2018; `South American Workshop on Cosmology in the LSST Era', ICTP-SAIFR, Brazil, Dec. 2018; seminar, EPFL, Switzerland, Jan 2019; Stanford physics colloquium, Feb. 2019~\cite{conference3}; conference `Accelerating universe in the dark', Kyoto, March 2019~\cite{conference1}; 2019 CCNU-USTC Junior Symposium, Wuhan, April 2019; conference `24th Rencontres Itzykson - Effective Field Theory in Cosmology, Gravitation and Particle Physics', Paris, June 2019; conference `The nature of Dark Matter and Large Scale Structure', Cyprus, June 2019, workshop `Large-Scale Structure Formation', Munich, July 2019, conference `EPS-HEP219', July 2019~\cite{conference2}.

We thank Matias Zaldarriaga for many conversations throughout the development of this project and for comments on some of the results. We thank Jeremy Tinker for allowing us to use the BOSS DR12 Challenge Boxes and Joe DeRose and Jeremy Tinker for helpful correspondence.  We thank Elisabeth Krause, Roman Scoccimarro and Risa Wechsler for helpful comments. We thank Mikhail Ivanov,  Marko Simonovic and Matias Zaldarriaga for comparing the results of their pipeline with our first results, which helped us to find a coding error in our pipeline which had small but not completely negligible consequences at quantitative level. We thank Thomas Colas for comparing his pipeline as well. We also thank Masahiro Takada and Takahiro Nishimichi for allowing us to compare our pipeline with their simulations.

NK thanks the LSSTC Data Science Fellowship Program, which is funded by LSSTC, NSF Cybertraining Grant 1829740, the Brinson Foundation, and the Moore Foundation.

 LS would like to additionally thank E. komatsu, A. Refregier, Uros Seljak, David Spergel and Ben Wandelt for many discussions and for overall encouragement, and also J.J. Carrasco, P. Creminelli, R. de Belsunce, M. Cataneo, S. Foreman, M. Lewandowski,  V. Mauerhofer, A. Maleknejad, E. Nadler, A. Nicolis, H. Perrier, R. Porto, R. Sgier, G. Trevisan, F. Vernizzi for support in the long journey to apply the EFTofLSS to data.
LS and GDA are partially supported by Simons Foundation Origins of the Universe program (Modern Inflationary Cosmology collaboration) and LS by NSF award 1720397. 

PZ would like to thank Ashley Perko for her patience in explaining the Mathematica code RedshiftBiasEFT~\footnote{See \href{http://stanford.edu/~senatore/}{EFTofLSS repository}.} and express his gratitude to the people of SLAC and SITP for their warm welcome where part of this work was done. LS and PZ would like to additionally thank Yi-Fu Cai for support and encouragement.

FB is a University Research Fellow. HGM acknowledges la Caixa Banking Foundation (LCF/BQ/LI18/11630024) for support.

The Boltzmann code CLASS was used to compute the matter linear power spectrum~\cite{Blas:2011rf}. Some plots were done using the Chainconsumer package~\cite{Hinton2016}. Part of the analysis was performed with the Mercury cluster at the University of Science and Technology of China, and PZ thanks Jie Jiang and Tian Qiu for support. Part of the analysis was also performed on the Sherlock cluster at the Stanford University and we thank the support team.

\appendix
 
 \section{\label{sec:galaxykernels} Galaxy kernels}
 We give the explicit expressions for the galaxy kernels appearing in the one-loop power spectrum and refer to \cite{Perko:2016puo} for details about their derivation. We choose to work in the basis of descendants (this is the first complete set of bias coefficient for LSS, established in \cite{Senatore:2014eva,Angulo:2015eqa} and with some typos corrected in \cite{Fujita:2016dne}; see~\cite{Senatore:2014eva,Angulo:2015eqa} for connection to former bases of bias coefficients, as for example~\cite{McDonald:2009dh}). For the one-loop power spectrum and tree-level bispectrum, all kernels can be described with 4 bias parameters~$b_i$.
 
The first and second order galaxy density kernel are:
 \begin{align}
     K_1 & = b_1, \\
     K_2(\q_1,\q_2) & = b_1\frac{\q_1\cdot \q_2}{q_1^2}+ b_2\left( F_2(\q_1,\q_2)- \frac{\q_1\cdot \q_2}{q_1^2} \right) + b_4 + \text{perm.} \ .
 \end{align}
The galaxy velocity kernels $G_n$ are simply the standard perturbation theory ones since the galaxy velocity field follows the dark matter velocity field, up to higher-derivative terms  which are degenerate with other counterterms that appear in the renormalization of the redshift space expression  (see e.g. \cite{Bernardeau:2001qr} for the expressions of $F_n$ and $G_n$).

The third-order galaxy density kernel has a much more involved expression. However, for the one-loop calculation, degeneracies appear in the one-loop diagram obtained from $\langle\delta^{(3)}\delta^{(1)}\rangle$, when UV-divergences are removed and the integral over the angular coordinates is performed, leading to the following simple expression:
\begin{align}
    K_3(k,q) & = \frac{b_1}{504 k^3 q^3}\left( -38 k^5q + 48 k^3 q^3 - 18 kq^5 + 9 (k^2-q^2)^3\log \left[\frac{k-q}{k+q}\right] \right) \nonumber \\
    &+ \frac{b_3}{756 k^3 q^5} \left( 2kq(k^2+q^2)(3k^4-14k^2q^2+3q^4)+3(k^2-q^2)^4 \log \left[\frac{k-q}{k+q}\right]  \right).
\end{align}

 \section{\label{sec:Loopeval} Evaluation of the loop integrals}

We present two strategies to evaluate the loop integrals appearing in the galaxy power spectrum~\eqref{eq:powerspectrum}: a simple numerical one and an analytical one relying on the decomposition of the linear power spectrum in $\log k$-space.

For the numerical evaluation we adopt the IR-safe and UV-safe integrand version of the 22-diagram + 13-diagram \cite{Lewandowski:2017kes}, such that most of the (unphysical) divergences that are hard to handle numerically are absent (see App.~D of \cite{Perko:2016puo} for an explicit expression)\footnote{The integrals are computed using the Cuhre routine of the CUBA library \cite{Hahn:2004fe}. This cubature algorithm happens to be very competitive in this low-dimension problem compared to Monte-Carlo algorithm such as Vegas. As the integral over one angular coordinate can be calculated analytically, the integrals are only two-dimensional.}.
To gain additional computational speed, we calculate only their differences to a precomputed cosmology \cite{Cataneo:2016suz}. 
The loop integrals are of the form:
\begin{equation}\label{eq:loopintegralform}
P_n(k) = \int \frac{d^3 q}{(2\pi)^3}\, P_{n,\text{integrand}}(\k,\q),
\end{equation}
which can be calculated with respect to a reference cosmology:
\begin{equation}
P_n^\text{target}(k) = P_n^\text{ref}(k)+\Delta P_n(k),
\end{equation}
where ``target'' and ``ref'' refer to a cosmology being evaluated and a fiducial (reference) cosmology, respectively. The difference $\Delta P_n(k)$ is the integral of the difference between the reference and the target integrands:
\begin{equation}\label{eq:difference}
\Delta P_n(k) = \int \frac{d^3 q}{(2\pi)^3} \, \left[ P_{n,\text{integrand}}^\text{target}(\k,\q) - \left( \frac{A_s^\text{target}}{A_s^\text{ref}} \right)^2 P_{n,\text{integrand}}^\text{ref} (\k,\q) \right],
\end{equation}
where the ratio of the scalar amplitudes $(A_s^\text{target}/A_s^\text{ref})^2$ is inserted to rescale the amplitude of the reference integrand to the target one such that $\Delta P_n(k)$ is as small as possible.

Once the loop integrals for a reference cosmology $ P_n^\text{ref}(k)$ are precomputed to high-enough precision, we are free to evaluate only the difference $\Delta P_n(k)$ instead of the full integral, to a relatively low accuracy \footnote{It can be shown that the allowed relative error $\epsilon_\Delta$ on the evaluation of $\Delta P_n(k)$ is inversely proportional to its size:
\begin{equation}
\epsilon_\Delta \approx \left| \frac{P_n^\text{target}(k)}{\Delta P_n(k)} \right| \epsilon_\text{target} \, ,
\end{equation}
where $\epsilon_\text{target}$ is the required precision on the evaluation of $P_n^\text{target}(k)$. A close inspection of the integrals involved in the 1-loop expression shows that the internal momenta that will contribute the most are of order $k$. The kernels being independent of the cosmology, we can estimate the integrals as follows with the corresponding linear power spectra evaluated at $k$:
\begin{equation}\label{eq:estimate}
\left| \frac{\Delta P_n(k)}{P_n^\text{target}(k)} \right| = \left| 1 - \left( \frac{A_s^\text{target}}{A_s^\text{ref}} \right)^2 \frac{\int d^3q \, P_{n,\text{integrand}}^\text{ref} (\k,\q)}{\int d^3 q \, P_{n,\text{integrand}}^\text{target} (\k,\q)} \right| \approx \left| 1 - \left( \frac{A_s^\text{target}}{A_s^\text{ref}} \right)^2 \left( \frac{P_{11}^\text{ref}(k)}{P_{11}^\text{target}(k)}\right)^2 \right|.
\end{equation}
This estimate has been shown to be good enough for the 2-loop matter power spectrum \cite{Cataneo:2016suz}. We thus confidently use this estimate in our 1-loop calculation and choose $\epsilon _\Delta \approx 5 \epsilon _\text{target}$.}. This simple manipulation speeds up the computation by at least a factor 2 without precision loss.

As we mentioned, following the procedure of~\cite{Carrasco:2013mua,Lewandowski:2017kes,Perko:2016puo}, we make our integrals UV-safe, by removing the UV-limit of the integrand. This contribution indeed, after integration, is degenerate with the counterterms. This has the effect to make the integrals smaller, and so easier to evaluate. It also diminishes the size of the `bare' part of the counterterms, so that they are, effectively, directly the renormalized ones, and therefore, expected to be of order one. 

One can also choose to evaluate the loop integrals on a set of complex power-law cosmologies produced by decomposing the linear power spectrum using an FFTLog algorithm \cite{Simonovic:2017mhp}.  This procedure has several advantages. First, the loop integrals in eq.~\eqref{eq:powerspectrum} are analytic for power-law cosmologies. In fact, they are formally equivalent to massless scalar propagator loop integrals in quantum field theory. As such, all UV divergences are automatically removed from the loop thanks to dimensional regularization and the counterterms needed to absorb these divergences are assured to be order $\mathcal{O}(1)$. The second advantage is practical. The loop integrals can be evaluate once and for all as they do not depend on the cosmology. Evaluating the loop integrals in eq.~\eqref{eq:powerspectrum} at a new cosmology reduces to decomposing the linear power spectrum in a sum of complex power-law cosmologies and a few matrix multiplications\footnote{We acknowledge the use of the C++ library Eigen~\cite{eigenweb}.}.

For $N$ sampling points, the linear power spectrum can be approximated using FFT in log-space by:
\begin{equation}
    P_{11}(k) = \sum_{-N/2}^{N/2} c_m k^{\nu+i \eta_m},
\end{equation}
where
\begin{equation}
    c_m = \frac{1}{N} \sum_{l=0}^{N-1} P_{11}(k)k_l^\nu k_\text{min}^{-i \eta_m} e^{-2\pi i m l /N}, \qquad \eta_m=\frac{2\pi m}{\log (k_\text{max}/k_\text{min})}.
\end{equation}
For $-3<\nu<-1.5$ the IR divergences in the loop cancel. Expanding the kernels in integer powers of $k^2$, $q^2$ and $|\k-\q|^2$, the loop integrals can be performed analytically on the power-law cosmologies $\eta_m$. The $22$-loop and $13$-loop diagrams become simple matrix multiplications:
\begin{align}
    P_{n,22}(k) & = k^3 \sum_{m_1,m_2}c_{m_1}k^{\nu+i\eta_{m_1}} \cdot M_{n,22}\left(-\tfrac{\nu+i\eta_{m_1}}{2}, -\tfrac{\nu+i\eta_{m_2}}{2}\right) \cdot k^{\nu+i\eta_{m_1}}, \\
    P_{n,13}(k) & = k^3 P_{11}(k)\sum_{m_1} c_{m_1} k^{\nu+i\eta_{m_1}} \cdot M_{n,13}\left(-\tfrac{\nu+i\eta_{m_1}}{2}\right),
\end{align}
where the matrix $M_{n,22}(\nu_1,\nu_2)$ and the vector $M_{n,13}(\nu_1)$ for each term $P_n$ (which is either of the 22 or the 13 form) in   the decomposition of the kernels as induced by~(\ref{eq:factorizedpowerspectrum})  can be obtained by decomposing the kernels in power laws of momenta, and performing the resulting momentum integral with the following expression:
\begin{equation}
    \frac{1}{2\pi^3}\int d^3q \, \frac{1}{q^{2\nu_1} |\k-\q|^{2\nu_2}}\equiv \frac{1}{8\pi^{3/2}}k^{3-2\nu_{12}} \frac{\Gamma(\tfrac{3}{2}-\nu_1)\Gamma(\tfrac{3}{2}-\nu_2)\Gamma(\nu_{12}-\tfrac{3}{2})}{\Gamma(\nu_1)\Gamma(\nu_2)\Gamma(3-\nu_{12})}.
\end{equation}
 In our analysis, we have implemented both procedures for the evaluation of the integrals, finding identical results for all practical purposes.
 
\section{\label{sec:IR-resummation} IR-resummation scheme}

The parameters parametrizing the tidal forces of long-wavelength modes $k$ on the short-wavelength modes and the effect of short-wavelength displacements are, respectively:
\begin{equation}
\epsilon_{\delta<} = \int^kd^3 q\, P_{11}(q), \qquad \epsilon_{s>} = k^2 \int_k^\infty  d^3 q\, \frac{P_{11}(q)}{q^2}.
\end{equation}
Both of these scale as powers of $k/k_\textsc{NL}$ and are thus small: the perturbative expansion is valid and the convergence is assured. However, long-wavelength displacements, which are parametrized by:
\begin{equation}
\epsilon_{s<} = k^2 \int^k d^3q \, \frac{P_{11}(q)}{q^2}.
\end{equation}
are generically of order one. They can not be treated perturbatively and therefore need to be resummed. We here follow the parametrically-controlled procedure developed in~\cite{Senatore:2014via}.  

The IR-resummation in redshift space given in terms of multipole moments reads \cite{Lewandowski:2015ziq} (see also~\cite{Senatore:2014vja}):
\begin{align}
P^\ell_n(k)|_N & = \sum_{j=0}^N \sum_{\ell'} \int \frac{dk'\, k'^2}{2\pi} M^{\ell\ell'}_{||N-j}(k,k')P^{\ell'}_n(k')_j, \label{eq:resumConvol}\\
M^{\ell \ell'}_{||N-j}(k,k') & = \int dq \, j_{\ell'}(k'q)	Q_{||N-j}^{\ell \ell'}(k,q), 
\label{eq:resumM2} \\
Q_{||N-j}^{\ell \ell'}(k,q) & =i^{\ell'}q^2\frac{2\ell+1}{2}\int_{-1}^{1}d\mu_k \, \int d^2 \hat{q} \, e^{-i\q \cdot \k} \, T_{0,r}(\k,\q)\times T_{0,r}^{-1}||_{N-j}(\k,\q) \mathcal{P}_\ell(\mu_k) \mathcal{P}_{\ell'}(\mu_q),\label{eq:resumQ}
\end{align}
where we denoted by $||_N$ a quantity that is expanded up to order $N$ in both $\epsilon_{\delta <},\; \epsilon_{s>} $ and $\epsilon_{s<}$, while with~$|_N$ a quantity is expanded \emph{up to} the order $N$ (loop level) in $\epsilon_{\delta <},\; \epsilon_{s>} $ and \emph{all} orders in $\epsilon_{s<}$\footnote{The careful reader might have noticed that the integral in eq.~\eqref{eq:resumConvol} runs on the true non-distorted wavenumbers and not on the measured ones. Therefore one need first to expand the power spectrum in multipoles without the AP effect, proceed to the IR-resummation scheme described here, then recombine the multipoles into the resummed power spectrum, and finally expand it again in multipoles with the AP effect this time included.}. 
$T_{0,r}(\k,\q)$ is given by
\begin{equation}\label{eq:RedshiftModulationExplicit}
T_{0,r}(\k,\q)=\exp \left\lbrace -\frac{k^2}{2} \left[ X_1(q) \left(1+2f\mu_k^2+f^2\mu_k^2 \right) + Y_1(q)\left( (\hat{k}\cdot\hat{q})^2 + 2f\mu_k\mu_q\hat{k}\cdot\hat{q}+f^2\mu_k^2\mu_q^2 \right) \right] \right\rbrace ,
\end{equation}
where
\begin{align}
X_1(q) & = \frac{1}{2\pi} \int_0^\infty dk \, \exp \left( -\frac{k^2}{\Lambda_\textsc{ir}^2}\right) P_{11}(k) \left( \frac{2}{3} - 2\frac{j_1(kq)}{kq} \right), \\
Y_1(q) & = \frac{1}{2\pi} \int_0^\infty dk \, \exp \left( -\frac{k^2}{\Lambda_\textsc{ir}^2}\right) P_{11}(k) \left( -2j_0(kq) + 6\frac{j_1(kq)}{kq} \right),
\end{align}
$j_i$ are the the spherical Bessel functions of type $i$ and $\mu_k=\hat{k}\cdot\hat{z}$. $\Lambda_\textsc{ir}$ is the scale to which we resum the linear IR modes, taken to be $\Lambda_\textsc{ir}= 0.066 \,h$ Mpc$^{-1}$ in this analysis. 

The $Q$ function can be approximated for efficient numerical evaluation \cite{Lewandowski:2015ziq}. The idea is to expand the exponential in eq.~\eqref{eq:RedshiftModulationExplicit} in powers of $k^2X_1 (\mu_k^2-1/3)$ and $k^2Y_1$. Up to order $3$ and $1$, respectively, the convergence has been checked to be good enough for a 1-loop calculation. Under this approximation, the integrals in eq.~\eqref{eq:resumQ} can be done exactly, hence allowing a fast numerical computation.
 
 We finally perform another numerical improvement. As it is well known (see for example the introduction of~\cite{Senatore:2014vja} and~\cite{Baldauf:2015xfa,Blas:2016sfa}), the IR-resummation is trustable, in the sense  that it resums terms that are larger than the rest, only for as long as it acts only on the oscillating part of the power spectrum. For this reason, we split the power spectrum at linear level and at loop level in a smoooth and an oscillating part. At linear level, the smooth part is defined with the Eisenstein and Hu no-wiggle power spectrum~\cite{Eisenstein:1997ik}, while the one-loop smooth part is defined by performing the loop integrals with the Eisenstein and Hu no-wiggle power spectrum. The oscillating part is what is left. Then, we act with the resummation matrices only on the oscillating part of the linear and loop power spectrum. This approach has the advantage that numerical inaccuracy in the resummation matrices and in the final resummation integral are relegated to the oscillating part, which is small. However, it has some disadvantages, such as, for example, the fact that whatever is oscillating and is accidentally included in the smooth part, as the split in oscillating and non-oscillating part of the power spectrum is somewhat arbitrary, is not resummed (see for example~\cite{Lewandowski:2018ywf} for a accurate discussion about this). We have checked that this numerically optimized procedure gives results that are practically undistinguishable from the original one, for the error bars we are dealing with now.

 The bispectrum should also in principle be resummed. However, given that we use the tree level expression and we stop at rather long wavenumbers, where the errors from cosmic variance are very large, we assume the effect of the IR-resummation can be neglected for the bispectrum.

 \section{\label{sec:bispectrum_likelihood} Technical aspects of bispectrum likelihood\label{sec:appendix_bispectrum_lik}}
 \def\BEFT{B^{({\rm EFT})}}
 \def\Bdata{B^{({\rm d})}_0}

We first re-derive the formula analogous to section~\ref{sec:technical-likl} with the inclusion of the bispectrum in the likelihood and then briefly discuss the application of the window function. The full covariance matrix can be decomposed into four blocks $C^{-1}_\text{PP}$,  $C^{-1}_\text{PB} \equiv C^{-1}_\text{BP}$ and  $C^{-1}_\text{BB}$, where P and B stand for power spectrum and bispectrum, respectively. The full likelihood reads:
 \begin{align}
& {\cal L}_{\rm FULL}(d|\{\vec\Omega,\vec b\}) = \nn {\cal L}_{\rm PP}(d|\{\vec\Omega,\vec b\}) + {\cal L}_{\rm PB}(d|\{\vec\Omega,\vec b\}) + {\cal L}_{\rm BB}(d|\{\vec\Omega,\vec b\}) \\
& = {\rm Exp}\left[-\frac{1}{2}(\PEFT_{\ell}{}^{(W)}(k,\{\vec\Omega,\vec b\})-\Pdata_\ell (k)) \cdot C_{\rm PP}^{-1}(k,k')_{\ell,\ell'}\cdot (\PEFT_{\ell'}{}^{(W)}(k',\{\vec\Omega,\vec b\})-\Pdata_{\ell'} (k'))\right] \nn \\
& \times {\rm Exp}\left[- (\PEFT_{\ell}{}^{(W)}(k,\{\vec\Omega,\vec b\})-\Pdata_\ell (k)) \cdot C_{\rm PB}^{-1}(k,k')_{\ell} \cdot (\BEFT_0(k',\{\vec\Omega,\vec b\}))-\Bdata (k'))\right] \nn \\
& \times {\rm Exp}\left[-\frac{1}{2}(\BEFT(k,\{\vec\Omega,\vec b\}))-\Bdata (k)) \cdot C_{\rm BB}^{-1}(k,k') \cdot (\BEFT_0(k',\{\vec\Omega,\vec b\})-\Bdata (k'))\right] ,
 \end{align}
 where on the second line we have the likelihood ${\cal L}_{\rm PP}(d|\{\vec\Omega,\vec b\})$ of the power spectrum only, eq.~\eqref{eq:likelyhoodone}, the third line represents the cross-likelihood ${\cal L}_{\rm PB}(d|\{\vec\Omega,\vec b\})$ of the power spectrum and the bispectrum, and the fourth line is the likelihood ${\cal L}_{\rm BB}(d|\{\vec\Omega,\vec b\})$ of the bispectrum only. The formulas with the explicit appearance of the window function and with the partial marginalization for ${\cal L}_{\rm PP}(d|\{\vec\Omega,\vec b\})$ are given in section~\ref{sec:technical-likl} and we do not repeat them here. In this analysis, following~\cite{Gil-Marin:2016wya}, we apply the window function to the bispectrum in a rather approximate way, where we apply the window function to each of the two power spectra that appear in the tree-level expression in eq.~(\ref{eq:treelevelbispectrum}). Such an approximate treatment should be sufficient  given the large error bars of the data at large wavenumbers.  Thus for ${\cal L}_{\rm BB}(d|\{\vec\Omega,\vec b\})$ the covariance matrix is unchanged while for the cross-likelihood, we have:
 \begin{align}
     {\cal L}_{\rm PB}(d|\{\vec\Omega,\vec b\}) & = {\rm Exp}\left[ - \PEFT_{\ell}(k,\{\vec\Omega,\vec b\}) \cdot C^{-1}_{{\rm PB}, W}(k,k')_{\ell} \cdot (\BEFT_0(k',\{\vec\Omega,\vec b\})-\Bdata (k')) \right. \nn \\
     & +  \left. \Pdata_\ell (k) \cdot C_{\rm PB}^{-1}(k,k')_{\ell} \cdot (\BEFT_0(k',\{\vec\Omega,\vec b\})-\Bdata (k')) \right],
 \end{align}
 where 
 \begin{equation}
      C^{-1}_{{\rm PB}, W}(k_1,k_2)_{\ell} = W(k_1,k')_{\ell',\ell}{}^\dagger \cdot C_{\rm PB}^{-1}(k',k_2)_{\ell'}.
 \end{equation}
 
Similarly to the power spectrum, we can perform a partial marginalization over some bias parameters. We now have $\vec b=\{\vec b_{NG}, \vec b_G\}$, with $\vec b_G=\lbrace b_3, c_{\rm ct}, c_{r,1}, c_{r,2}, c_{\epsilon ,2},c_{\epsilon ,3}, c_{\epsilon,4} \rbrace$ and $\vec b_{NG}= \lbrace b_1, b_2, b_4, c_{\epsilon ,1} \rbrace$.
Writing, in analogy to what we did for the case of the power spectrum only:
 \begin{align}
\PEFT_{\ell}(k,\{\vec\Omega,\vec b_G\}) & =\sum_i b_{G,i}\; \PEFT_{\ell,\, {\rm lin,\, } i}(k,\{\vec\Omega,\vec b_{NG}\})+\PEFT_{\ell,\, {\rm const}}(k,\{\vec\Omega,\vec b_{NG}\}), \\
\BEFT_0(k,\{\vec\Omega,\vec b\}) & = \sum_i b_{G,i}\; \BEFT_{0,\, {\rm lin,\, } i}(k,\{\vec\Omega,\vec b_{NG}\})+\BEFT_{0,\, {\rm const}}(k,\{\vec\Omega,\vec b_{NG}\}),
 \end{align}
 we get:
\bea
&&{\cal L}_{\rm PB}(d|\{\vec\Omega,\vec b_G,\vec b_{NG}\}) =\\ \nn
&&\qquad= {\rm Exp}\left[ -\frac{1}{2}\sum_{ij}  b_{G,i} \, G_{2,ij}(\{\vec\Omega,\vec b_{NG}\}) b_{G,j} + \sum_i  b_{G,i} \, G_{1,i}(\{\vec\Omega,\vec b_{NG}\}) \, + G_{0}(\{\vec\Omega,\vec b_{NG}\}) \right]\ ,
\eea
where
 \begin{align}
G_{2,ij} & = 2 \PEFT_{\ell,\,{\rm lin,\, } i}(k,\{\vec\Omega,\vec b_{NG}\}) \cdot C^{-1}_{{\rm PB}, W}(k,k')_{\ell}\cdot \BEFT_{0,\,{\rm lin,\, }j}(k',\{\vec\Omega,\vec b_{NG}\})\ , \\
G_{1,i}  & = -\PEFT_{\ell,\,{\rm lin,\,}i} (k,\{\vec\Omega,\vec b_{NG}\}) \cdot C^{-1}_{{\rm PB}, W}(k,k')_{\ell} \cdot (\BEFT_{0,\,{\rm const}}(k',\{\vec\Omega,\vec b_{NG}\}) - \Bdata(k'))  \nn \\
& - \PEFT_{\ell,\,{\rm const}}(k,\{\vec\Omega,\vec b_{NG}\}) \cdot C^{-1}_{{\rm PB}, W}(k,k')_{\ell} \cdot \BEFT_{0,\,{\rm lin,\,} i}(k',\{\vec\Omega,\vec b_{NG}\}) \nn \\
& + \Pdata_{\ell}(k) \cdot C^{-1}_{{\rm PB}}(k,k')_{\ell} \cdot \BEFT_{0,\,{\rm lin,\, }i}(k',\{\vec\Omega,\vec b_{NG}\}), \\
G_{0} & = - \PEFT_{\ell,\,{\rm const}}(k,\{\vec\Omega,\vec b_{NG}\}) \cdot C^{-1}_{{\rm PB}, W}(k,k')_{\ell}\cdot ( \BEFT_{0,\,{\rm const}}(k',\{\vec\Omega,\vec b_{NG}\}) - \Bdata(k')) \nn \\
& + \Pdata_{\ell}(k) \cdot C^{-1}_{{\rm PB}}(k,k')_{\ell} \cdot (\BEFT_{0,\,{\rm const}}(k',\{\vec\Omega,\vec b_{NG}\}) - \Bdata(k')),
 \end{align}
and
\bea  
&&  {\cal L}_{\rm BB}(d|\{\vec\Omega,\vec b_G,\vec b_{NG}\}) = \\ \nn
&&\qquad = {\rm Exp}\left[  -\frac{1}{2}\sum_{ij}  b_{G,i} \, H_{2,ij}(\{\vec\Omega,\vec b_{NG}\}) b_{G,j} + \sum_i  b_{G,i} \, H_{1,i}(\{\vec\Omega,\vec b_{NG}\}) \, + H_{0}(\{\vec\Omega,\vec b_{NG}\}) \right]\ ,
\eea where
 \begin{align}
H_{2,ij}(\{\vec\Omega,\vec b_{NG}\}) & =\BEFT_{0,\,{\rm lin,\, } i}(k,\{\vec\Omega,\vec b_{NG}\}) \cdot C^{-1}_{{\rm BB}}(k,k') \cdot \BEFT_{0,\,{\rm lin,\, }j}(k',\{\vec\Omega,\vec b_{NG}\})\ , \\
H_{1,i}(\{\vec\Omega,\vec b_{NG}\}) & = -\BEFT_{0,\,{\rm lin, \,}i }(k,\{\vec\Omega,\vec b_{NG}\}) \cdot C^{-1}_{{\rm BB}}(k,k') \cdot (\BEFT_{0,\,{\rm const}}(k',\{\vec\Omega,\vec b_{NG}\}) - \Bdata(k'))\, , \\
H_{0}(\{\vec\Omega,\vec b_{NG}\}) & = - \frac{1}{2} (\BEFT_{0,\,{\rm const}}(k,\{\vec\Omega,\vec b_{NG}\}) -\Bdata(k))  \cdot C^{-1}_{{\rm BB}}(k,k') \cdot ( \BEFT_{0,\,{\rm const}}(k',\{\vec\Omega,\vec b_{NG}\}) - \Bdata(k')).
 \end{align}
 
Defining $T_a = F_a + G_a + H_a$, the marginalized likelihood of the power spectrum plus bispectrum is given by:
\begin{align}\label{eq;marginalizedbispLikli}
    &{\cal L}_{\rm FULL}(d|\{\vec\Omega, b_{NG}\}) = \int d b_G\; {\cal L}_{\rm FULL}(d|\{\vec\Omega,\vec b_G,\vec b_{NG}\})  \nn \\
    &= {\rm Exp}\left[\frac{1}{2} T_{1,i}(\{\vec\Omega,\vec b_{NG}\})\cdot T_{2}(\{\vec\Omega,\vec b_{NG}\})^{-1}{}_{ij} \cdot T_{1,j} (\{\vec\Omega,\vec b_{NG}\}) \right. \nn \\
    & \left. +  T_0(\{\vec\Omega,\vec b_{NG}\})-\frac{1}{2}\log\left[\det\left(T_2(\{\vec\Omega,\vec b_{NG}\})\right)\right]\right] ,
\end{align}
where we dropped an irrelevant constant.

We finish by commenting on the application of the window function to the bispectrum monopole.  As mentioned, unfortunately, we do not have at our disposal an exact formula for the application of the window function to the bispectrum, contrary to what we  have in eq.~(\ref{eq:windowzation}) for the power spectrum. Therefore, we use an approximate procedure that was used in~\cite{Gil-Marin:2014sta} and that we described above. It is unclear how accurate this procedure is, and, in order to limit the systematic error induced by this approximation, we start analyzing the data  for the bispectrum at a minimum wavenumber $k_{\rm min}=0.04\hinvMpc$. It would be preferable to have a more accurate way to implement the window function. We leave this to future work.

\section{Sound horizon at decoupling\label{app:sound_horizon}}

The sound horizon $r_d$ at decoupling at epoch $z_d$ reads:
\begin{equation}\label{eq:rdexpression}
    r_d = \int_{z_d}^\infty \frac{c_s(z)}{H(z)}dz, 
\end{equation}
where $c_s$ is the sound speed in the primordial photon-baryon plasma given by:
\begin{equation}
    c_s^2(z) =\frac{c^2}{3}\left[ 1+\frac{3}{4}\frac{\rho_b(z)}{\rho_\gamma(z)} \right]^{-1},
\end{equation}
where $\rho_b$ and $\rho_\gamma$ are the enery density of baryons and radiation respectively. The integral in~(\ref{eq:rdexpression}) can be evaluated exactly as:
\begin{equation}
    r_d = \frac{2c}{H_0 \sqrt{3 R \Omega_m}} \log \left[ \frac{\sqrt{1+z_d+R}+\sqrt{(1+z_d) R \tfrac{\Omega_\text{rad}}{\Omega_m}+R}}{\sqrt{1+z_d}\left(1+\sqrt{R \tfrac{\Omega_\text{rad}}{\Omega_m}}\right)} \right],
\end{equation}
where $R = \tfrac{3}{4}\tfrac{\Omega_b}{\Omega_\gamma}$, the normalized radiation density today is $\Omega_\text{rad} = \left[1+N_\text{ur}\tfrac{7}{8}(\tfrac{4}{11})^{4/3}\right]\Omega_\gamma h^2$ with $N_\text{ur}$ the number of ultra-relativistic species and $\Omega_\gamma h^2= 2.47282 \cdot 10^{-5}$, $H_0$ is, as usual, the Hubble constant at present time, and $c$ is the speed of light. 

When we use the Planck prior on $r_d$, we use the following value from~\cite{Aghanim:2018eyx}
Planck2018 TT,TE,EE+lowP+lensing (Table 2) to analyze the CMASS sample:
\be
r_d = 147.09 \, \pm \, 0.26 \ {\rm Mpc}, \qquad z_d = 1059.94 \, \pm \, 0.30,
\ee
while for the simulations we use instead $r_d$ computed at their fiducial cosmology, Table~\ref{tab:simulspec}, and use the same error.
In our analysis, we take $z_d$ at its fixed Planck fiducial value considering that for the cosmologies under probe, $r_d(z_d+\sigma (z_d))-r_d(z_d-\sigma (z_d)) < 0.2 \, \sigma (r_d)$, and so the effect is marginal.

\section{Correlation matrices\label{app:covariance}}

\begin{figure}[h!]
\centering
\includegraphics[width=0.99\textwidth,draft=false]{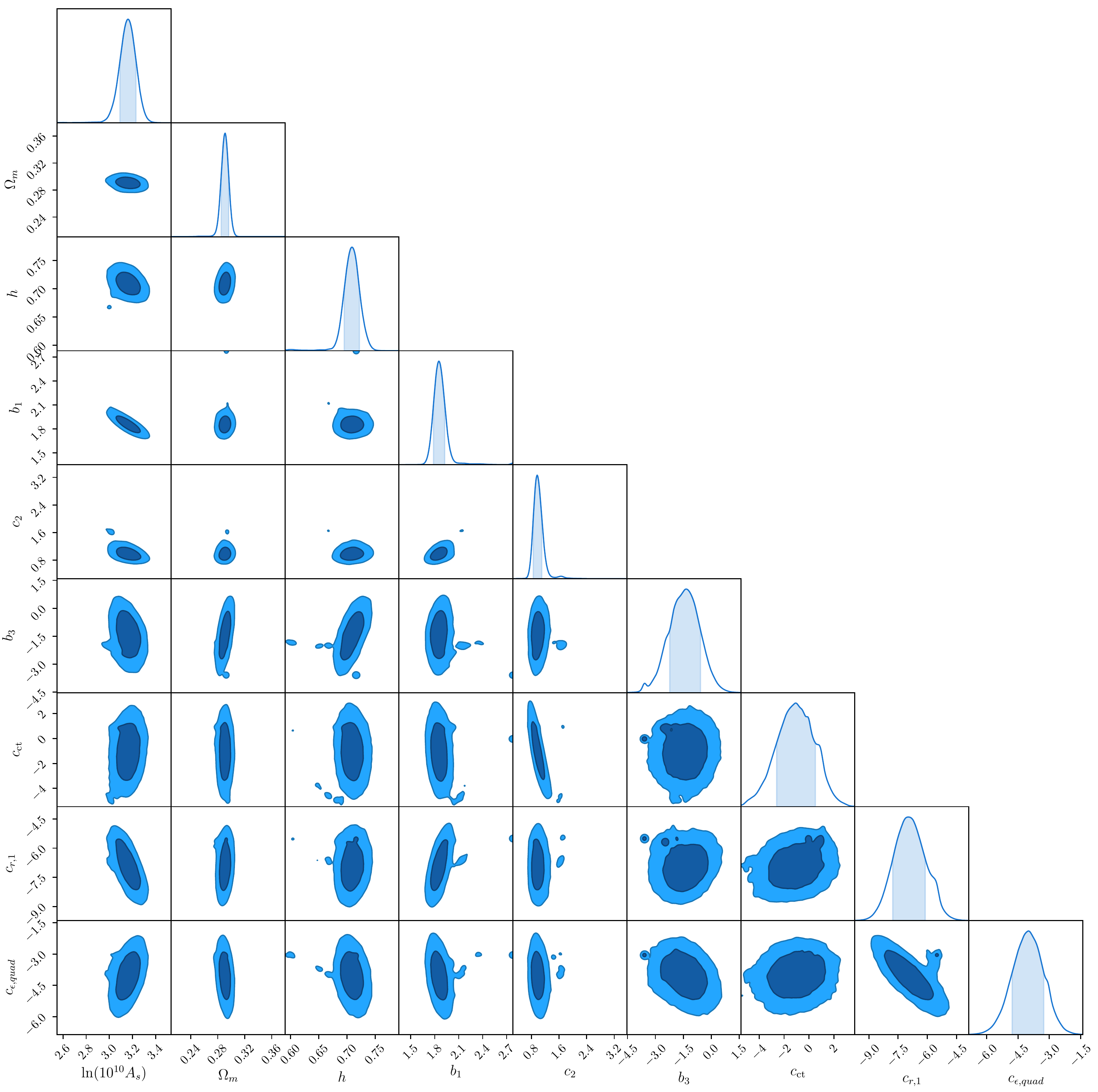}
\caption{
\small 2D posterior distributions of all cosmological and EFT parameters, obtained from fitting the power spectrum up to $\kmax = 0.20 \hinvMpc$ of the challenge box D with the non-marginalized likelihood.
 \label{fig:contournonmargChallengeD} }
\end{figure}

\begin{figure}[h!]
\centering
\includegraphics[width=0.99\textwidth,draft=false]{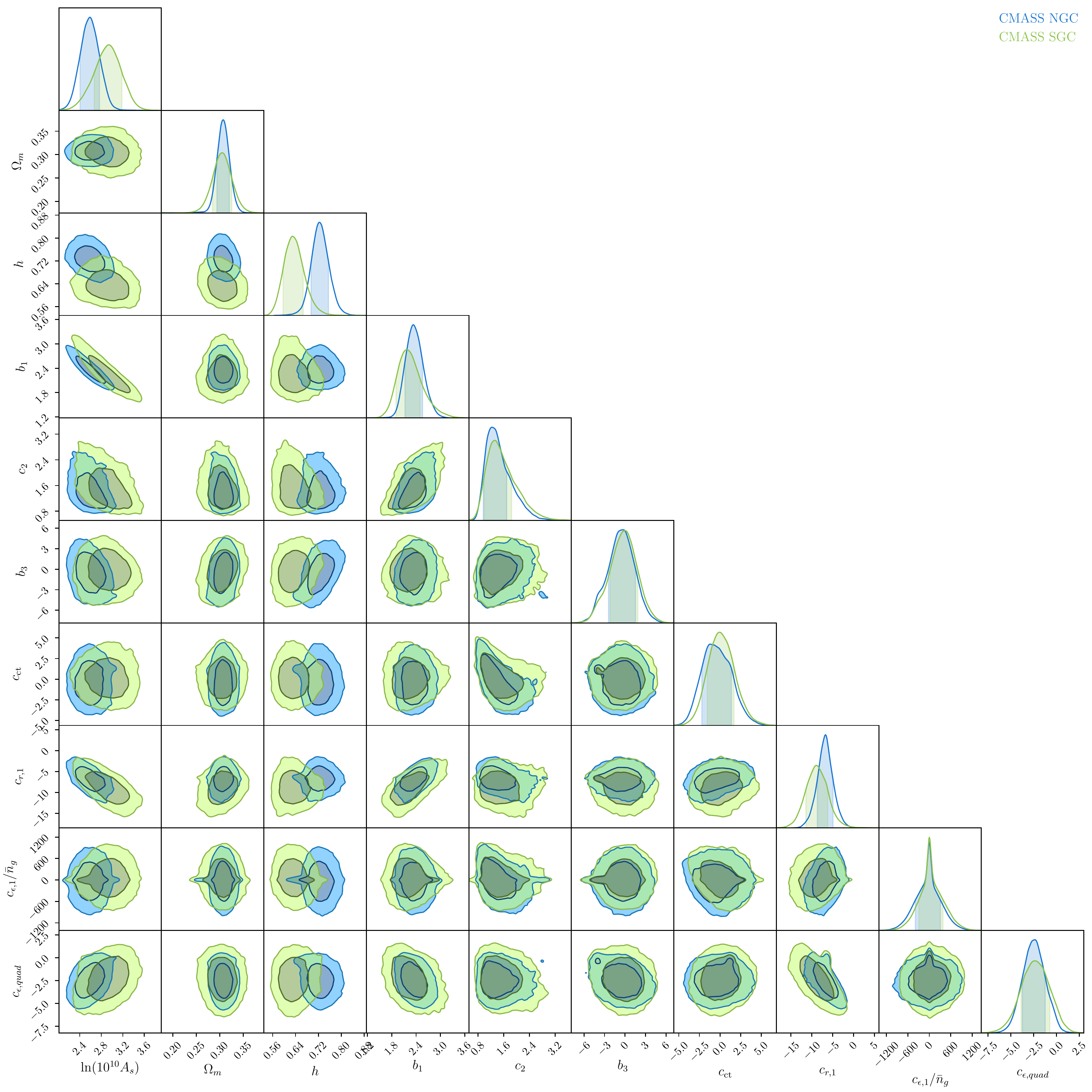}
\caption{\small 2D posterior distributions of all cosmological and EFT parameters, obtained from fitting the CMASS NGC and SGC power spectra independently up to $\kmax = 0.23 \hinvMpc$ with the non-marginalized likelihood. 
 \label{fig:contournonmargData} }
\end{figure}

{In Tables~\ref{tab:parameter_correlations_challengeD}  and~\ref{tab:parameter_correlations_data} we provide the correlation matrices for the parameters entering the likelihood obtained by evaluating the non-marginalized posterior of the power spectrum. We present results at $\kmax = 0.2 \hinvMpc$ for Challenge Box D, and for the CMASS data. In Fig. ~\ref{fig:contournonmargChallengeD} and~\ref{fig:contournonmargData} we plot the  2D posterior distributions for challenge D and for CMASS NGC and SGC.}

\begin{table}
    \centering
    \caption{Parameter Correlations Challenge D, $k_{\rm max}=0.20$ \label{tab:parameter_correlations_challengeD}}
   \begin{tabular}{c|ccccccccc}
         & $\ln (10^{10} A_s)$ & $\Omega_m$ & $h$ & $b_1$ & $c_2$ & $b_3$ & $c_{\rm ct}$ & $c_{r,1}$ & $c_{\epsilon,quad}$\\ 
        \hline
        $\ln (10^{10} A_s)$ &  1.00 & -0.10 & -0.10 & -0.89 & -0.80 & -0.01 &  0.09 & -0.57 &  0.19 \\ 
                 $\Omega_m$ & -0.10 &  1.00 &  0.34 & -0.14 & -0.02 &  0.40 & -0.10 &  0.05 & -0.25 \\ 
                        $h$ & -0.10 &  0.34 &  1.00 & -0.29 & -0.15 &  0.50 & -0.06 & -0.02 & -0.28 \\ 
                      $b_1$ & -0.89 & -0.14 & -0.29 &  1.00 &  0.82 & -0.13 & -0.06 &  0.53 & -0.06 \\ 
                      $c_2$ & -0.80 & -0.02 & -0.15 &  0.82 &  1.00 & -0.14 & -0.33 &  0.16 &  0.05 \\ 
                      $b_3$ & -0.01 &  0.40 &  0.50 & -0.13 & -0.14 &  1.00 &  0.03 &  0.07 & -0.33 \\ 
               $c_{\rm ct}$ &  0.09 & -0.10 & -0.06 & -0.06 & -0.33 &  0.03 &  1.00 &  0.25 &  0.15 \\ 
                  $c_{r,1}$ & -0.57 &  0.05 & -0.02 &  0.53 &  0.16 &  0.07 &  0.25 &  1.00 & -0.69 \\ 
        $c_{\epsilon,quad}$ &  0.19 & -0.25 & -0.28 & -0.06 &  0.05 & -0.33 &  0.15 & -0.69 &  1.00 \\ 
        \hline
    \end{tabular}
    \end{table}

 \begin{table}
    \centering
        \caption{Parameter Correlations CMASS NGC, $\kmax = 0.2 \hinvMpc$  \label{tab:parameter_correlations_data}}
    \begin{tabular}{c|cccccccccc}
         & $\ln (10^{10} A_s)$ & $\Omega_m$ & $h$ & $b_1$ & $c_2$ & $b_3$ & $c_{\rm ct}$ & $c_{r,1}$ & $c_{\epsilon,1}/\bar{n}_g$ & $c_{\epsilon,quad}$\\ 
        \hline
               $\ln (10^{10} A_s)$ &  1.00 &  0.08 & -0.23 & -0.90 & -0.46 & -0.18 &  0.12 & -0.48 &  0.10 &  0.30 \\ 
                        $\Omega_m$ &  0.08 &  1.00 & -0.10 & -0.13 & -0.18 &  0.28 &  0.05 &  0.04 & -0.05 & -0.01 \\ 
                               $h$ & -0.23 & -0.10 &  1.00 & -0.16 & -0.13 &  0.40 & -0.07 & -0.11 & -0.10 &  0.04 \\ 
                             $b_1$ & -0.90 & -0.13 & -0.16 &  1.00 &  0.57 &  0.04 & -0.10 &  0.49 & -0.10 & -0.31 \\ 
                             $c_2$ & -0.46 & -0.18 & -0.13 &  0.57 &  1.00 &  0.14 & -0.56 & -0.21 & -0.34 & -0.11 \\ 
                             $b_3$ & -0.18 &  0.28 &  0.40 &  0.04 &  0.14 &  1.00 &  0.02 & -0.10 & -0.01 & -0.06 \\ 
                      $c_{\rm ct}$ &  0.12 &  0.05 & -0.07 & -0.10 & -0.56 &  0.02 &  1.00 &  0.21 & -0.08 &  0.12 \\ 
                         $c_{r,1}$ & -0.48 &  0.04 & -0.11 &  0.49 & -0.21 & -0.10 &  0.21 &  1.00 &  0.30 & -0.67 \\ 
        $c_{\epsilon,1}/\bar{n}_g$ &  0.10 & -0.05 & -0.10 & -0.10 & -0.34 & -0.01 & -0.08 &  0.30 &  1.00 &  0.02 \\ 
               $c_{\epsilon,quad}$ &  0.30 & -0.01 &  0.04 & -0.31 & -0.11 & -0.06 &  0.12 & -0.67 &  0.02 &  1.00 \\ 
        \hline
    \end{tabular}
\end{table}

\section{Joint-EFT analysis of CMASS NGC$\times$SGC  \label{app:CMASSNGCSGCsameEFT}}
\begin{figure}[h!]
\centering
\includegraphics[width=0.496\textwidth,draft=false]{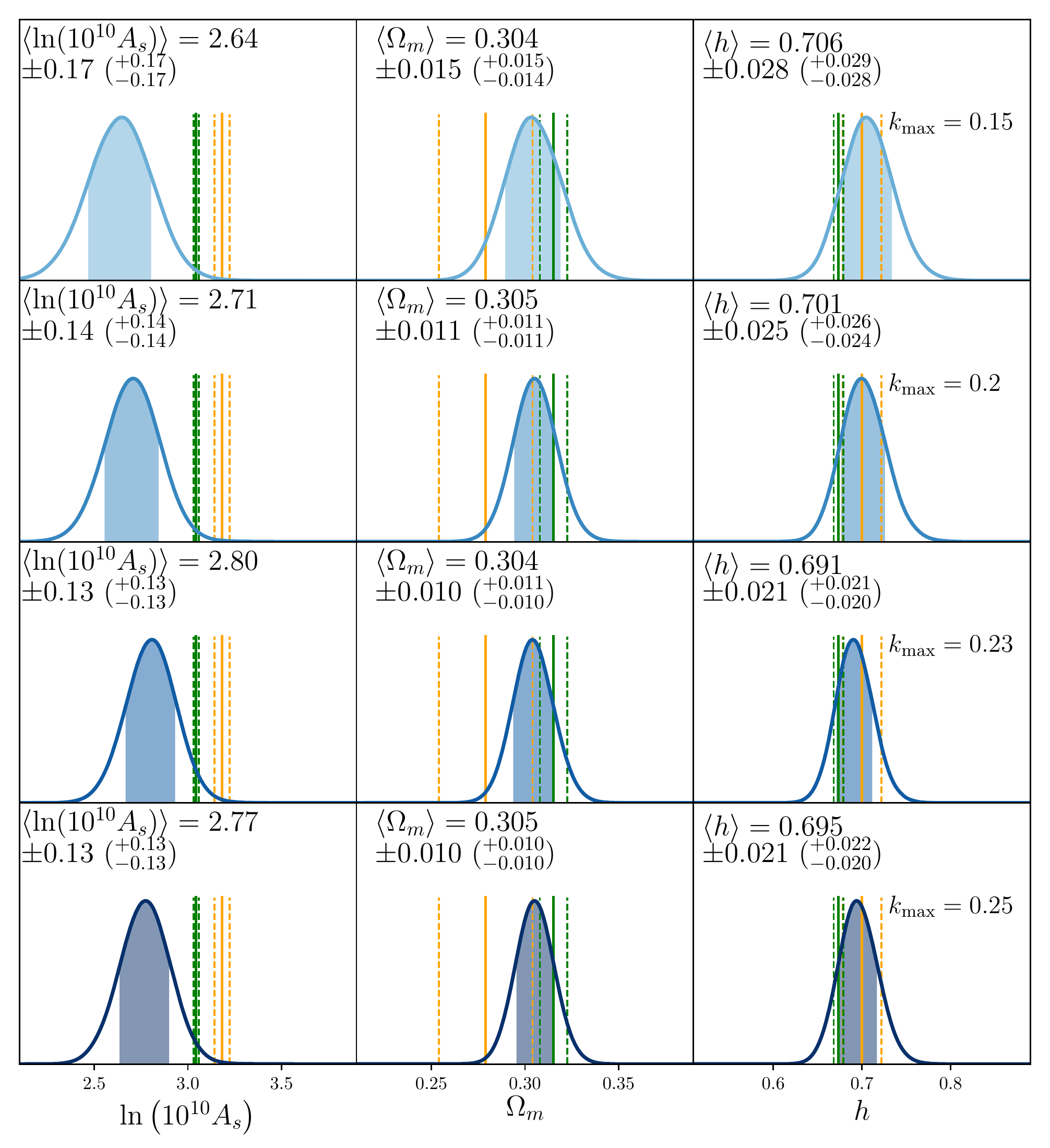}
\caption{\small  Marginalized posterior distributions of the cosmological parameters obtained from fitting the power spectrum of CMASS NGC and SGC power spectra with a same set of parameters but for the shot noises for the two skycuts. 
 \label{fig:posteriorJointEFT} }
\end{figure}

For our main results, when analyzing the combination CMASS NGC and SGC, we conservatively kept two independent set of EFT parameters between the two skycuts in the fit. This is the same choice as done by the BOSS collaboration to account for different selection effects~\cite{Beutler:2016arn}. Here, we provide 1D marginalized posterior distributions of the cosmological parameters in Fig.~\ref{fig:posteriorJointEFT} if we choose to fit with one set of EFT parameters but the shot noises for the whole CMASS sample. We find that the error bars on the cosmological parameters are roughly improved by $10\%$ at $\kmax = 0.23 \hinvMpc$ with respect to the former case, and the central values are shifting at most by about half a $\sigma_{\rm stat}$. The minimum $\chi^2 / {\rm d.o.f.}$ is $1.05$. From Fig.~\ref{fig:contournonmargData} we do not find evidence that the two samples NGC and SGC detect different EFT parameters, but we cautiously opt for the most conservative choice of two independent sets of parameters for our main results.

\bibliography{references}


\end{document}